\documentclass[a4paper,12pt]{article}
\usepackage{setspace}
\usepackage[utf8]{inputenc} 
\usepackage[backend=bibtex,giveninits=true,sorting=none,style=numeric-comp,doi=false,date=year]{biblatex}
\bibliography{references} 

\usepackage{amssymb}
\usepackage{authblk} 
\usepackage{tikz} 
\usepackage{hyperref,graphicx} 
\usepackage{amsmath} 
\usepackage{textcomp}
\usepackage{pgfplots}
\usepackage{caption}
\usepackage{multicol}
\usepackage{url}
\usepackage{breakurl}
\usepackage{todonotes}
\usepackage{subfigure}
\usepackage{cellspace}
\usepackage[a4paper,
            bindingoffset=0.2in,
            left=0.6in,
            right=0.6in,
            top=0.6in,
            bottom=0.6in,
            footskip=.25in]{geometry}

\usepackage{xcolor}
\usepackage{soul}

\usepackage[normalem]{ulem}

\newcommand{\del}[1]{\ifmmode\text{{\color{red}\sout{\ensuremath{#1}}}}\else{\color{red}\sout{#1}}\fi}

\title{\textbf{Nucleation patterns of polymer crystals analyzed by machine learning models}}

\author[1,2]{\textbf{Atmika Bhardwaj}}
\author[1,2]{\textbf{Jens-Uwe Sommer}}
\author[1]{\textbf{Marco Werner}}
\affil[1]{\textit{Institute Theory of Polymers, Leibniz-Institut für Polymerforschung Dresden e.V., Hohe Strasse 6, 01069 Dresden, Germany.
}}
\affil[2]{\textit{Institute of Theoretical Physics, Technische Universität Dresden, Zellescher Weg 17, 01069 Dresden, Germany.}}

\date{\color{black}\today}
\pgfplotsset{compat=1.15}

\begin{document} 
\maketitle

\begin{abstract}
We use machine learning algorithms to detect the crystalline phase in undercooled melts in molecular dynamics simulations.
Our classification method is based on local conformation and environmental fingerprints of individual monomers.
In particular, we employ self-supervised auto-encoders to compress the fingerprint information and a Gaussian mixture model to distinguish ordered states from disordered ones.
The resulting identification of crystalline monomers agrees to a large extent with human-defined classifiers such as the stem-length-based classification scheme as developed in our previous work [C. Luo and J.-U. Sommer, \textit{Macromolecules} \textbf{44} (2011), 1523], but does not require any foreknowledge about the structure of semi-crystalline polymers.
Because of its local sensitivity, the method allows the resolution of detailed time patterns of crystalline order before an apparent signature of the transition is visible in thermodynamic properties such as for the specific volume.
At a pre-transition point, we observe the highest crystallization efficiency using the fraction of monomers being conserved in the crystalline phase as compared to the number of monomers joining that phase.
\end{abstract}

\section{Introduction}

Crystallization in polymers is a longstanding challenging problem in polymer science and despite intense research over decades, no principal theoretical understanding has been achieved so far.
While crystallization in polymers takes a classical route of nucleation and growth~\cite{Rieger2003}, the constraints imposed by the chain connectivity in long polymers make the process far more complex than in low molecular weight materials since in addition to spatial order also conformational order is necessary.
This requires transitioning from the high-entropy coiled configuration into a low-entropy extended state, the equilibrium state of an ideal polymer crystal~\cite{sommer:piuopc:2007}.
Polymer materials are typically found in semi-crystalline states, characterized by the coexistence of amorphous and crystalline regions, where the crystalline parts comprise back-and-forth folded stems that are stacked into quasi-two-dimensional lamellar sheets with a thickness depending on the degree of supercooling~\cite{keller1957note, armitstead1992polymer, strobl1997physics} and thus correspond to non-equilibrium meta-stable states.
Predicting the crystallinity in polymers as well as resulting material properties is delicate~\cite{reiter::2003} as many factors, such as temperature, cooling rate, pressure, and entanglement state of the melt can influence it~\cite{xiao:m:2017}.
To explain the frozen-in non-equilibrium of semi-crystalline polymers kinetic models have been proposed which are based on simplified nucleation pathways~\cite{hoffman:jrnbssa:1961, hoffman:k:1969, sadler:prl:1986} which cannot reflect the full complexity of the non-equilibrium transition.
More recently, multi-stage nucleation pathways have been proposed~\cite{olmsted1998spinodal, strobl2000melt} which predict mesophases of a lower degree of order.
However, no direct evidence has been obtained for these transient states so far.

Computer simulations may help investigate the non-equilibrium nature of polymer crystallization since we can explore all molecular details of conformations, and access the trajectories of all monomers in their journey throughout the crystallization and melting processes.
These studies could, in particular, show the entanglement-controlled selectivity for early nuclei~\cite{CH_ent2014, luo:m:2016, luo:aml:2016, xiao:m:2017} which are also concluded from experimental studies~\cite{saalwachter:mcp:2023, wang:pnas:2023}.
Also, evidence for folding patterns has been obtained in advanced NMR-studies~\cite{hong:prl:2015,yuan:aml:2015} that are in agreement with entanglement-controlled length scales~\cite{luo:m:2016}.
On the other hand, the search for unique order parameters that characterize and classify the possibly short-living precursor states was despite some efforts~\cite{luo:m:2011} not yet successful.

Machine learning (ML) algorithms can potentially investigate nucleation pathways based on hidden conformation patterns that may have limited representation in existing concepts for the nucleation of chain-like molecules.
For low molecular weight materials, ML methods have already shown potential in identifying phases and transition states without human-defined order parameters in simulation data that allow for investigation of nucleation and crystal growth~\cite{Zheng2022, Christos2022}.
Coli and Dijkstra~\cite{coli2021artificial} investigated binary colloidal mixtures with two colloid sizes, where, at diameter ratios (small to large hard sphere) between $0.48$ and $0.62$, an interesting crystal structure called icosahedral AB$_{13}$ is stable; this structure is composed of large spheres (A) and clusters of 13 small spheres (B)~\cite{eldridge:n:1993}.
They showed that, through supervised learning, artificial neural networks can classify the phase to which each colloidal particle belongs based on local bond order fingerprints.
This approach helps distinguish the formation of the AB$_{13}$ crystal structure from the binary fluid state, as well as from the fcc crystal structure, even at the level of individual colloid particles, with high accuracy~\cite{coli2021artificial}.
Pereira investigated ML tools to build classification models to predict the likelihood of small organic molecules (new or existing) undergoing crystallization~\cite{pereira2020machine}.
Effects of the local structure on crystal growth can be seen in~\cite{freitas2020uncovering} for the cases of two distinct material families: semiconductors and metals.
Here, the authors combine molecular dynamics (MD) simulations and employ ML algorithms to demonstrate that the modified liquid structure near solid-liquid interfaces reduces liquid atom mobility, a key factor for constructing a predictive model that considers temperature-dependent growth rates~\cite{freitas2020uncovering}.
The structure of polymers requires a similar paradigm of fingerprinting a local state to account for chain connectivity as well.

In this study, we aim to characterize the environment of each monomer based on an unsupervised learning method to obtain a measure of crystallinity with monomer resolution.
To decide whether each monomer belongs to the crystalline or the amorphous phase, we consider both the local conformation along the backbone as well as its alignment with an eventually ordered environment.
In contrast to ad-hoc classification methods based on the picture of extended chain parts (stems) which occur with a very low probability in the molten state only, we do have to make any presumptions about the conformational state of the corresponding sequence.
As we will discuss, different emphases on the two aspects, local conformation, and alignment of the immediate environment influence the crystalline fraction and the shape of the resulting cluster surfaces.
The high local resolution for crystallinity is used in testing the pre-transition conditions, where clusters are small and where possible precursor states are anticipated~\cite{Rieger2003, strobl2000melt,xu2021concepts, zhang2016review}.

The rest of this work is organized as follows: in section~\ref{sec:md}, the MD simulation model and details of the cooling run for the coarse-grained polyvinyl alcohol model are described.
In section~\ref{sec:classification}, we introduce fingerprints for describing each monomer's environment as well as machine-learned-based classification methods of these fingerprints to attribute each monomer to a crystalline and amorphous phase.
In section~\ref{sec:results}, we evaluate the data-driven classifiers in comparison to stem-length and orientational order-based methods and analyze the exchange dynamics between crystalline and amorphous phases during the cooling.
Our results are summarized and discussed in section~\ref{sec:conclusion}.

\section{Molecular Dynamics (MD) Simulations}\label{sec:md}

We performed coarse-grained (CG) molecular dynamics (MD) simulations using the LAMMPS package~\cite{LAMMPS} to obtain a cooling trajectory from a polymer melt state.
The force field is closely aligned with the coarse-grained poly(vinyl alcohol) (CG-PVA) model as proposed earlier~\cite{meyer2002formation, meyer2001formation}.
We consider \(1000\) polymer chains each of which has \(1000\) (\(10^6\) total) monomers or CG beads~\cite{CF2011macro} and use a polymer melt state relaxed for $\sim 4\times 10^7$ MD steps in the context of a previous work~\cite{luo:aml:2016}.
The bonded interactions are harmonic (spring constant, $\mathrm{k}/2=1.0\times 10^3~\mathrm{k_BT/\sigma^2}$) and the non-bonded interactions are approximated by Lennard-Jones 9-6 (LJ96) potential.
We used a slightly reduced bond bead-bead spring constant as compared to $\mathrm{k}/2=1.352\times 10^3~\mathrm{k_BT/\sigma^2}$~\cite{vettorel2007, CH_lammps2009}.
Thus, we maintain the physical essence by fixing bond lengths near $0.5\sigma$, with a slightly increased standard deviation of $0.031\sigma$ compared to $0.027\sigma$, allowing for lower gradients in the MD integration.
Reduced units are employed for length as \(\sigma = 0.52~\mathrm{nm}\).
Throughout the paper, spatial dimensions are given in units of $\sigma$ if not noted otherwise.
The bond length kept close to \(0.5 ~(\widehat{=}~0.26~\mathrm{nm})\).
The reduced temperature of \(T = 1\) corresponds to \(550 ~ \mathrm{K}\)~\cite{CH_lammps2009}.
Time is estimated as \(3.5~\mathrm{ps}\) from Rouse relaxation time.
The time step of (MD) integration is \(0.01~(\widehat{=}~ 35~\mathrm{fs})\).
We employ periodic boundary conditions within the NPT ensemble at 1 atm pressure, utilizing a Berendsen barostat and Nosé-Hoover thermostat.
Continuous cooling is performed from \(T = 0.9~(\widehat{=}~495 ~\mathrm{K})\) to \(T = 0.75~(\widehat{=}~412.5 ~\mathrm{K})\) during \(7.5 \times 10^7\) MD steps \((\widehat{=}~2.6~\mathrm{\mu s})\).

\begin{figure}
    \begin{center}
        \includegraphics[width=0.8\hsize]{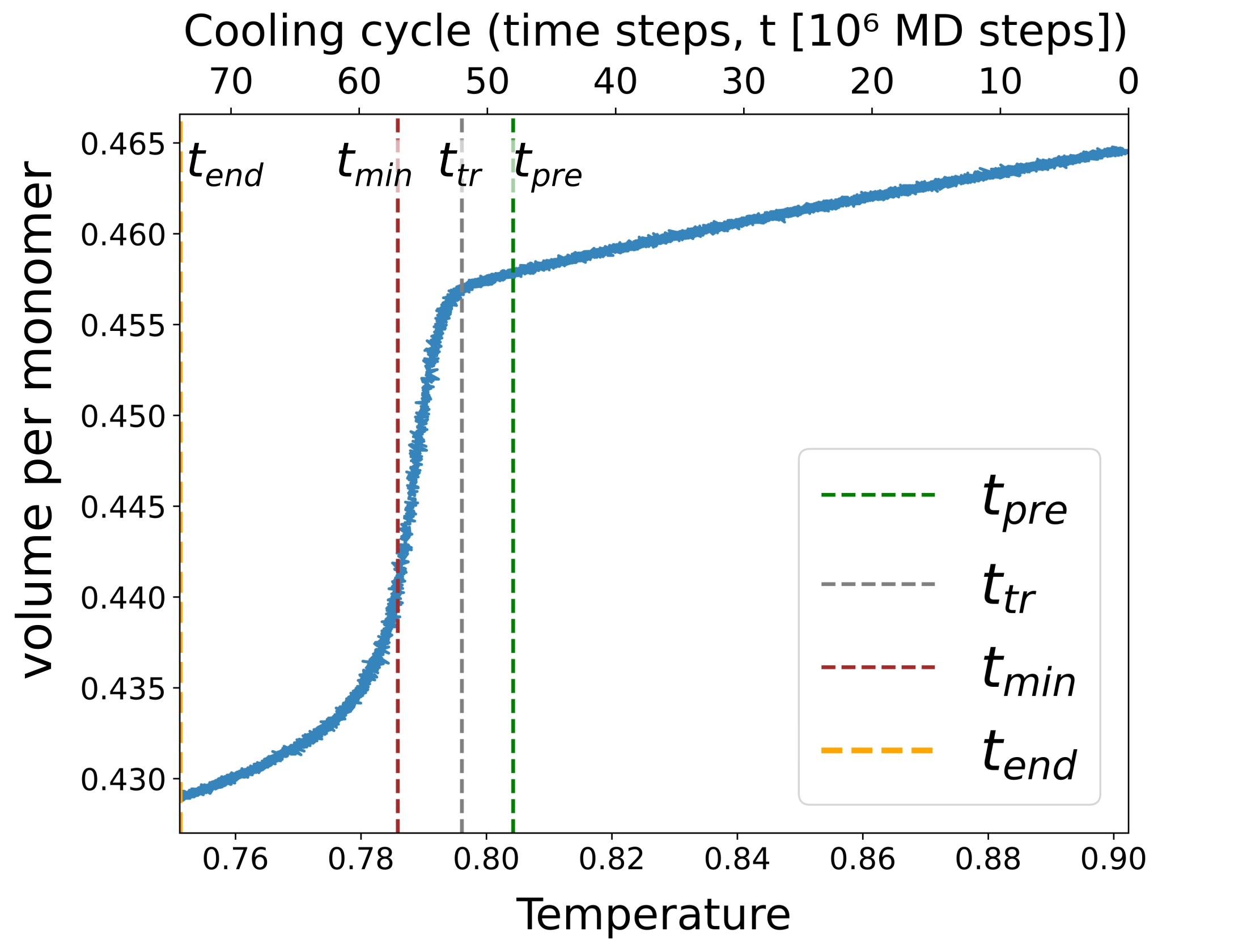}
    \end{center}
    \caption{We analyze the NPT cooling cycle of a coarse-grained model of poly(vinyl alcohol), CG-PVA, where the temperature decreases from 0.9 to 0.75 (in Lennard-Jones units), and each time step corresponds to $10^6$ molecular dynamics steps.
    The specific volume on the y-axis is determined by $L^3/10^6$.} \label{fig:tempvstime}
\end{figure}

In figure~\ref{fig:tempvstime}, we show the specific volume, $v$, of the monomer during the cooling cycle.
The four characteristic times are labeled there and are summarized in table~\ref{tab:times}.
At the time $t_{tr}$, the specific volume shows a turning point from a linear decrease with respect to temperature (constant expansion coefficient) into a flank of rapid decrease with time that indicates a beginning of the transition into a semi-crystalline phase.
As we will show in section~\ref{sec:cooling}, an earlier point, $t_{pre}$, is characteristic for the beginning stabilization of early nuclei.
In the same section section~\ref{sec:cooling}, we will discuss a time point, $t_{min}$, where a qualitative change of the exchange dynamics of monomers between crystalline and amorphous phases is observed before the long-term growth/ripening dominates.

\begin{table}
    \begin{tabular}{ | Sc | Sc | Sc | Sc | }
	\hline 
        \textbf{Symbol} & \textbf{Time point, t}[$10^6$ MD steps] & \textbf{Temperature} & \textbf{Description} \\
        \hline
        $t_{pre}$ & 48 & 0.805 & Beginning of stable early nuclei\\
        $t_{tr}$  & 52 & 0.797 & Apparent transition point\\
        $t_{min}$ & 57 & 0.787 & Crossover to ripening \\
        $t_{end}$ & 74 & 0.752 & End of simulation\\
	\hline
    \end{tabular}
    \caption{Overview of characteristic reference time points.}\label{tab:times}
\end{table}

\section{Classification Methods}\label{sec:classification}

\subsection{Structural fingerprints (SF)}\label{sec:parameters}

In polymers, crystallization is characterized by the local alignment of polymer chains into a conformational ground state, such as an all-trans or helical structure.
The patterns of crystal ordering result in distinctive signatures, both in the local arrangement of monomers from different chains, such as the lateral packing of aligned stems, and in the conformation of individual chain contours, which can be either stretched out as stems and/or folded.
When picking out a single monomer, its environment and contour information can be relevant to decide if it belongs to a crystal domain or the amorphous phase of the semi-crystal.
This motivates the creation of a local monomer-based fingerprint that includes alignment information concerning the environment and the neighboring contour, while the decision boundary remains open for a machine-learned classification scheme.

\begin{figure}[!ht]
    \centering
    \subfigure[]{
        \includegraphics[width=0.49 \hsize]{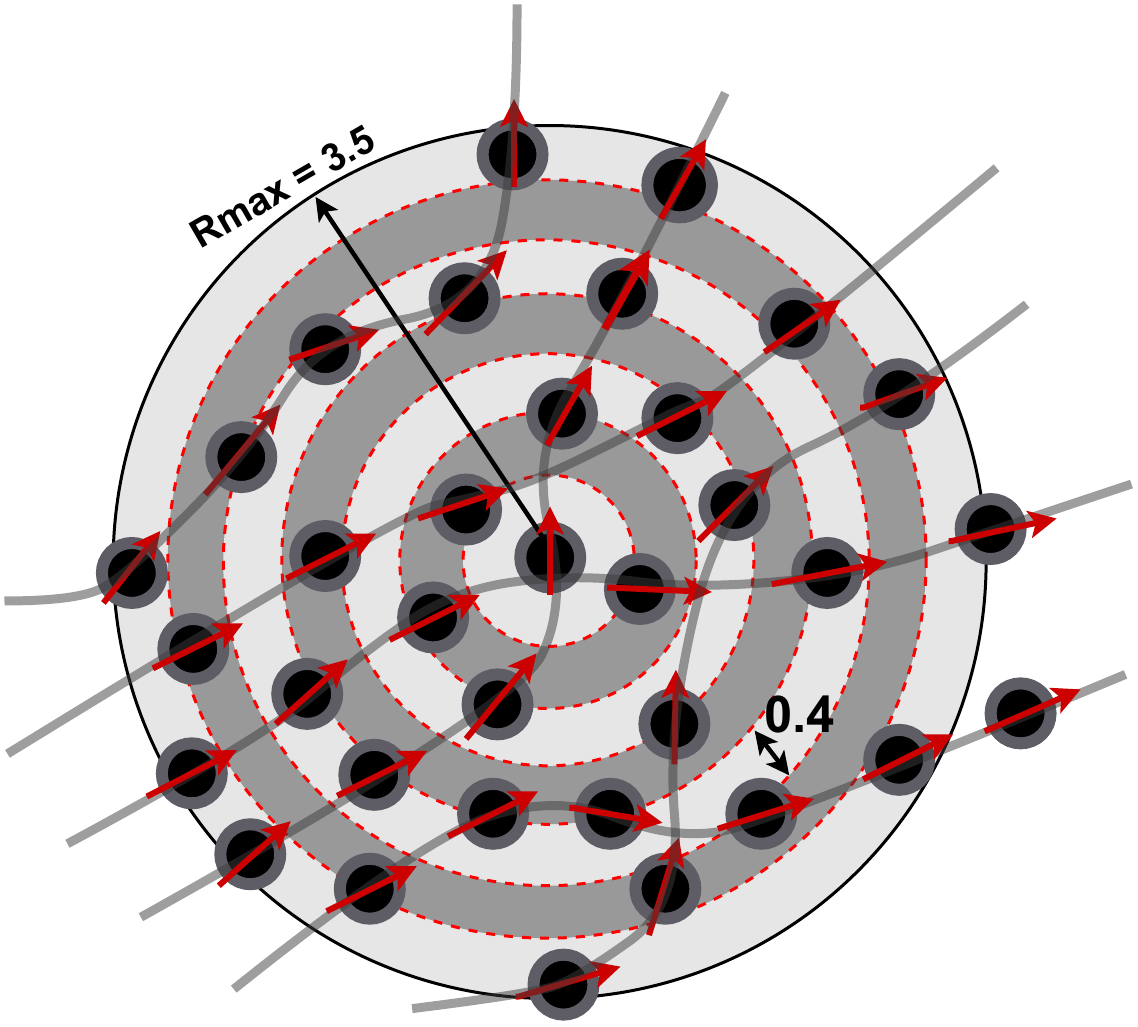}
        \label{fig:Avg_orientation}
        }
    \subfigure[]{
        \includegraphics[width=0.45 \hsize]{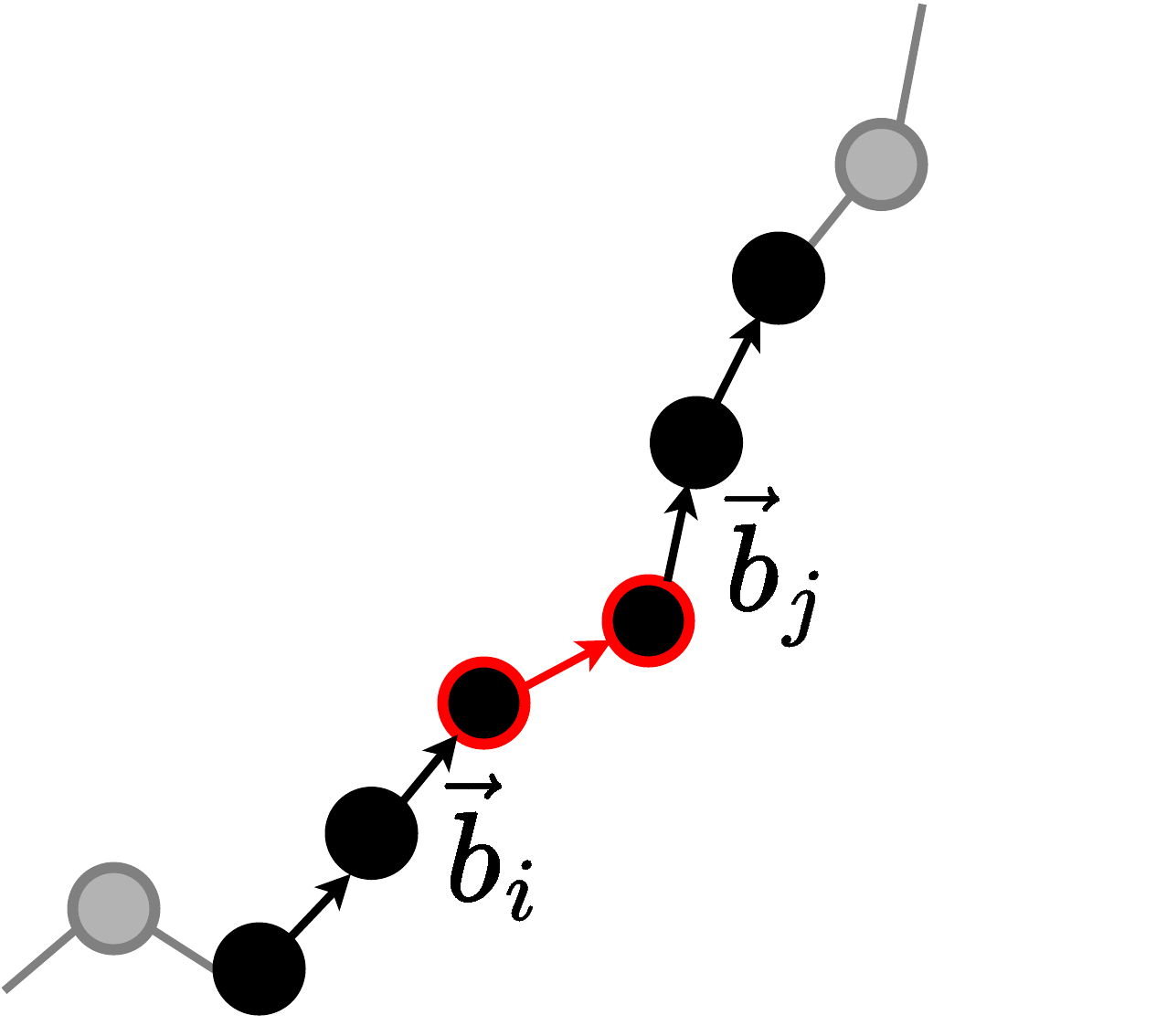}
        \label{fig:Conformation1}
        }
    \caption{(a) Mean alignment ($\langle\cos^2\Theta_{ij}\rangle_j$; $i$ = bond vector of the reference monomer, $j$ = bond vectors of monomers in a shell, $\cos\Theta_{ij}=(\vec b_i \cdot \vec b_j)~/~(\|\vec b_i\|~\|\vec b_j\|)$) of the reference bond vector with the neighboring bond vectors present in the next seven concentric shells.
    We get an array of length $7$ for each monomer.
    (b) Arrays of all possible scalar products between two bond vectors in the local conformation, capturing alignment along the contour, are represented as $\cos^2\Theta_{ij}$; $i \neq j$.
    This gives an array of length $10$ corresponding to each monomer.}\label{fig:X}
\end{figure}

To define a structural fingerprint for each monomer, we refer to the alignment between groups of bond vectors found in its vicinity.
In particular, we associate each monomer $i$ along a chain with the bond vector $\vec b_i$ to the succeeding neighbor monomer $i+1$ along the chain contour.
For this reference vector $\vec b_i$, we define two groups of other bond vectors $j$ as shown in the figure~\ref{fig:X}.
The first group considers all bond vectors within a spherical environment of radius $R_{max}$ around the monomer.
The second group is defined along the chain contour by the neighboring monomers within indices $i-n_{max},\dots, i,\dots i + n_{max}$, where $i$ represents the index of the monomer being considered.
To identify crystalline domains, it is beneficial to have fingerprint information defined for all monomers throughout the space.
Hence, for incomplete groups at the chain ends, we define them as replicas of adjacent groups near the chain end.
We consider the effect negligible, with a range of at most $0.3\%$ over- or under-representation of disordered conformations in the resulting fingerprint.
The assessed neighborhood can be further expanded/contracted upon variation of $R_{max}$ and $n_{max}$, respectively.
There is a compromise between keeping it small to emphasize essential local details and yet not too small which might lead to statistically insignificant local patterns due to the limited number of monomers involved.

\textbf{Local environment fingerprint, $\vec X_{env}$}: 
For the environment group, we define circular shells around the reference monomer and identify all bond vectors $j$ positioned within each shell.
For each shell $K$, we calculate the component
\begin{equation} \label{eqn:Xenv}
X_{env,~K}=\langle \cos^2 \Theta_{ij}\rangle~_{j\in K}~~,
\end{equation}
of a fingerprint vector, ${\vec X}_{env}=\{X_{env,~K}\}$, where, $\cos(\Theta_{ij})=(\vec b_i \cdot \vec b_j)~/~(\|\vec b_i\|~\|\vec b_j\|)$ is given by the angle $\Theta_{ij}$ between the reference bond vector, $\vec b_i$, and the bond vectors attributed to monomers $j$ that are located in $K$.
Here, average $\langle \rangle$ runs for all monomer indices $j$ in shell $K$.

We use the pair-correlation information to decide the value of $R_{max}$ by the inclusion of the third correlation shell (see appendix figure~\ref{fig:rdf}).
A radius \(R_{max} = 3.5\sigma\) (\(\widehat{=} ~ 1.82~\mathrm{nm}\)) around each reference monomer is considered with the thickness of each concentric shell as \(\Delta r = 0.4\) (\(\widehat{=}~0.26 ~\mathrm{nm}\)) as shown in figure~\ref{fig:Avg_orientation}.
The bin size $\Delta r$ is chosen such that the peaks in the radial distribution are barely resolved (see appendix figure~\ref{fig:rdf}).

For the selected setup, $\Delta r=0.4$ and $R_{max}=3.5$, we obtain a total of $7$ entries.
Here, we excluded the first shell from $R=0$ to $R=0.4$, which only contains the monomer itself.
The entries $X_{env,~K}$ may take values between $0$ (perpendicular alignment), and $1$ (parallel alignment), and a random distribution of $\vec b_j$ would approach $1/2$ in the limit of large bond vector count.

\textbf{Local conformation fingerprint, $\vec X_{conf}$}:
The local chain conformations are taken into account by choosing a neighborhood of a group of six consecutive monomers, i.e. five consecutive bond vectors $b_{i-n_{max}}\dots b_{i+n_{max}}$, here, $i$ labels the bond vector attributed to the considered monomer $i$.
Then, we look for all the bond angles formed by different combinations of the bond vectors, taking two at a time.
A conformation-related fingerprint,
\begin{equation}\label{eqn:Xconf}
X_{conf,~jk} = \cos^2 {\Theta_{jk}}~~,
\end{equation}
for each combination $(j,k)$, where, $i-n_{max}\leq j < i+n_{max} $ and $j < k \leq i+n_{max}$ is calculated based on the squared cosine angles between bond vectors $j$ and $k$.

For the considered case, $n_{max}=2$, the $ {\vec X}_{conf} =\{X_{conf,jk}\}$ has 10 entries.
Figure~\ref{fig:Conformation1} illustrates the set of five consecutive bonds, two immediate neighbors both at the front and the back, of the reference bond vector.

A \textbf{combined structural fingerprint (SF)} is defined as a concatenated vector from both groups:
\begin{equation}\label{eqn:X}
    \vec X = \left({\vec X}_{conf},{\vec X}_{env}\right)~~.
\end{equation}

The fingerprint vector $\vec X$ of every monomer has a dimension of $17$ and will be referred to as the set of structural fingerprints (SF) throughout the paper.
A matrix, $\hat X_{train}=\{\vec X_1^T,\vec X_2^T,\dots\}$ (with "T" denoting the transpose of the column vectors), is formed by arranging this information row-wise for all training examples.

\subsection{AutoEncoder (AE)}\label{sec:ae_description}
\begin{figure}[!ht]
    \begin{center}
    \includegraphics[width= 0.8\hsize]{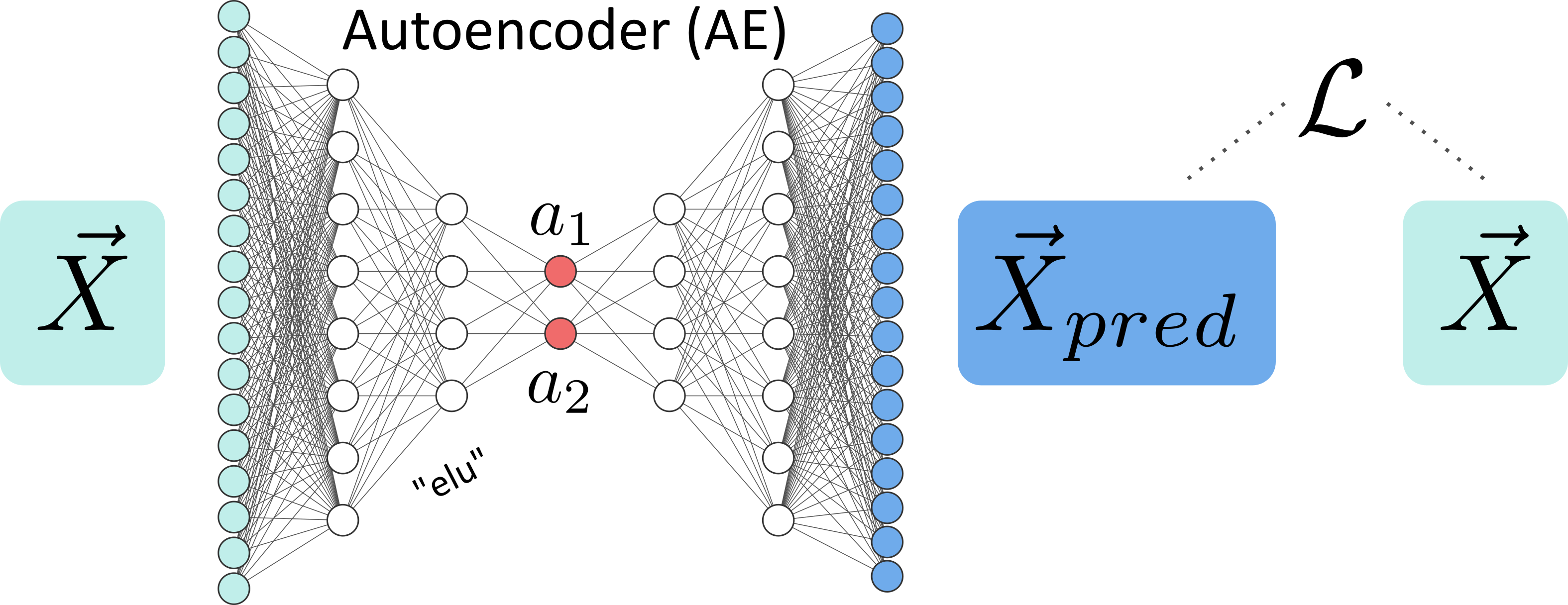}
    \end{center}
    \caption{A schematic diagram of an autoencoder (AE).
    It consists of an encoder section that encodes the input data, \(\vec X\), into compressed latent space data ($\vec a$) and a decoder section that decompresses the latent space data to regenerate the input.
    This means that \(\vec X\) is taken in for self-supervised learning and the crucial information is stored in the two bottle-neck neurons.
    This crucial information is then used to regenerate \(\vec X\).
    The generated output, \(\vec X_{pred}\), tries to match \(\vec X\) through several iterations.
    The number of nodes per layer deviates for illustration purposes (see table~\ref{tab:ae_hyper}).} \label{fig:Autoencoder}
\end{figure}

We use an artificial neural network in the form of an autoencoder~\cite{rumelhart:n:1986,michelucci::2022} (AE) to compress SF to a required dimensional array that we call the latent space vector, $\vec a$ (figure~\ref{fig:Autoencoder}).
After forward passing a new fingerprint example through a trained AE, the latent space vector is expected to contain the crucial information that the model has used to reproduce the input at the output $\vec X_{pred}$.
We qualify the reconstruction error by the loss,
\begin{equation}
\mathcal{L}(\vec X_{pred},\vec X)= \frac{1}{N}\sum(\vec X - \vec X_{pred})^2~~,
\label{eq:loss}
\end{equation}
which is employed for optimization via backpropagation.
Here, $N$ refers to the number of training examples.

We provide the input data $\vec X$ at the $17$ neurons of the input layer and use $\sim 10^4$ training examples ($N$) for the learning.
As input (and target) values we use the components $X_i$ of the vector $\vec X$ remapped to the interval $[-1/2,1]$ as reflected by the second Legendre polynomial $P_2(X_i) = \frac{1}{2}(3 X_i -1)$.
We use a variable learning rate while training with a patience monitor on the validation loss such that after 25 epochs of the non-decreasing validation loss, the learning rate decreases by \(0.5\) of its current value.
In figure~\ref{fig:AE_training}, in the appendix, we show the training and validation losses for the AE.
The table~\ref{tab:ae_hyper} (appendix) contains a list of the hyperparameters used during our model training.
Once the model is trained, one can easily use it to predict the new unseen dataset.

\subsection{Hierarchical clustering (HC)}
The hierarchical clustering~\cite{murtagh:wdmkd:2017} is an unsupervised technique using the agglomerative clustering approach (bottom-up).
In the beginning, each data point is treated as a separate cluster.
Then, the nearest data point(s) are identified and the corresponding clusters are merged.
This step is performed as one moves up the hierarchy until the formation of the desired number of clusters is achieved.
This implies that one can access multiple partitions, each linked to a specific level within the hierarchy, along with its corresponding dendrogram representation.
The HC method used in this work finds two clusters with 'Euclidean' linkage and 'ward', also known as the minimum variance method, as a minimizing linkage criterion between the sets of observations.
The minimum variance method generates clusters that meet both compactness and isolation criteria \cite{murtagh:wdmkd:2017}.
So, it separates the data points based on their density in space.

We directly input the SF for $N$ training examples, that is a $17*N$ -dimensional matrix, $\hat X_{train}$, into the HC classifier to predict the binary labels.
We use these labels as color labels and assign each of them to the compressed versions of the $N$ examples, denoted as $\hat a_{train}$, for visualizing the decision boundary, which is independent of the compressed space.

\subsection{Gaussian mixture model (GMM)}
Complex data distribution can be represented as a combination of multiple Gaussian distributions~\cite{day1969estimating}.
The key idea behind a Gaussian mixture model (GMM) is to estimate the parameters of each Gaussian distribution also known as a component, i.e. its mean and standard deviation, and the weights assigned to each component.
These weights represent the relative importance or contribution of each component to the overall distribution.
By finding the optimal values for these parameters, the model can represent the underlying data distribution~\cite{mclachlan1988mixture, astonpr373}.

We employ a GMM to assign a set of AE-compressed fingerprints $\vec a$ to two distributions, which, as we will demonstrate, represent ordered and disordered states.
We utilize the GMM implementation in the sklearn library (scikit-learn 1.2.2) using the 'full' mode.
Thereby, we suppose the point cloud to be composed of two Gaussian distributions with full freedom in the orientation and asphericity of their ellipsoidal shape as given by four covariance matrix components for each distribution~\cite{scikit-learn}.

\subsection{Multilayer perceptron (MLP)} \label{sec:mlp_description}
The labels defined at the lowest temperature ($T = 0.75, ~ t_{end} = 74$), see figure~\ref{fig:tempvstime}, are based on the HC algorithm applied to the $\hat X_{train}$.
To reuse the resulting decision boundary to conformations at different temperatures (time steps), we train an MLP taking into account the SF matrix, $\hat X_{train}$, as an input, and the HC-based labels as a training target.
Appended table~\ref{tab:mlp_hyper} contains a list of all the hyperparameters used during its training for this work.

\subsection{Labelling Workflow} \label{sec:workflow}
We start by reading the coordinate files for a cooling run of a polymer melt towards a semi-crystalline state as obtained using MD simulations described in section \ref{sec:md}.
The simulation box is divided along the z-axis into 100 slices of equal thickness.
We use the SF of $N$ monomers ($\sim 10^4$) contained in the 15th slice ($0.14L_z < z \leq 0.15 L_z$) of the simulation box at the last cooling time step $t_{end} = 74$ (lowest $T = 0.75$) to compose a training data set, $\hat X_{train}$.
The rationale for selecting this slice is discussed in section~\ref{sec:choiceoftrainingdataset}, with a comparison to randomly selecting monomers throughout the box.
Likewise, the SF for all monomers at all time steps is computed to evaluate and apply the models after training.
We denote the complete, time-dependent fingerprint as $\vec X_{all}(t)$.

\begin{figure}[!ht]
    \begin{center}
    \includegraphics[width=\hsize]{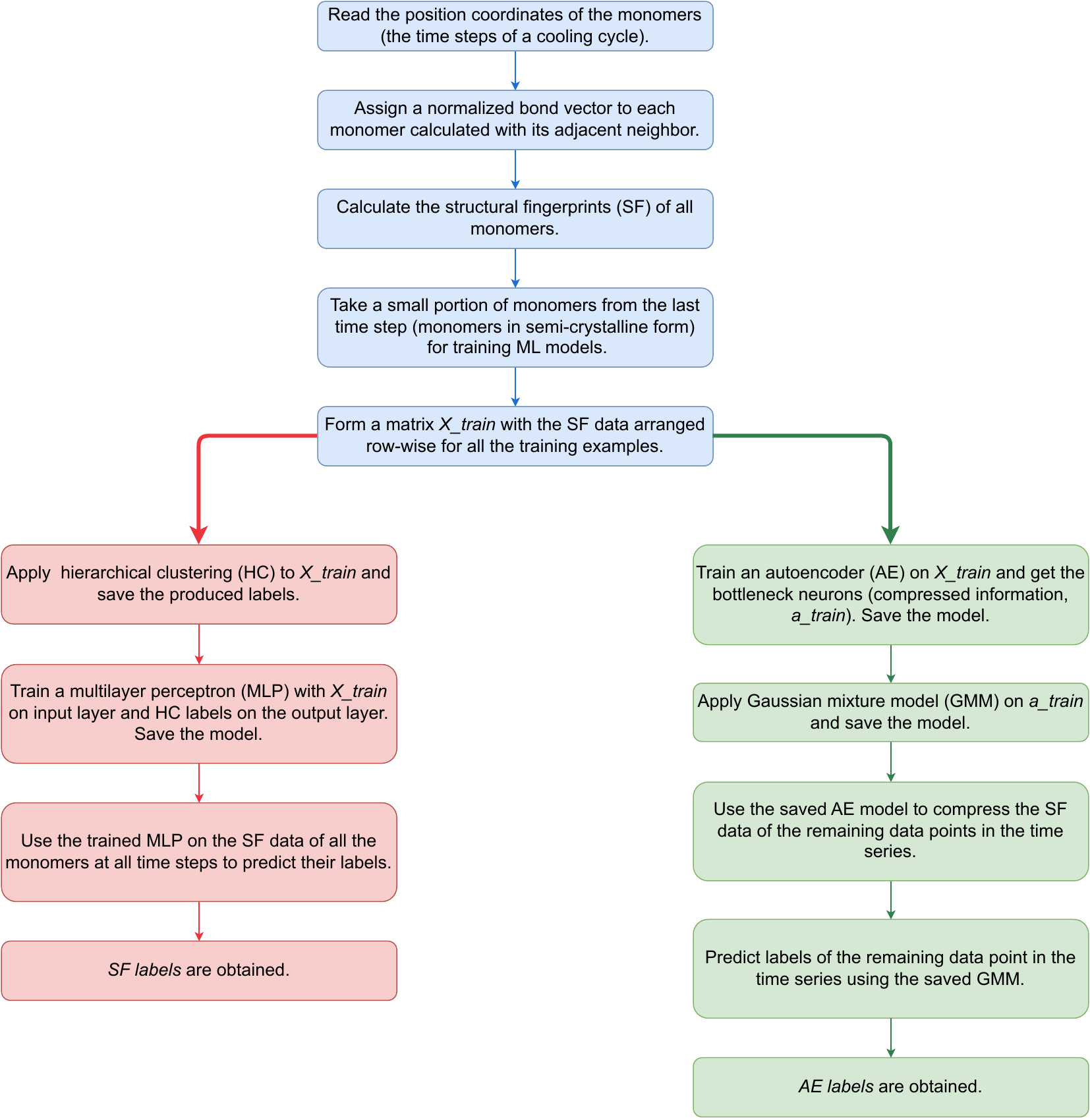}
    \end{center}
    \caption{A schematic diagram illustrating the steps involved in training and making predictions using the machine learning models.
    For the first step, hierarchical clustering (HC) can be applied directly to the concatenated list of structural fingerprints (SF): red-colored boxes.
    Additionally, the SF data can be compressed to obtain encoded (AE) data, and then the Gaussian mixture model (GMM) is applied exclusively to the most critical (compressed) information: green-colored boxes.
    The blue-colored boxes contain the steps that are common to both algorithms.} \label{fig:methods}
\end{figure}
We apply the HC algorithm on $\hat X_{train}$ and the GMM on its compressed version, $\hat a_{train}$, that defines a set of latent vectors $\vec a$ for each data example in $\hat X_{train}$ forwarded through the AE that was trained on $\hat X_{train}$.
This way, we obtain two sets of labels that we denote as \textit{SF labels} and \textit{AE labels}, respectively.

To attribute \textit{SF} and \textit{AE labels} to all monomers during the time corresponding to the dataset $\vec X_{all}(t)$, we need to transfer the effective decision boundaries discovered in the point clouds $\hat X_{train}$ and $\hat a_{train}$ to the unseen full and latent fingerprints in $\vec X_{all}(t)$ and $\vec a_{all}(t)$, respectively.
For this purpose, we train an MLP with $17$ input neurons to predict the class based on the $17$-dimensional full SF.
The MLP is trained using $\hat X_{train}$ and then employed to predict labels for all rows in $\vec X_{all}(t)$.
The decision boundary of sigmoid output is taken as $1/2$.
To pass the GMM decision boundary, the "GaussianMixture" of sklearn~\cite{scikit-learn} provides us with a "predict" function to predict the labels of $\vec a_{all}(t)$ based on the model used to estimate the labels of $\hat a_{train}$.
For the predicted labels throughout time (temperature), we stick to the notation \textit{SF labels} and \textit{AE labels} for simplicity.

\subsection{Stem length (SL) classification strategy}\label{sec:stem_len}

A human-based classification approach used in previous work is based on the local arrangement of monomers along the contour.
For this, one looks for the number of consecutive parallelly aligned or trans-trans (tt) monomers.
To check the alignment state of a monomer, one takes its immediate back and front neighbors and the two bond vectors associated with them.
If the bond angle between these consecutive bond vectors is greater than \(\Theta \geq 150^{\circ}\), then the central monomer is said to be in a tt-state.
The number of successive tt-state monomers, \(d_{tt}\), describes stems of variable lengths.
Based on the distribution of $d_{tt}$ in the semi-crystalline state, a monomer is defined to belong to the crystalline phase, if its respective stem has a length $d_{tt}\geq15$~\cite{CF2011macro}.

\subsection{Average orientation classification strategy (AO)}\label{sec:avg_orientation}
A human-designed alternative for classifying the locally ordered state can be based on the orientational order of a local environment.
Here, we consider the alignment of each monomer with all other bond vectors within the radius $R_{max}=3.5$ (\(\widehat{=}~1.82~\mathrm{nm}\)) identical with the radius taken into account for $\vec X_{env}$ (equation~(\ref{eqn:Xenv})).
For this purpose, we take the average $\langle \cos^2 \Theta_{ij}\rangle_j$ of the angles $\Theta_{ij}$ between the bond vector of the reference monomer $i$ with all the surrounding bond vectors $j$ within the cutoff distance.
This is similar to the local environment fingerprint calculation, just without further dividing the sphere around each monomer into concentric shells of equal thickness.
Hence, we apply equation~(\ref{eqn:Xenv}) without the subdivision into groups $K$ and get an orientation parameter associated with each monomer.
Based on the distribution of order parameters in the semi-crystalline state, we define a classification boundary manually and assign the more ordered monomers with $\langle\cos^2 \Theta_{ij}\rangle_j>0.6$ defined as crystalline, and others as amorphous.

\begin{figure}[!ht]
    \centering
    \subfigure[]{
        \includegraphics[width=0.47\hsize]{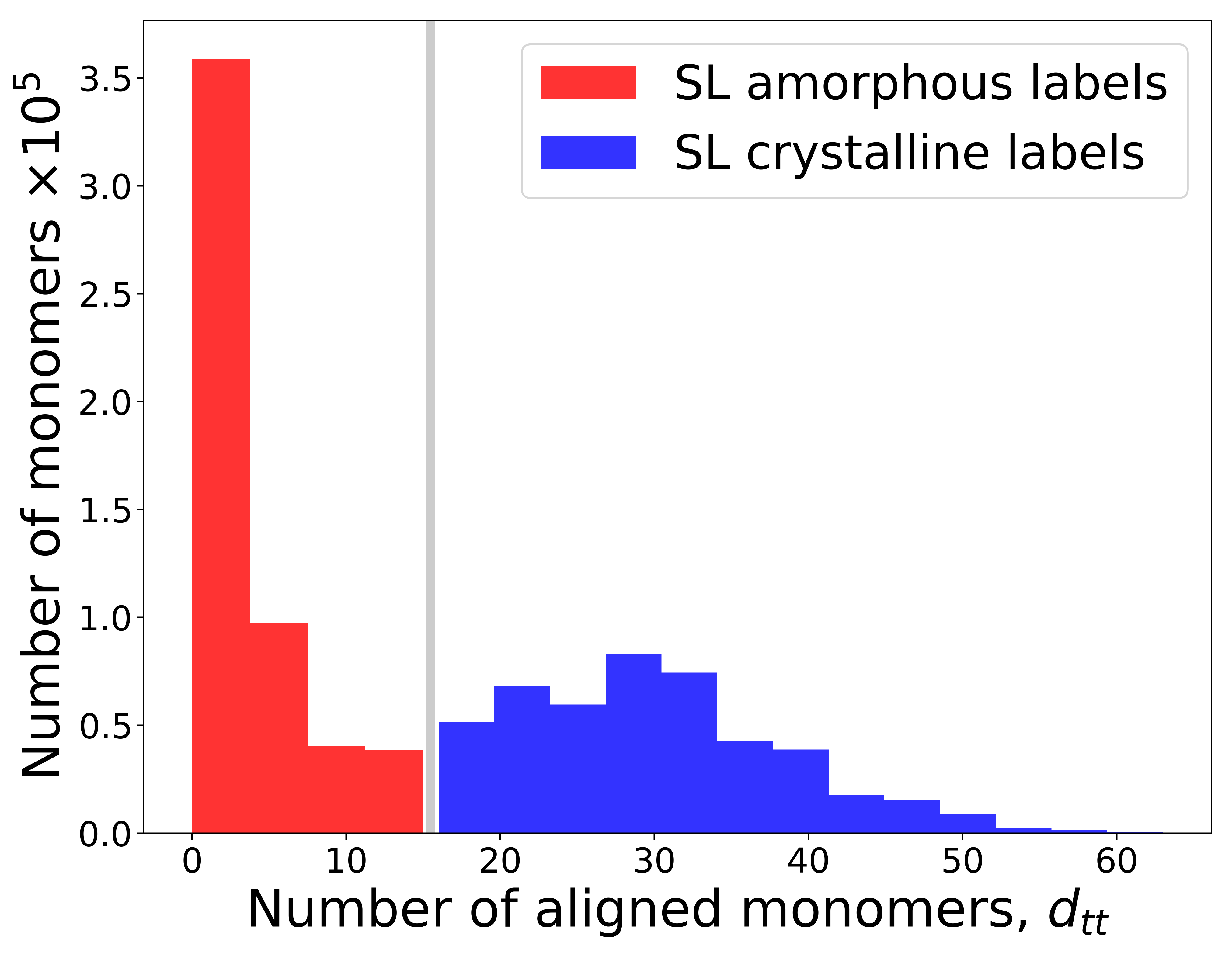}
        \label{fig:SL_hist}
        }
    \subfigure[]{
        \includegraphics[width=0.47 \hsize]{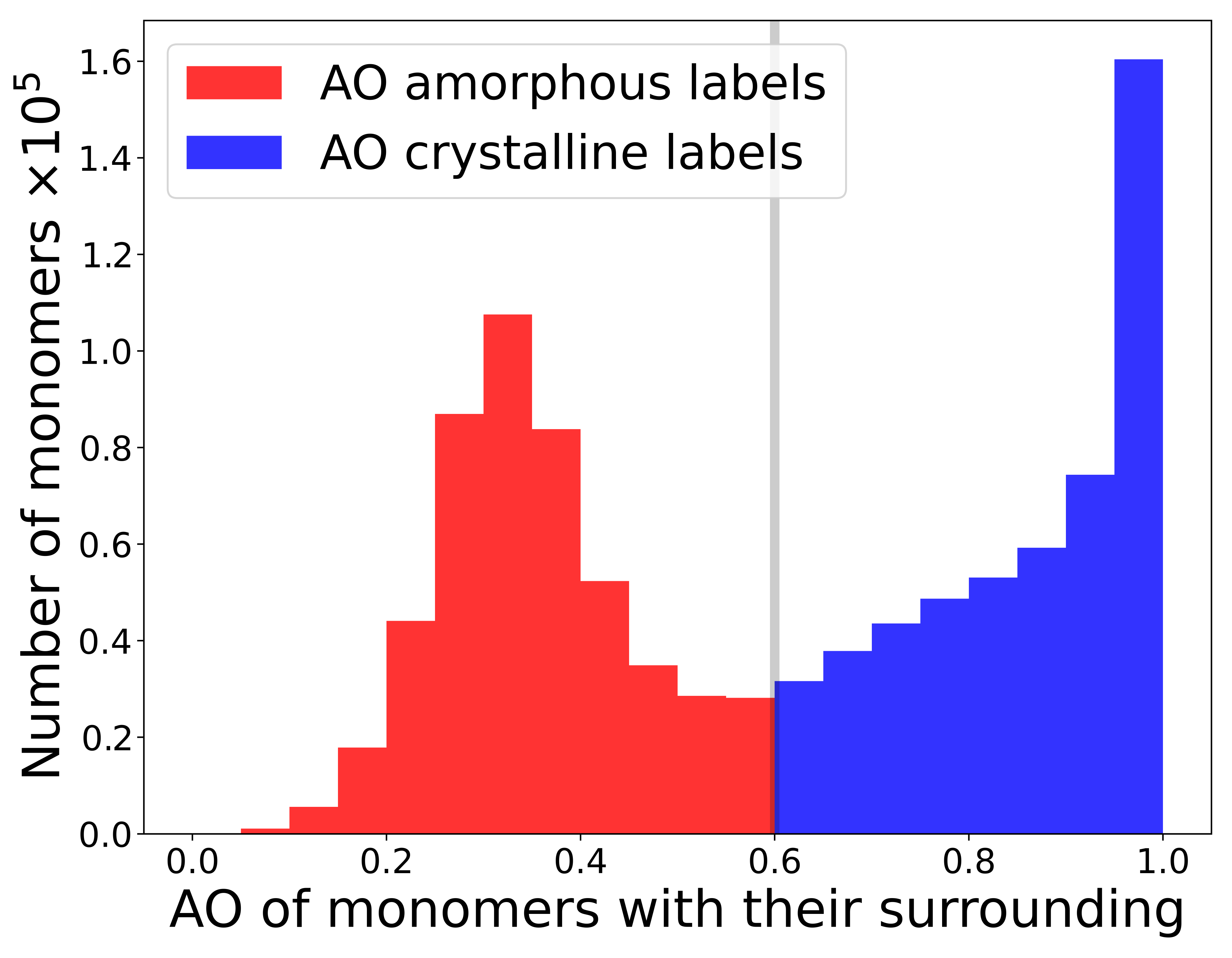}
        \label{fig:ao_hist}
        }
    \caption{(a) A histogram of all the stem lengths (SL) seen at the $t_{end} = 74$ ($T = 0.75$) of a cooling cycle.
    The grey solid line is the manually created decision boundary at $15$ between the amorphous and crystalline labels.
    The set of monomers with \(d_{tt} < 15\) falls in the amorphous phase while the ones with \(d_{tt} >= 15\) fall in the crystalline phase~\cite{CF2011macro}.
    (b) A histogram of the average orientation (AO) of each monomer with its neighboring monomers within a sphere of cutoff radius $3.5$ (\(\widehat{=}~1.82~\mathrm{nm}\)).
    The decision boundary is manually created at $0.6$.
    The number of monomers on the y-axis refers to the number of coarse-grained beads.}\label{fig:hist}
\end{figure}

In figure~\ref{fig:hist} we show the distribution of SL (\ref{fig:hist}), and AO (\ref{fig:ao_hist}) to evaluate the chosen decision boundaries.
In both cases, the boundary was chosen near a minimum that divides two leading modes.

\section{Results and Discussions}\label{sec:results}

\subsection{Classification of crystalline and amorphous monomers}

\begin{figure}[!ht]
    \centering
    \subfigure[]{
        \includegraphics[width=0.31\hsize]{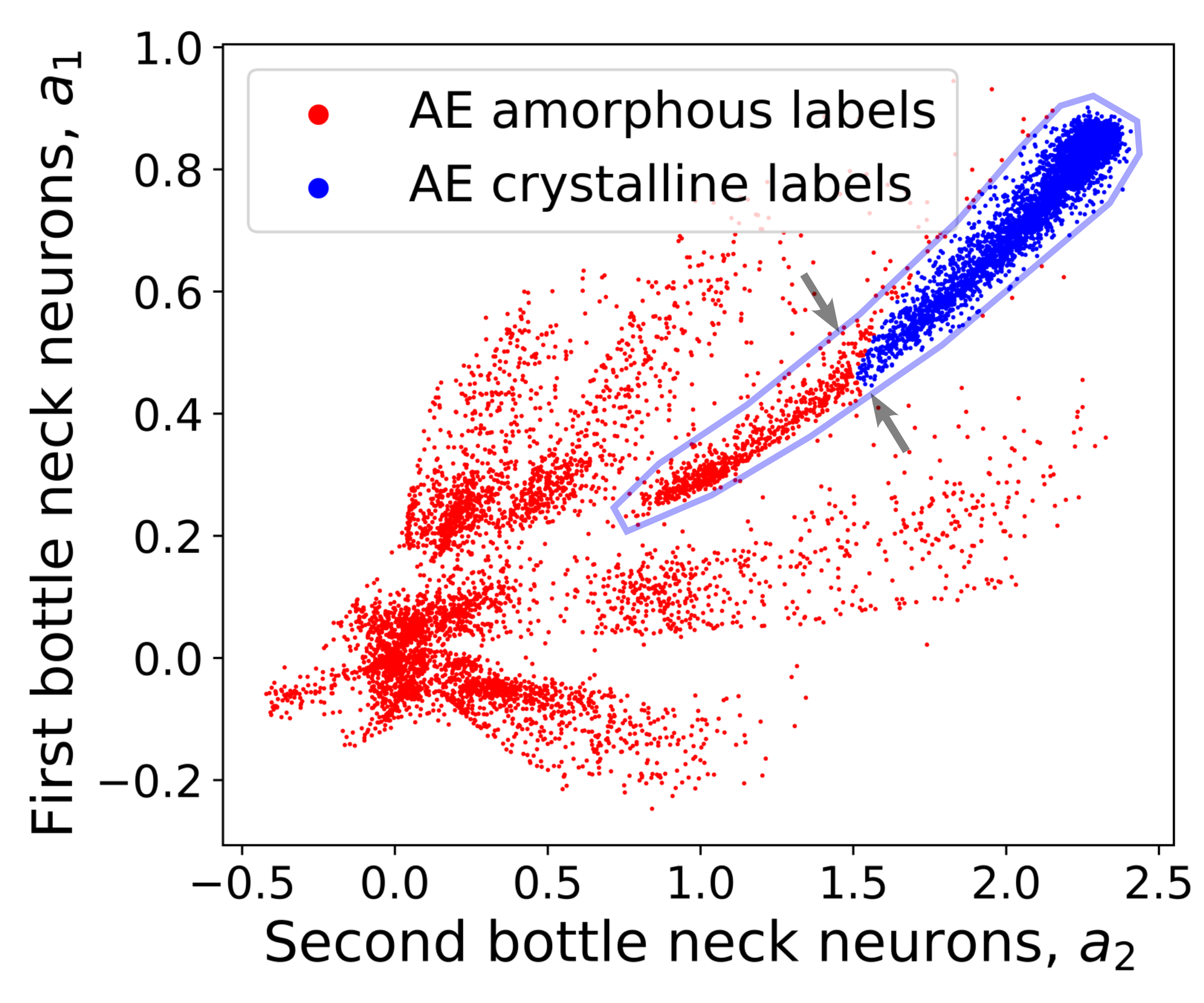}
        \label{fig:AE_PC_74}
        }
    \subfigure[]{
        \includegraphics[width=0.31\hsize]{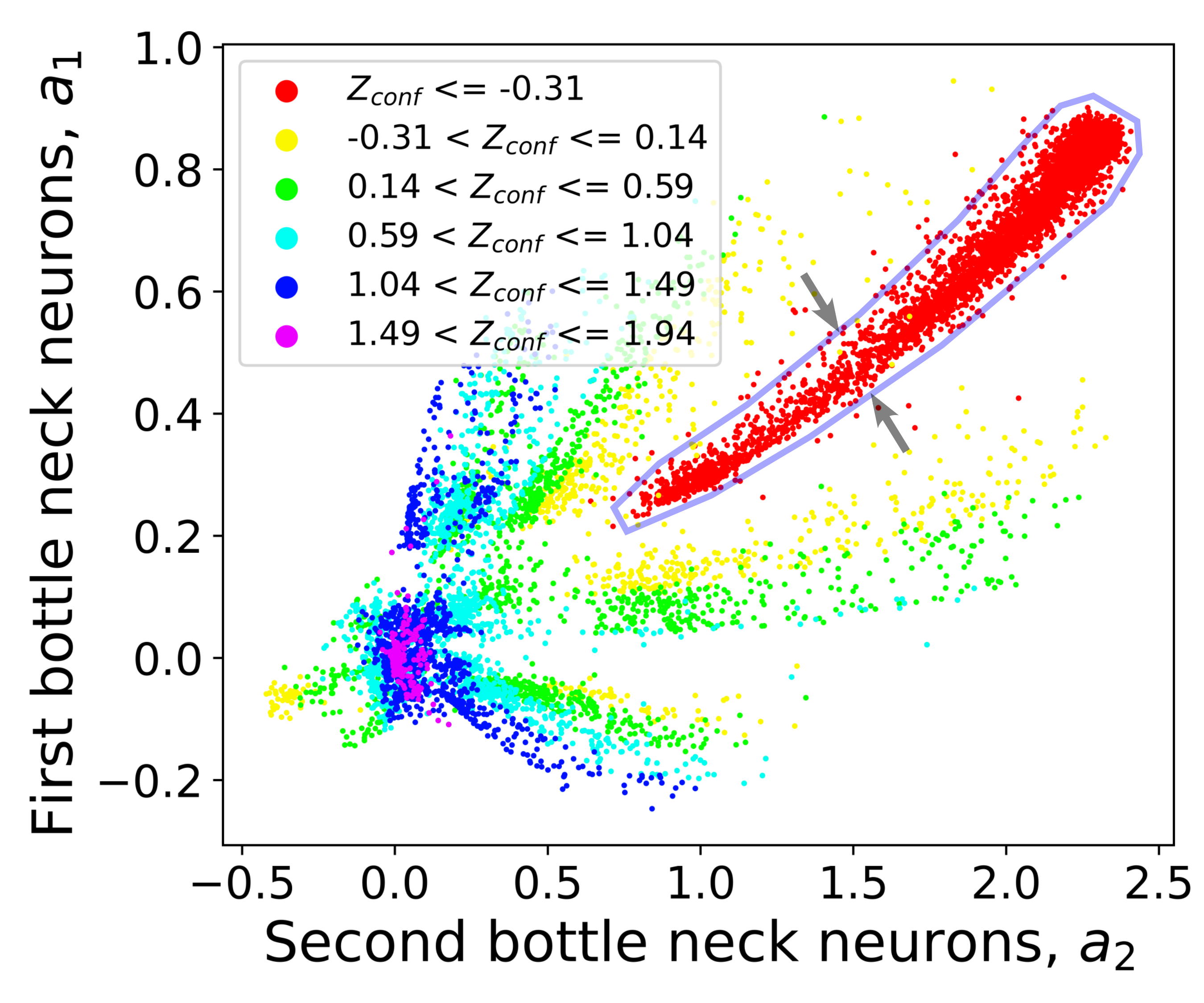}
        \label{fig:conf_multi}
        }
    \subfigure[]{
        \includegraphics[width=0.31\hsize]{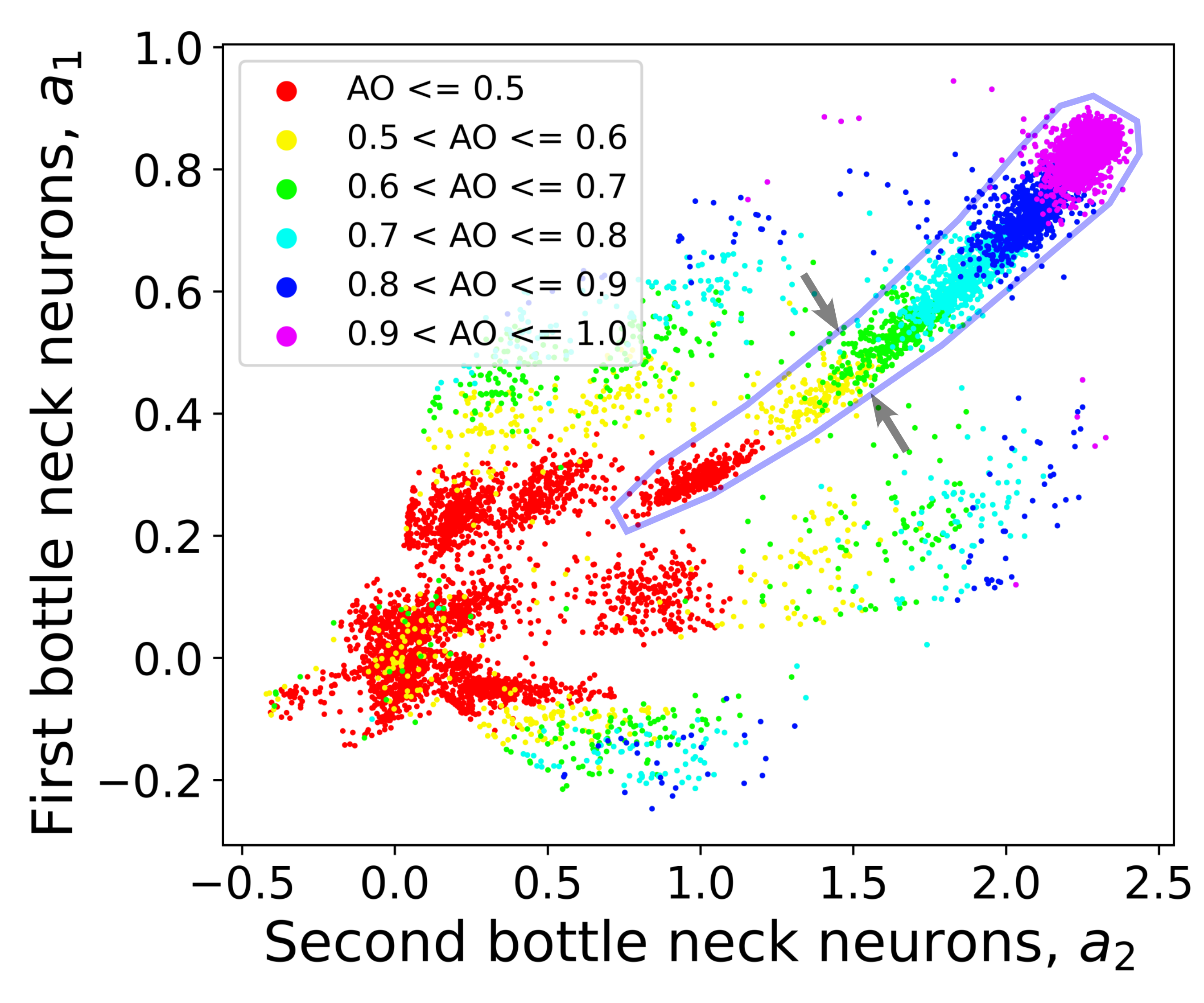}
        \label{fig:AO_multi}
        }
    \caption{(Auto)Encoded representation, $\hat a=\{a_1,a_2\}$, of $\sim 10^4$ structural fingerprints, $\hat X$, in the 15th slice ($0.14L_z < z \leq 0.15L_z$) at $t_{end}=74$.
    (a) The two phases "crystalline" and "amorphous" are colored according to \textit{AE labels}.
    Notice the dense protruding middle branch-like (\textit{main branch}) structure that is highlighted by a thin blue-colored encircling line.
    The two arrows indicate approximately the boundary of the \textit{main branch} between crystalline and amorphous labels.
    They are repeated in all the subplots here and in figures \ref{fig:latent_in_time}, and \ref{fig:PC_74}, at the same position.
    (b) One-dimensional principal component analysis (PCA) is performed on the conformation fingerprint, $\vec X_{conf}$, and the compressed version is referred to as $Z_{conf}$.
    The values of $Z_{conf}$, referring to six intervals, are color-coded.
    See also appendix section~\ref{sec:pca_conf}.
    (c) Values of the alignment order parameter (AO) (section~\ref{sec:avg_orientation}) referring to six intervals are color-coded.
    }
    \label{fig:PC_74_joint}
\end{figure}

In figure~\ref{fig:PC_74_joint}, we show the compressed version of the structural fingerprints, $\hat a_{train}$, for the monomers from the selected $15$th slice ($\hat X_{train}$).
Note that the specific shape of the latent space point cloud depends on the selection of the training data set and the initialization of weights.
Therefore, in table~\ref{tab:dataset} (appendix), we assessed the consistency of the two classes identified by GMM.
The selected slice $15$ falls within the mid-field, showing a mismatch of only $1\%$ to $2\%$ in predicted labels.
This evaluation was conducted using randomly chosen sets of $10^4$ and $2\times10^4$ monomers for training.
Based on the point cloud for slice $15$, the GMM identified two classes labeled as blue and red, which are shown in the figure~\ref{fig:AE_PC_74}.

For comparison, the labels from the other classification methods \textit{SF, SL} and \textit{AO} are presented equivalently in the appendix, figure~\ref{fig:PC_74}.
For the interpretation of the latent space, in figure~\ref{fig:PC_74_joint}, we also display two order parameters based on the conformation fingerprint ($\vec X_{conf}$) in figure~\ref{fig:conf_multi} and the average orientation (AO) in figure~\ref{fig:AO_multi} (see section~\ref{sec:avg_orientation} and figure~\ref{fig:ao_hist}).
These are presented as color codes overlaid onto the point cloud.
This gives us an insight into how to associate the latent space information with the physical parameters.
For figure~\ref{fig:conf_multi}, the set of $\vec X_{conf}$ has been compressed to a single dimension using principal component analysis (PCA).
Here, we call the 1D compressed conformation fingerprint as $Z_{conf}$.

To justify the choice of the class attributions "crystalline" and "amorphous", in figure~\ref{fig:AE_PC_74}, we identify the corresponding crystal domains in the simulation box with these labels and examine their order in relation to the domain's principal director.
This is exemplified for the \textit{AE labels} in section~\ref{sec:domains}.
Since \textit{SL} and \textit{AO labels} refer to manually defined boundaries towards states of higher order for stem conformations and orientational order, respectively, they can be used as additional guidelines for the attribution of \textit{SF} and \textit{AE labels}.
In table~\ref{tab:comparison}, an inter-scheme comparison is shown that quantifies the number of those common monomers at the last time step, $t_{end} = 74$, based on all monomers.
We observe that the majority of labels are shared between data-driven crystalline labels (\textit{SF} and \textit{AE}) and both manually defined crystalline labels (\textit{SL} and \textit{AO}).
Hence, we conclude that the chosen attribution of \textit{AE labels} is well justified.

\begin{table}
    \begin{tabular}{ | Sc | Sc || Sc | }
	\hline
        \textbf{Scheme 1} & \textbf{Scheme 2} & \textbf{\(\frac{\text{Common crystalline labels in both schemes}}{\text{Number of crystalline labels in Scheme 1}}\)} \\
	\hline
	SF & AE & 0.84 \\
	SF & SL & 0.81 \\
	SF & AO & 0.86 \\
	\hline
	AE & SF & 1.00 \\
	AE & SL & 0.91 \\
	AE & AO & 0.99 \\
	\hline
	SL & SF & 0.89 \\
	SL & AE & 0.84 \\
	SL & AO & 0.93 \\
	\hline
	AO & SF & 0.87 \\
        AO & AE & 0.84 \\
        AO & SL & 0.85 \\
        \hline
    \end{tabular}
    \caption{\label{tab:comparison} Inter-scheme comparison of the labeled crystalline indices generated by different classification schemes at \(t_{end} = 74\) for all the coarse-grained monomers.
    We calculate a fraction of common crystalline labels between Scheme 1 and Scheme 2.
    Values have been rounded to the second digit.
    We can observe that the Gaussian mixture model (GMM) applied to compressed or encoded (AE) data shares all of its indices with hierarchical clustering (HC) on the concatenated structural fingerprints (SF), yielding a similarity score of $100\%$.
    On the other hand, when HC is applied to SF, it labels some additional indices as crystalline that matches only \(84\%\) with GMM on AE.}
\end{table}

It could also be argued that, in the case of higher spatial and conformational order, we anticipate reduced exploration of degrees of freedom in the local environment and conformation fingerprint vectors, resulting in smaller clusters in the latent space.
This can be inspected via the order parameters in figures~\ref{fig:conf_multi} and \ref{fig:AO_multi}.
In figure~\ref{fig:conf_multi}, we see a group with the lowest values of $Z_{conf}$ associated with a central branch-like cluster within the point cloud.
In the following, we define this area as the \textit{\textbf{main branch}}, which is encircled with a bluish line in the latent space plots.
Note that the \textit{main branch} holds large fractions of "crystalline" monomers throughout all classification schemes (see also figure~\ref{fig:AE_PC_74}).
Consistently, in the appendix (figure~\ref{fig:conf_order}), it is demonstrated that low values of $Z_{conf}$ correspond to stretched conformations, as indicated by the bond vector correlation along the local contour.
When comparing figures~\ref{fig:conf_order} and ~\ref{fig:conf_multi}, it is also evident that the degree of conformational order, represented by $Z_{conf}$, is encoded perpendicular to the fanned-out branch structure of the latent map.
In figure~\ref{fig:AO_multi}, it is visible that as the threshold for the orientational order parameter decreases, a larger area at the tips of the branch-like clusters is encompassed.
We conclude that the degree of orientational order is (mostly) encoded as the distance from an apparent center near $\vec a = \vec 0$ along the branches.
A comparison between figures~\ref{fig:AE_PC_74} and~\ref{fig:conf_multi} reveals that the boundary in the case of \textit{AE labels} within the \textit{main branch} divides monomers with ordered conformation (figure~\ref{fig:conf_multi}) further into a group of large alignment (crystalline) and small alignment (amorphous), approximately matching the division in figure~\ref{fig:ao_hist}.

It is interesting to see that the GMM discovers a boundary between crystalline and amorphous monomers that unites both criteria for the crystalline monomer to belong to an ordered conformation as well as being located within an aligned environment.
Note that this is achieved only by attributing the latent points to one wide and one narrow distribution.
Hence, the distinction between the different degrees of exploration of latent space as a fundamental imprint of order and disorder transitions is clearly detectable using a combination of the AE and the GMM once the ordered labels are sufficiently populated.

\begin{figure}
    \centering
        \subfigure[]{
        \includegraphics[width=.31\hsize]{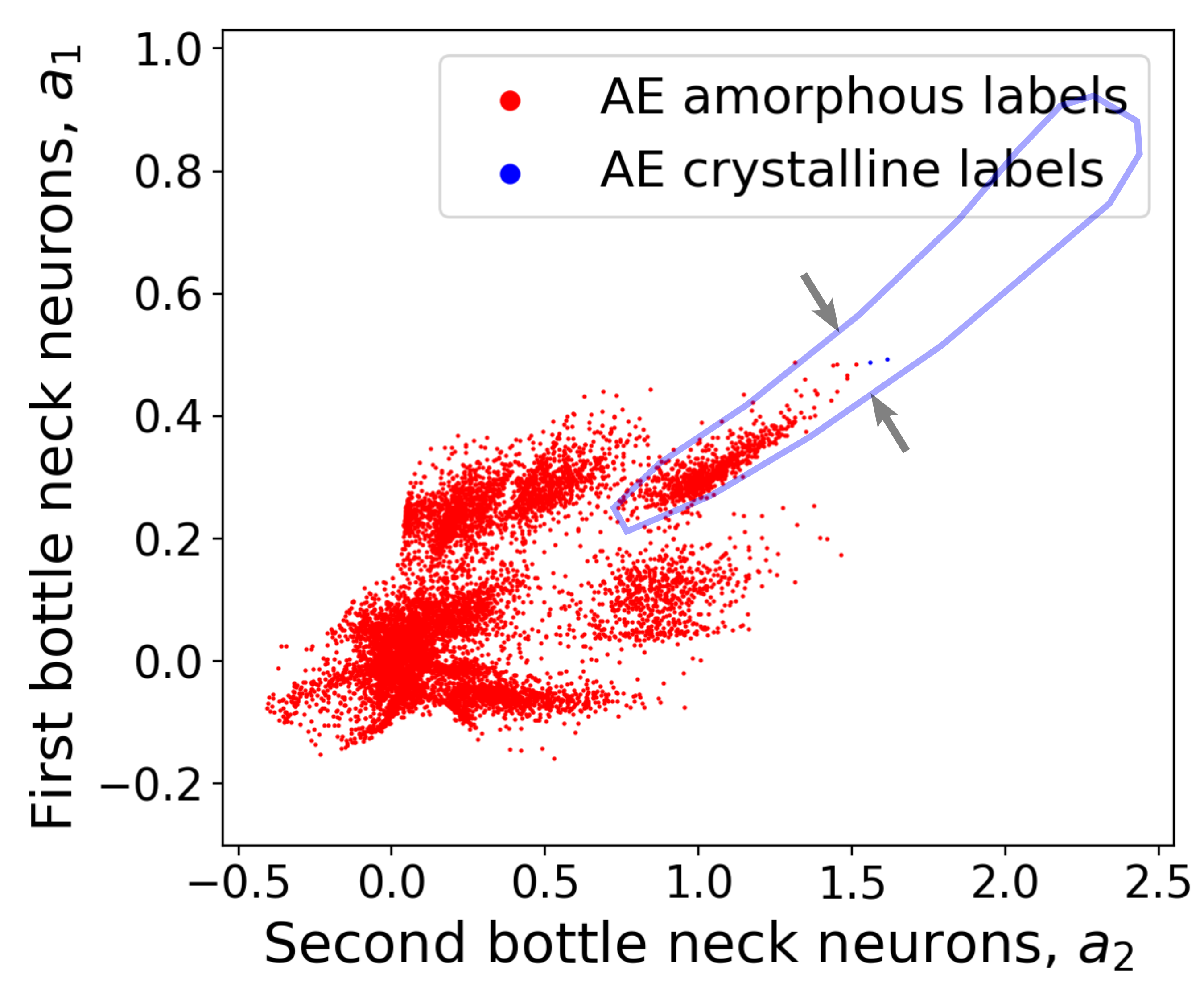}
        \label{fig:time_48}
        }
        \subfigure[]
        {
        \includegraphics[width=0.31\hsize]{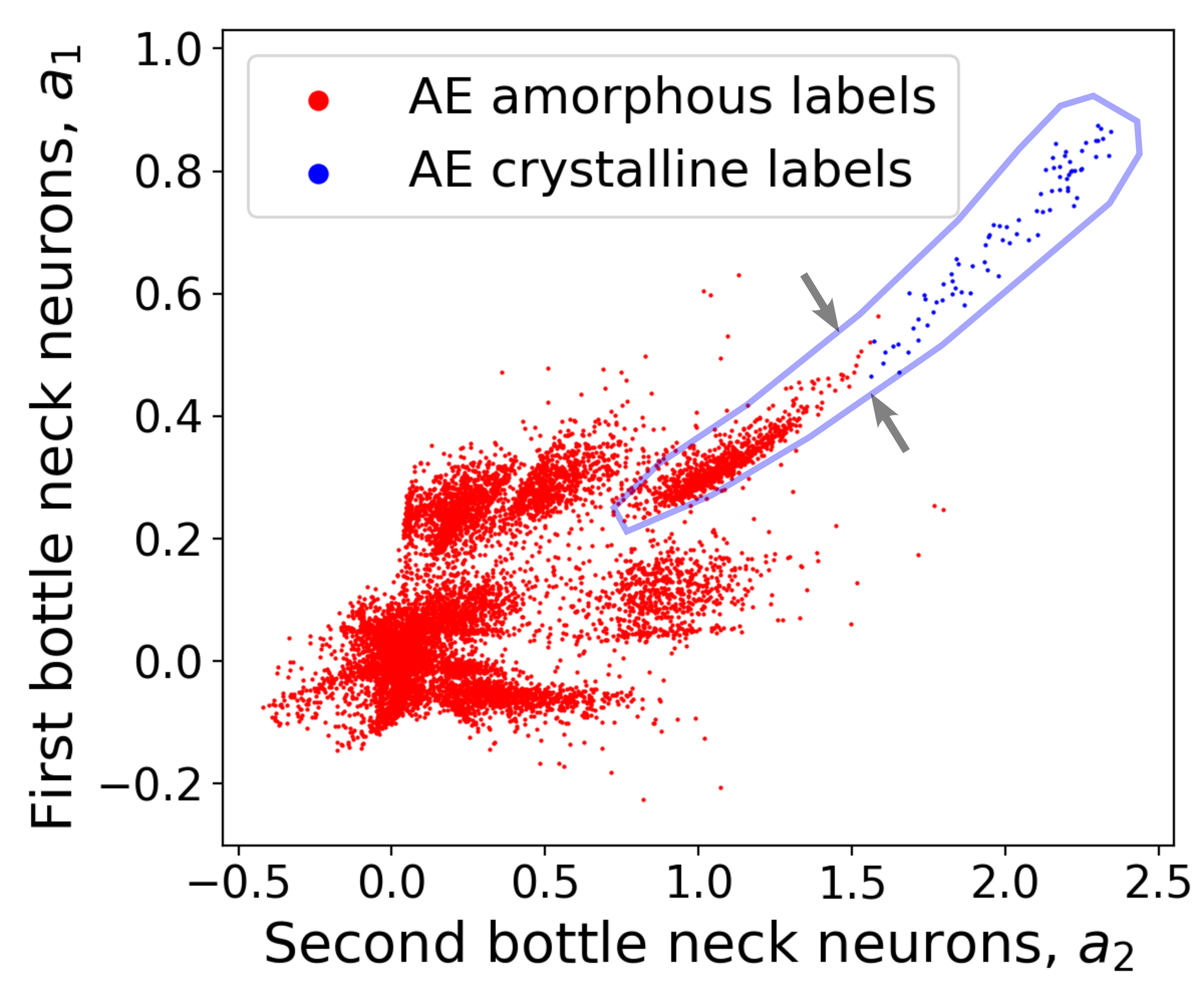}
        \label{fig:time_52}
        }
        \subfigure[]
        {
        \includegraphics[width=0.31\hsize]{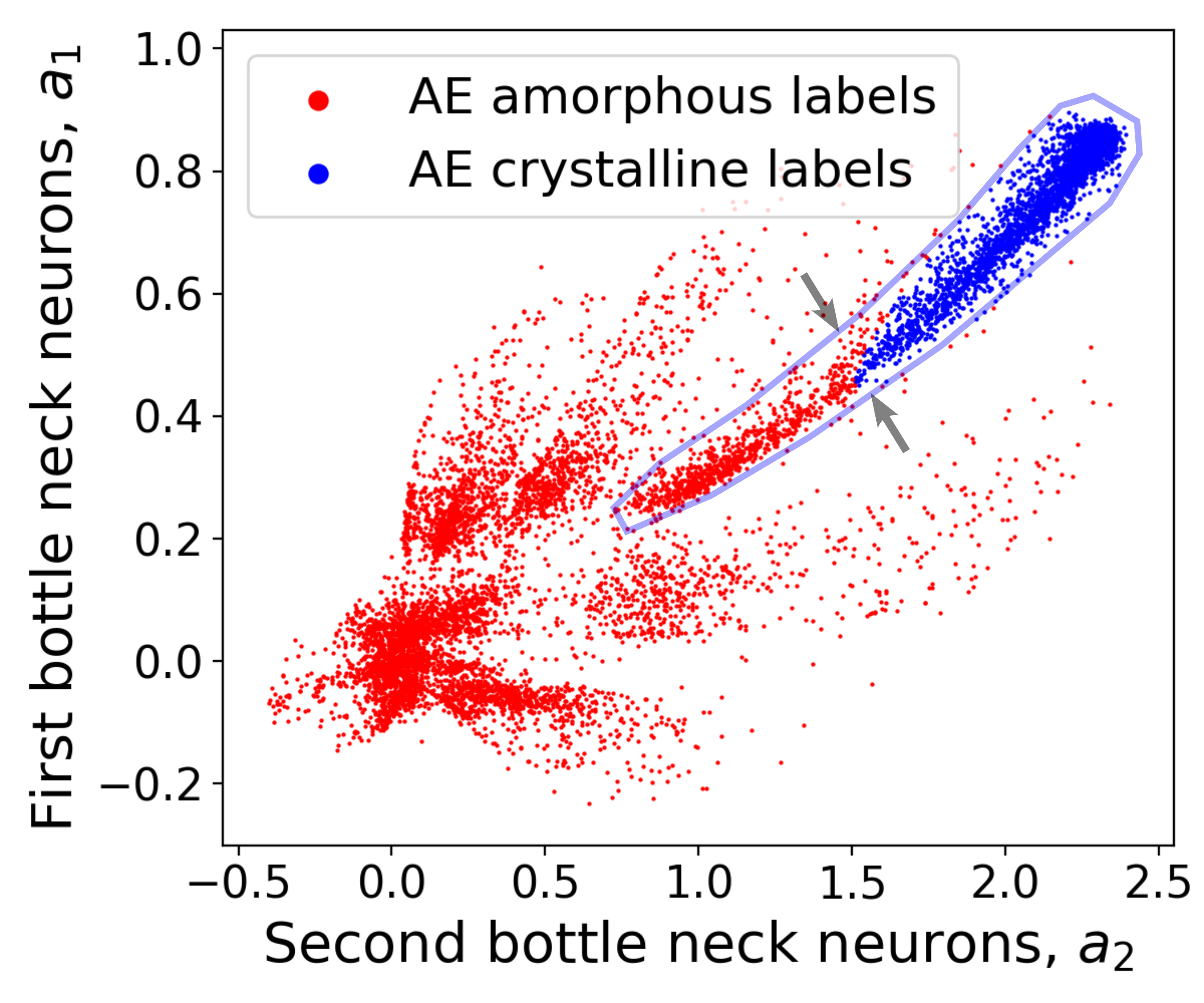}
        \label{fig:time_57}
        }
    \caption{The crystalline and the amorphous phases classified via \textit{AE labels} are assigned to the AE-compressed version of the fingerprints in latent space $\vec a$ at various time steps of the cooling cycle: (a) at $t_{pre} = 48$, (b) at $t_{tr} = 52$, (c) at $t_{min} = 57$.
    Also see figure~\ref{fig:Timeseries_3}.} \label{fig:latent_in_time}
\end{figure}

In figure~\ref{fig:latent_in_time}, we present the temporal evolution of the exploration of latent space through snapshots taken at four different time points (table~\ref{tab:times}).
During the process of cooling, the number of crystalline labels starts to grow after $t_{pre} = 48$.
Before this time step, the monomers classified via GMM were predominantly amorphous, with little fluctuations in the appearance of the blue points.
After $t_{tr} = 52$, the \textit{main branch} and the other branches extend over larger distances from $\vec a = 0$, revealing an increasing fraction of monomers in an aligned environment (figure~\ref{fig:time_52}).
At $t_{pre}$, some monomers occupy lower parts of the \textit{main branch} (figure~\ref{fig:time_48}) indicating a stretched conformation.
Only after $t_{tr}$, the common crystallinity criterion - ordered conformation in an aligned environment - is fulfilled for an increasing number of monomers (figure~\ref{fig:time_57}).

After transition time $t_{min}>t_{tr}$, in figure~\ref{fig:time_57}, we see that the latent space gets occupied in regions outside of the \textit{main branch} that are dedicated to larger orientation order (compare with figure~\ref{fig:AO_multi}).
There is an increasing number of monomers with an aligned environment without being attributed to the crystalline state.
One may expect this condition to be fulfilled mostly by monomers near the boundary between the crystalline and the amorphous phase (compare with figure~\ref{fig:SF_AE_avg_74} below).
In this context, it is interesting to note that the so-called rigid-amorphous fraction (RAF) ~\cite{wunderlich:pips:2003,schick:ta:2003} also is assumed to have contributions near the interface region~\cite{lee:jpspbpp:2017}.
These are the monomers in the amorphous phase that show a decrease in monomer mobility as well as an enhanced glass transition temperature as compared to the mobile fraction.
For future investigation, it will be therefore interesting to analyze the spatially resolved monomer mobility and its relationship with structural patterns, as observed, for instance, in latent space (fig~\ref{fig:AE_PC_74}).

\begin{figure}[!ht]
    \centering
    \subfigure[]{
    \includegraphics[width=0.4\hsize]{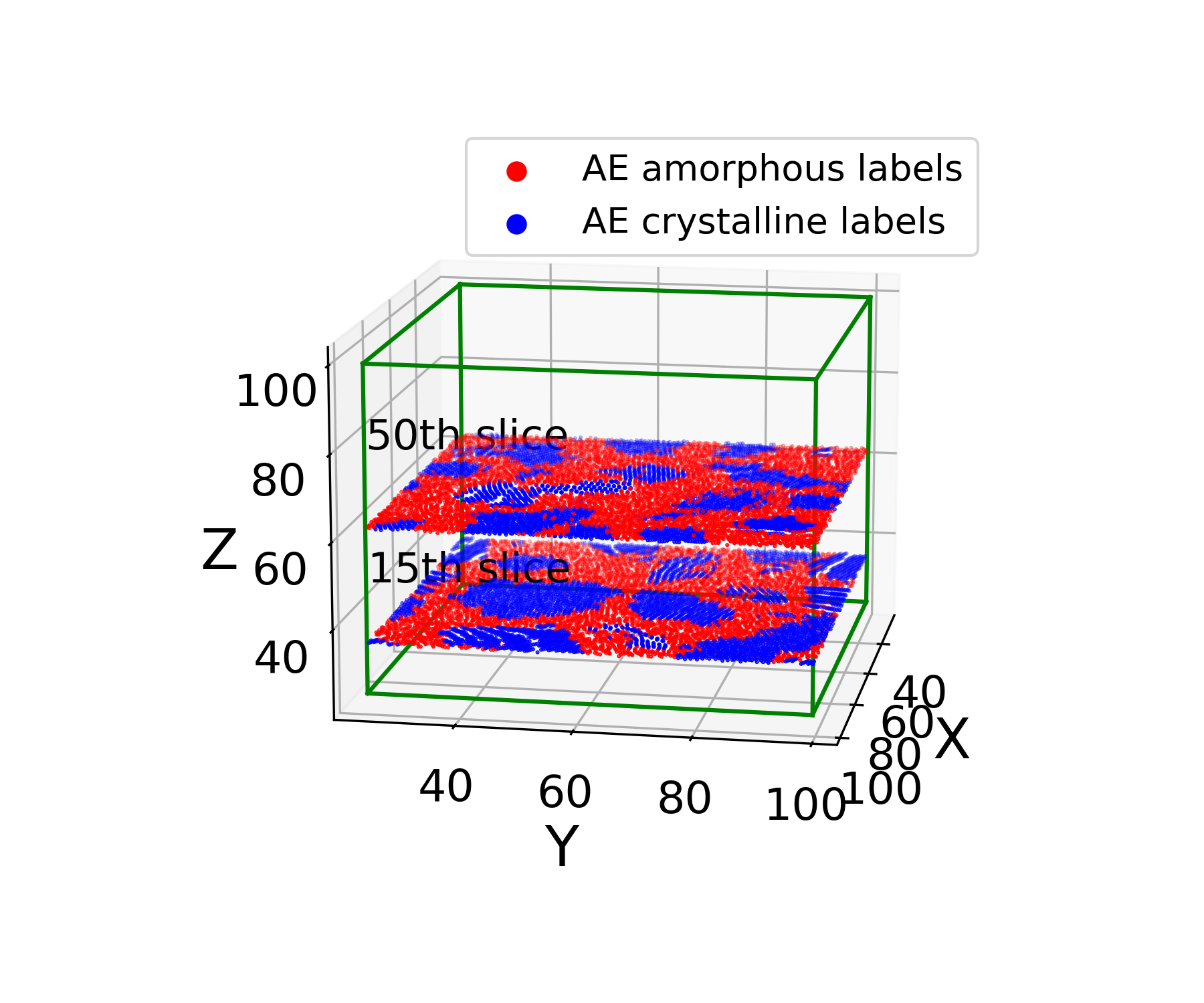}
        \label{fig:box_slices}
        }
    \subfigure[]{
        \includegraphics[width=0.4\hsize]{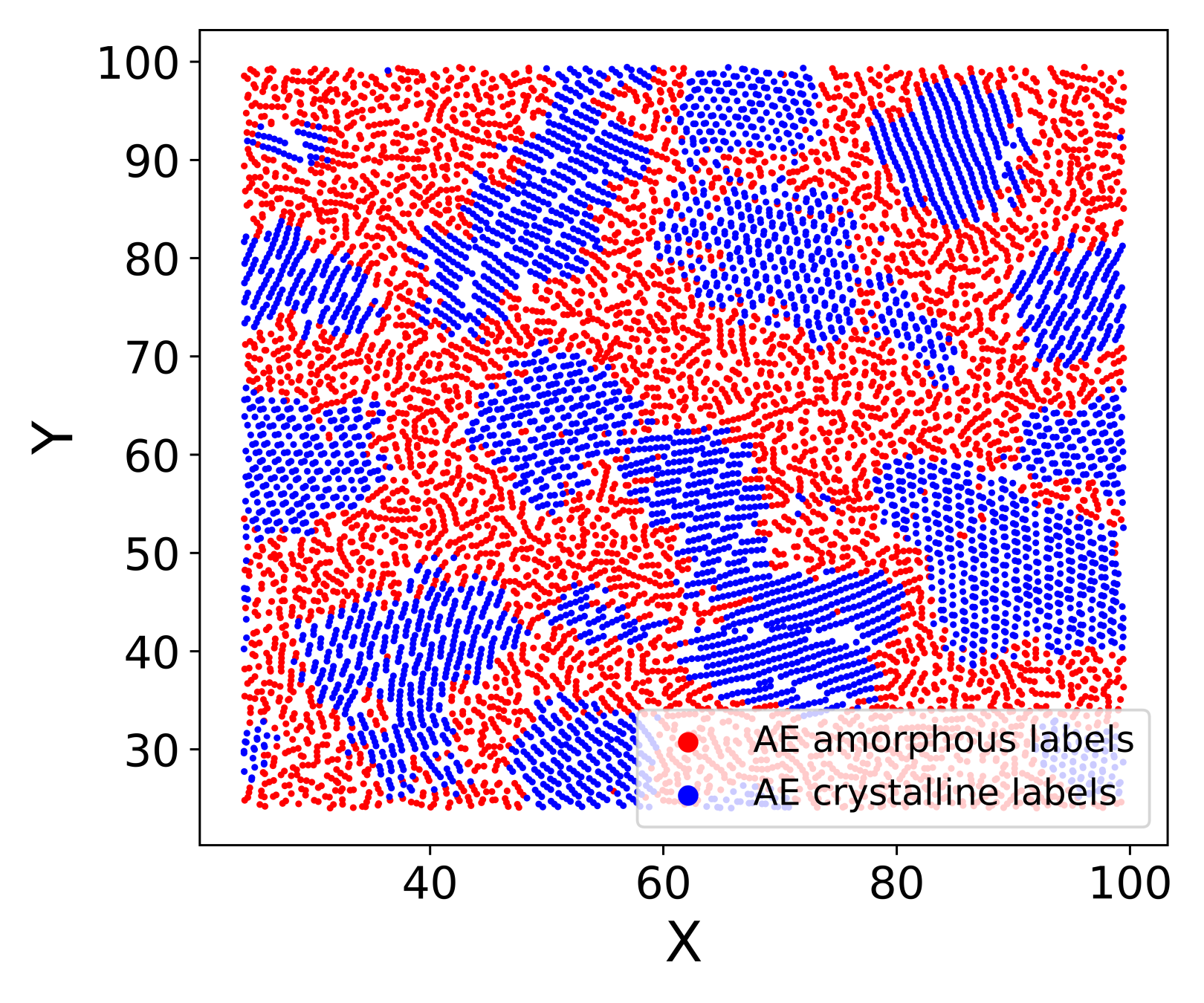}
        \label{fig:AE_Sl_74}
        }
    \centering
    \subfigure[]{
        \includegraphics[width=0.4\hsize]{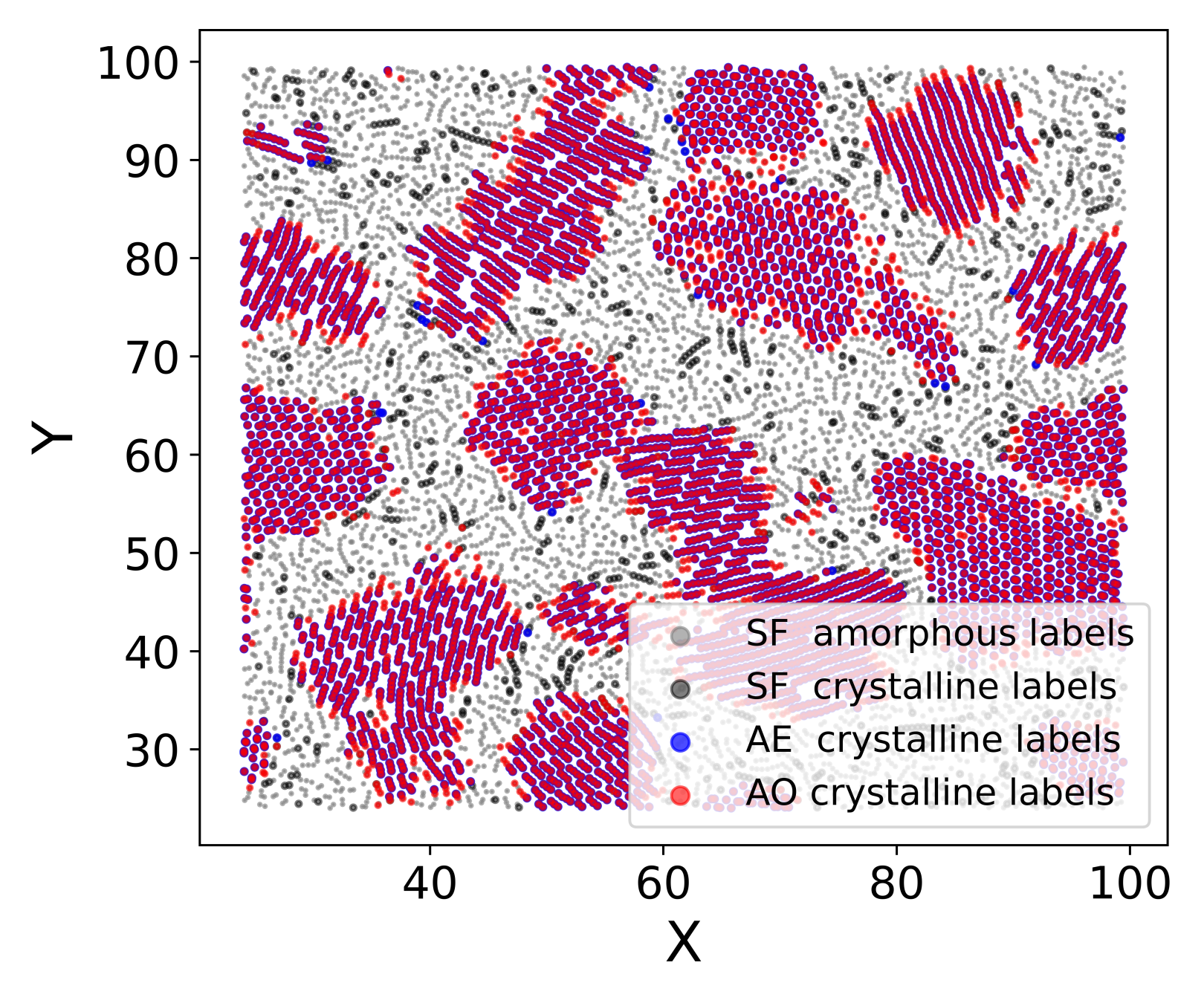}
        \label{fig:SF_AE_avg_74}
        }
    \subfigure[]{
        \includegraphics[width=0.4\hsize]{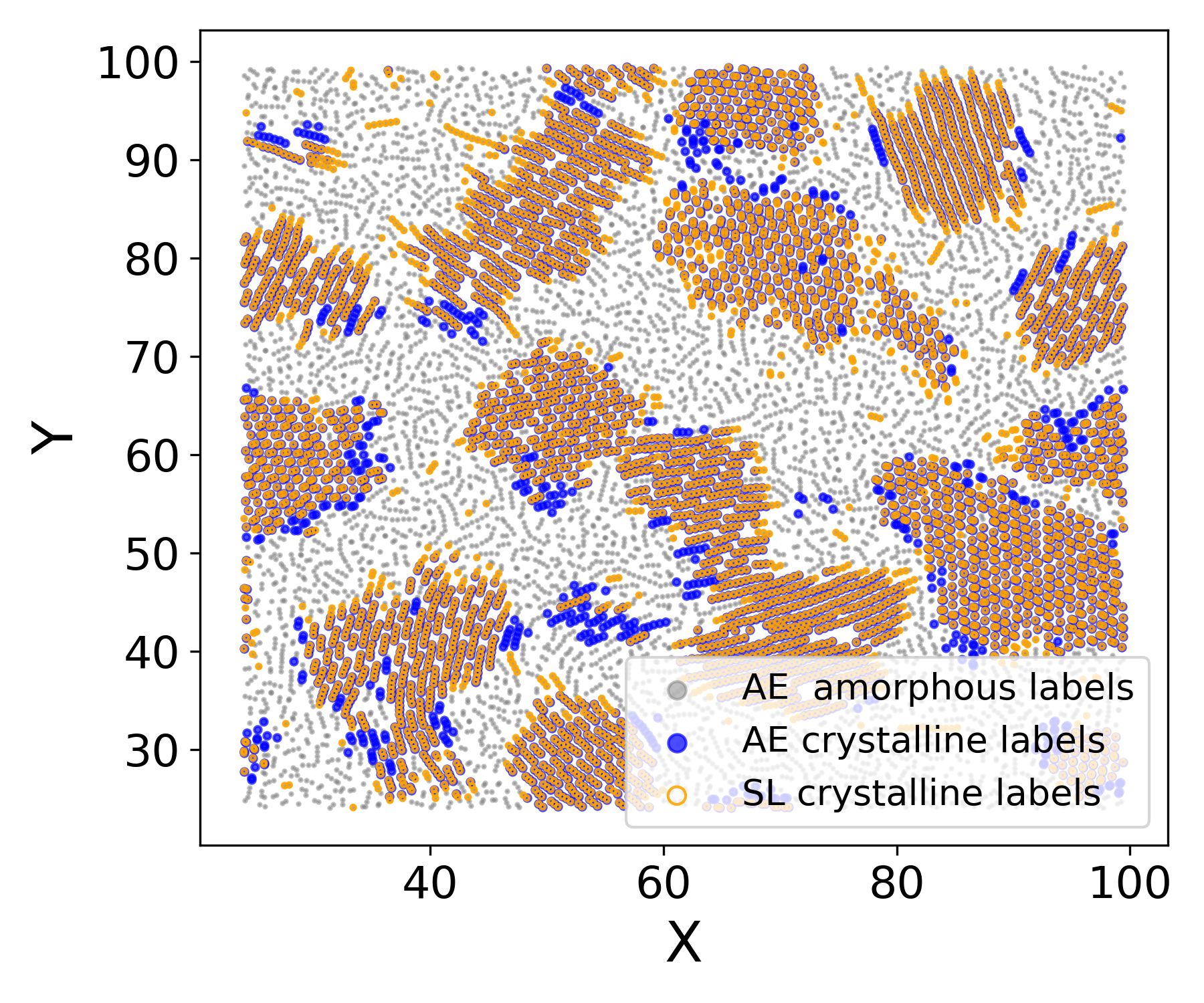}
        \label{fig:AE_SL_74}
        }
    \caption{(a) Monomers of the 15th ($0.14 L_z < z \leq 0.15 L_z$), and 50th slice ($0.49 L_z < z \leq 0.5 L_z$) colored by their \textit{AE labels} at $t_{end} = 74$.
    The data points on the 15th slice are used as a training set for the autoencoder and Gaussian mixture model (GMM) resulting in \textit{AE labels}.
    (b) Top view of monomers of the 50th slice colored using \textit{AE labels}.
    (c) Comparison of \textit{SF} and \textit{AO labels} to \textit{AE labels}.
    (d) Comparison between \textit{SL} and \textit{AE labels}.}\label{fig:scheme_comparison}
\end{figure}

To visualize the classification results in the simulation box, we show in figure~\ref{fig:scheme_comparison} a 50th cross-section of the box ($0.49 L_z < z \leq 0.5 L_z$).
In figures~\ref{fig:box_slices} and~\ref{fig:AE_Sl_74}, we show a 3D view and a top view of the slice, respectively, and color the monomers according to the \textit{AE labels} as shown in figure~\ref{fig:AE_PC_74}.
In figures~\ref{fig:SF_AE_avg_74} and~\ref{fig:AE_SL_74}, the \textit{AE labels} are compared to the \textit{SF} and the \textit{AO labels} as well as to \textit{SL labels}, respectively.
We observe that the \textit{AE labels} lead to compact crystal domains, while the \textit{SF labels} result in the appearance of additional monomers in a crystalline state, apart from the compact domains of \textit{AE labels}.
These additional monomers can be seen to be located in the areas between the AE-based compact crystal domains, and appear as stretched conformations.
This is expected as the crystalline \textit{SF labels} are mostly located within the \textit{main branch} in figure~\ref{fig:SF_PC_74} (similar to figure~\ref{fig:conf_multi}) where some points do not fulfill the condition of an aligned neighborhood (see figure~\ref{fig:AO_multi}).
In figure~\ref{fig:AE_SL_74}, we observe that in the case of the \textit{SL labels}, single stems surrounded by amorphous areas—those without an aligned environment—are noticeable.
On one hand, in some cases, the manually defined cutoff ($d_{tt}\geq 15$) does not limit the crystalline labels to fall inside a domain of neighboring stems but allows it to stretch into an amorphous environment giving rise to not-so-smooth domain surfaces.
On the other hand, stems that are smaller than the defined cutoff are ignored (purely blue points in figure~\ref{fig:AE_SL_74}).
In figure~\ref{fig:SF_AE_avg_74}, we see the domains defined solely based on the alignment parameter (\textit{AO labels}) to be relatively compact with smooth boundaries.

\begin{figure}[!ht]
    \centering
    \subfigure[]{
        \includegraphics[width=0.47\hsize]{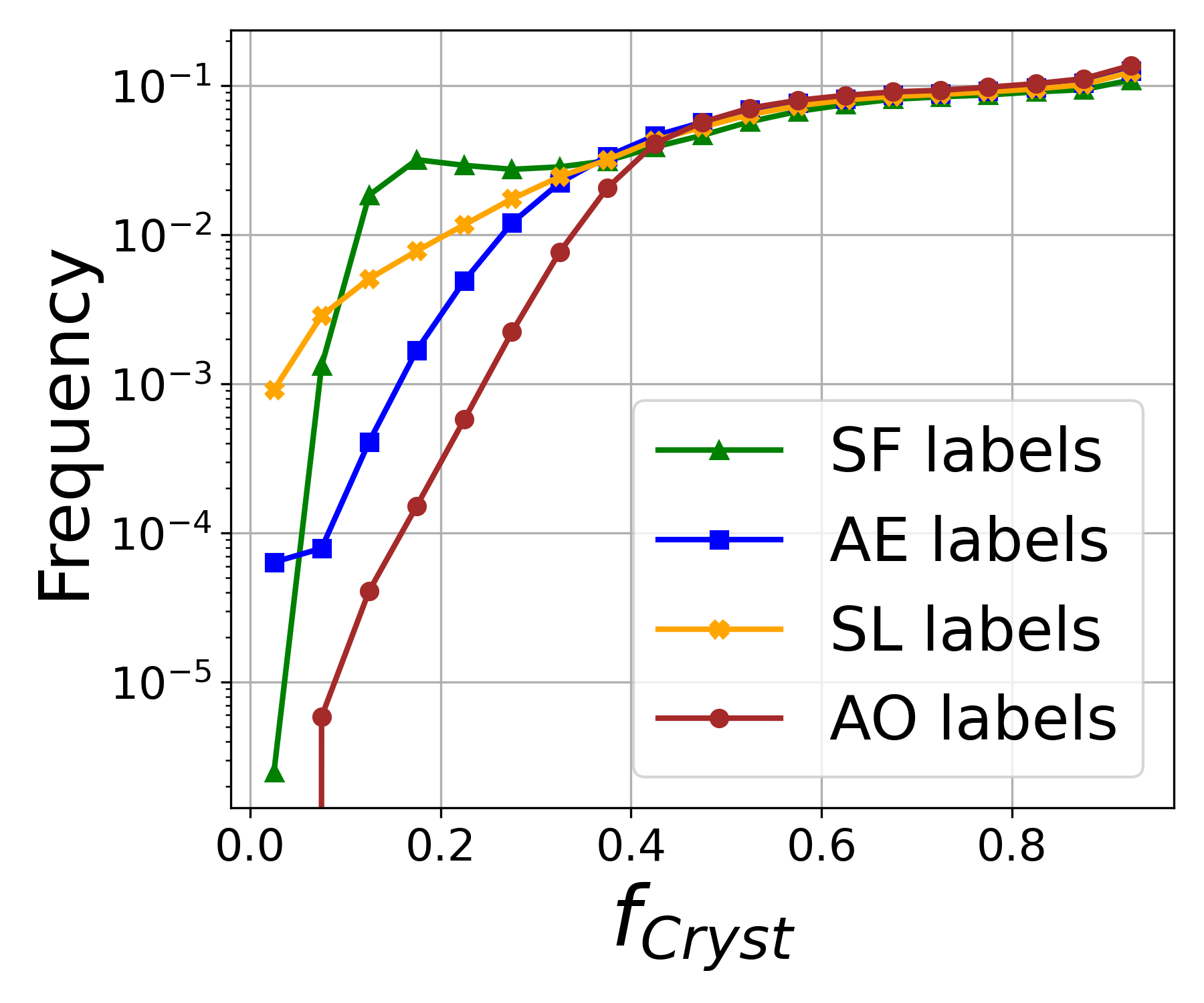}
        \label{fig:crys_frac_num}
        }
    \subfigure[]
        {
        \includegraphics[width=0.45 \hsize]{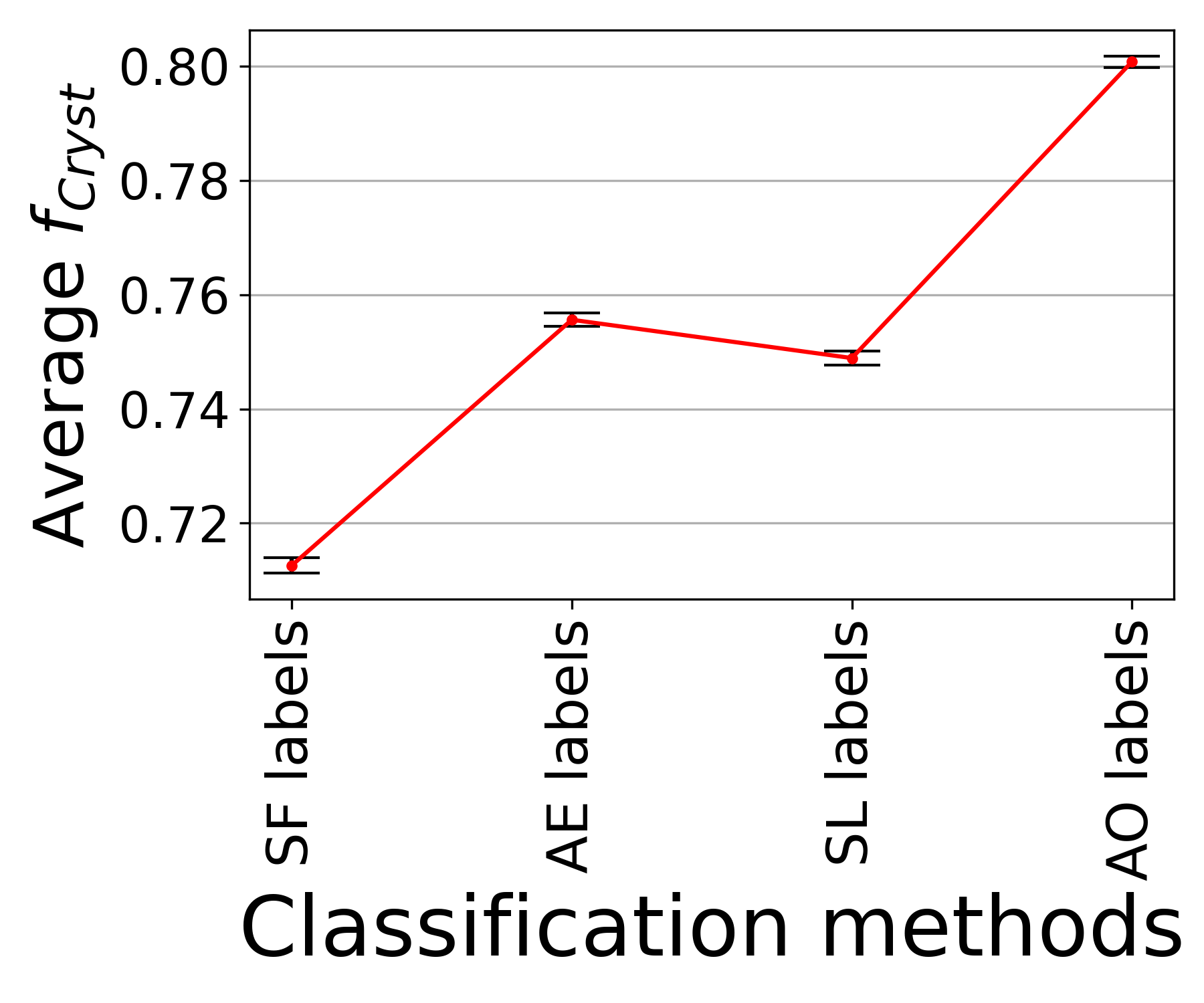}
        \label{fig:avg_crys_frac}
        }
    \caption{A comparison of the classification schemes concerning a fraction of crystalline monomers, $f_{Cryst}$, that are found in a neighborhood radius $3\sigma$ (\(\widehat{=} ~1.56~\mathrm{nm}\)) around each crystalline monomer.
    (a) A histogram representation of $f_{Cryst}$ for all monomers.
    (b) The mean values of $f_{Cryst}$ for each classification scheme.}\label{fig:crys_frac}
\end{figure}

To quantify the degree of cluster compactness, we consider a neighborhood of each label in the crystalline class.
Here, we choose a neighborhood of \(3.0\) (\(\widehat{=} ~1.56~\mathrm{nm}\)) around each such crystalline monomer.
Amongst its neighbors, we calculate the fraction of crystalline monomers, $f_{Cryst}$.
In figure~\ref{fig:crys_frac}, we present a histogram of the distribution of $f_{Cryst}$ (figure~\ref{fig:crys_frac_num}), ranging from $0$ to $1$, for all the classification schemes with a uniform bin size of $0.05$.
The mean values of $f_{Cryst}$ are shown in figure~\ref{fig:avg_crys_frac}.
We see that the \textit{AO labels} are the most compact crystal domains as they have the smallest frequencies at the lower $f_{Cryst}$ in figure~\ref{fig:crys_frac_num}, and the largest mean value in figure~\ref{fig:avg_crys_frac}.
The \textit{SF labels} show a small local maximum at the lower values of $f_{Cryst} \approx 0.175$ in  figure~\ref{fig:crys_frac_num}.
This indicates the presence of crystalline labeled monomers that are not part of grown crystal domains but are spread in the amorphous regions outside the domains.
In figure~\ref{fig:crys_frac}, we observe that the compactness of the \textit{AE labels} falls between that of the \textit{SL} and \textit{AO labels}.
This can be attributed to the combination of decision boundaries influenced by both conformation and environment, as discussed earlier.

\subsection{Cooling cycle}\label{sec:cooling}

\begin{figure}[!ht]
    \begin{center}
        \includegraphics[width=0.7\hsize]{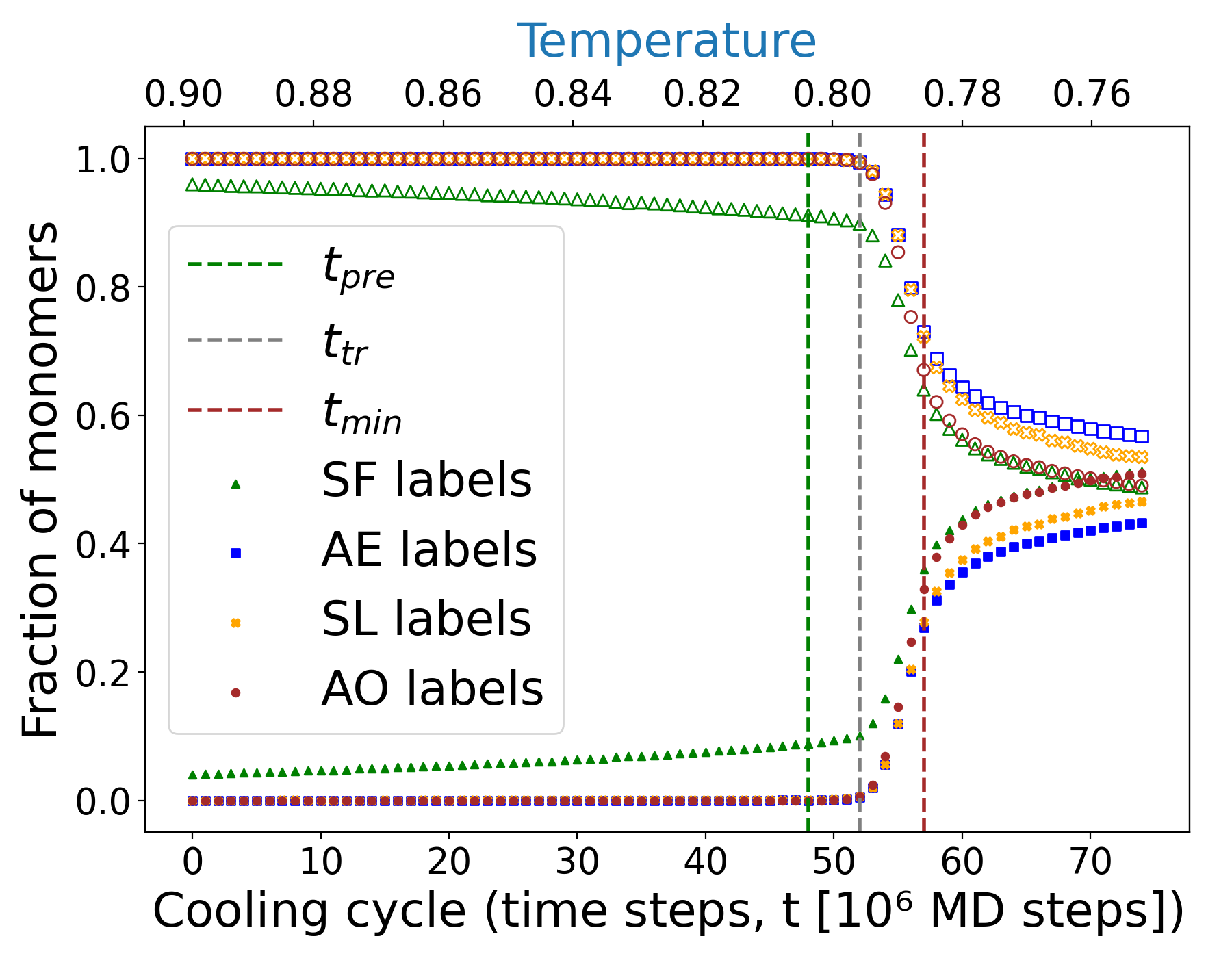}
    \end{center}
    \caption{A fraction of the crystalline (closed symbols) and the amorphous labels (open symbols) for each of the classification schemes as a function of the simulation time.
    Vertical dashed lines show the characteristic times (table~\ref{tab:times}).
    For instance, the transition point, $t_{tr}$, from where we start to observe an increase in the crystallinity fraction.}\label{fig:Timeseries_3}
\end{figure}

We have trained an MLP and a GMM (in section~\ref{sec:mlp_description}) to predict the \textit{SF} and the \textit{AE labels} based on $\hat X_{train}$ and $\hat a_{train}$, respectively, at $t_{end} = 74$ $(T = 0.75)$ and applied the trained neural nets to make predictions at each step of the cooling cycle time series.
Figure~\ref{fig:Timeseries_3} contains the total fraction of the crystalline and the amorphous labels at each time step.
We can see that the fraction of crystalline/amorphous \textit{AE labels} is approximately equal to the fraction of \textit{SL labels} at all times around the transition point, $t_{tr} = 52$ (grey line).
The threshold for the \textit{SL labels} can be slightly increased to require a sequence of $17$ consecutive monomers to better align with the \textit{AE labels} after the transition.
Additionally, the \textit{SF labels} yield more crystalline monomers compared to the other labeling schemes, especially before the transition.
Consequently, when using \textit{SF labels}, it becomes challenging to distinguish early nuclei from spontaneously formed ordered patterns.
We can conclude that the SF labels are not sufficient for a clear discrimination between the crystalline and amorphous phases.

It is interesting to study the dynamics of the crystalline/amorphous labels at different temperatures (time steps).
For this purpose, we look into the number of the crystalline monomers at each step of the time series that survived at the last time step ($t_{end}$ = 74), meaning, the number of common crystalline indices at \(t\) and \(t_{end} = 74\).
We call this number at any time step t, $N_{Surviving} (t, t_{end})$, and repeat this process for the \textit{SF, AE, SL,} and \textit{AO labels}.
We also consider
\begin{equation} \label{eqn:f_sur}
    f_{Surviving}(t,t_{end}) = \frac{N_{Surviving}(t,t_{end})}{N_{Cryst}(t)}\times \frac{N_{tot}}{N_{Cryst}(t_{end})}
\end{equation}
as the fraction of "surviving" crystalline monomers at every step $t$.
In equation~(\ref{eqn:f_sur}), we multiply the surviving fraction by the inverse crystalline fraction at the reference time point $t_{end}$ with respect to the total number of monomers, which is $N_{tot}=10^6$ in the simulation box.
This allows us to compare the result to $f_{Surviving}(t,t_{end})=1$ in the case of random matches only.

\begin{figure}[!ht]
    \subfigure[]
        {
        \includegraphics[width=0.45\hsize]{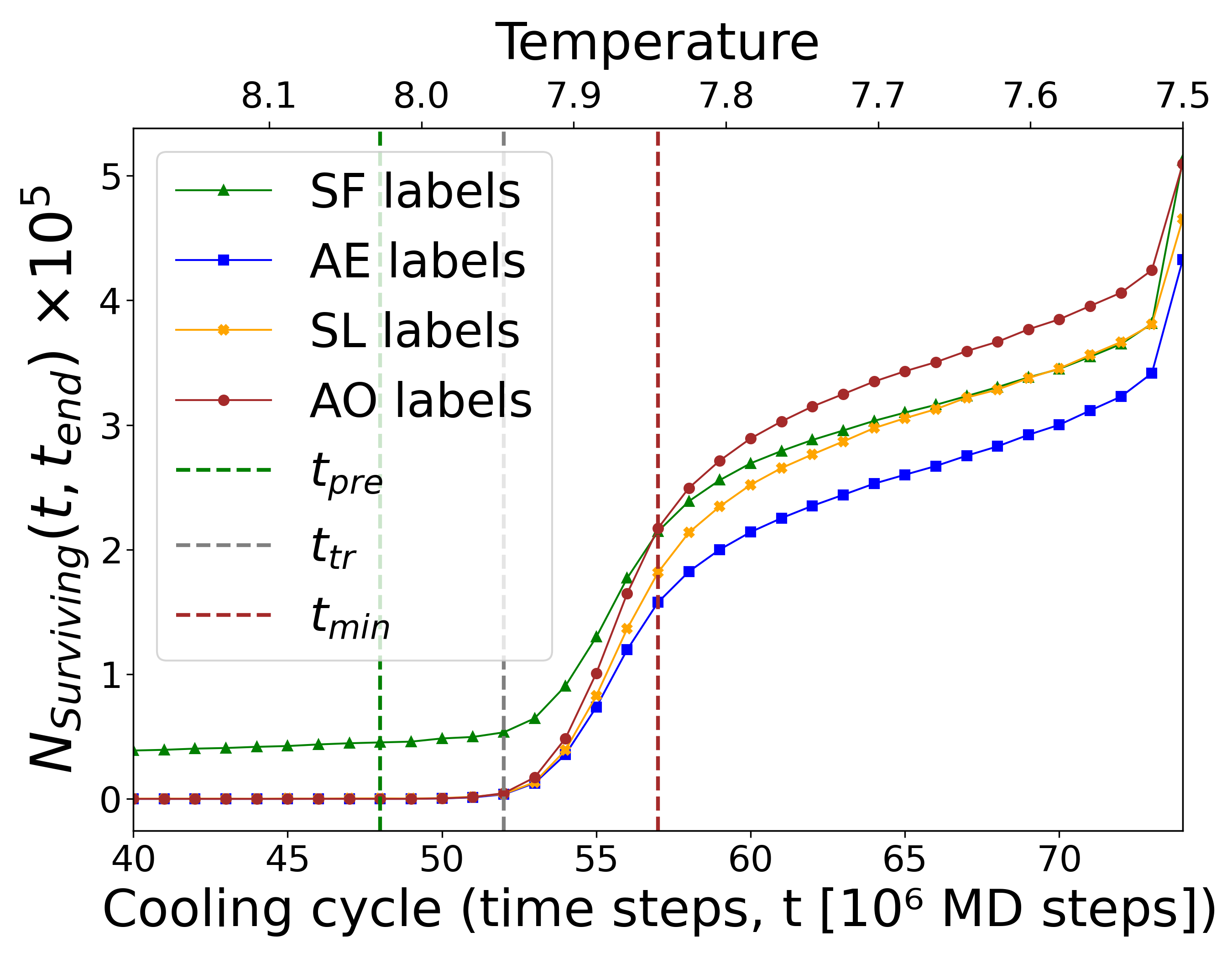}
        \label{fig:Nsurv}
        }
    \subfigure[]
        {
        \includegraphics[width=0.45\hsize]{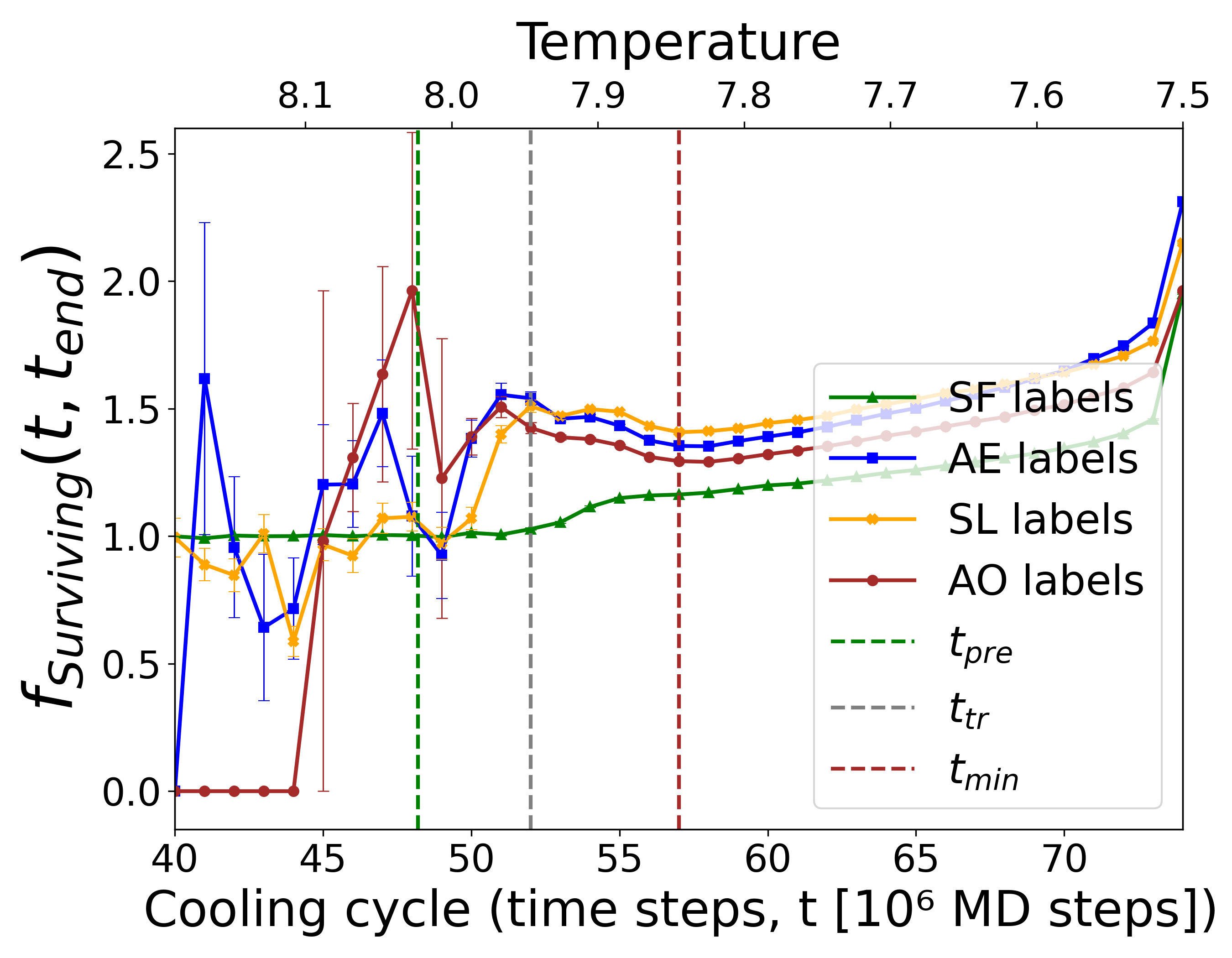}
        \label{fig:fsurv}
        }
    \subfigure[]
        {
        \includegraphics[width=0.31\hsize]{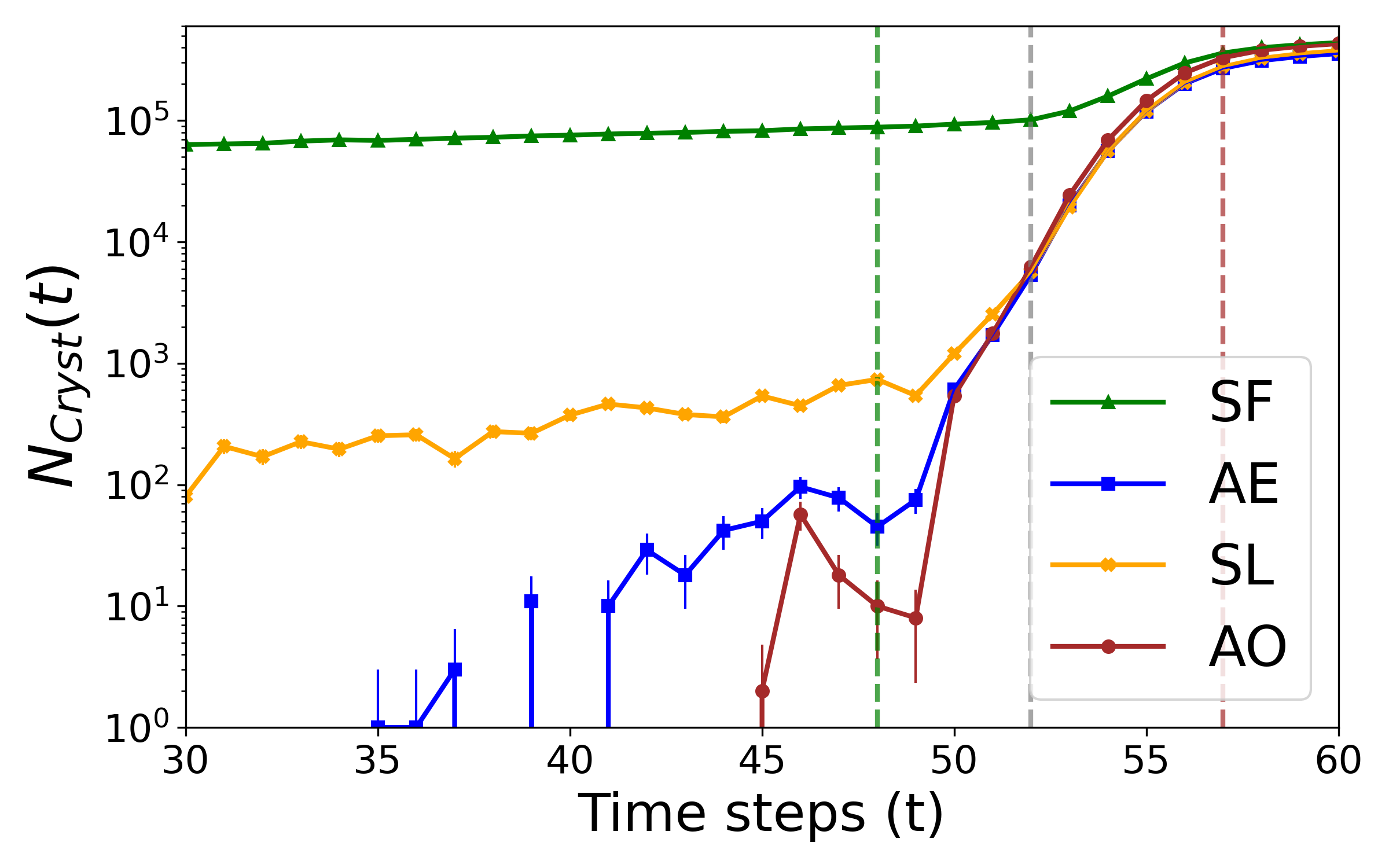}
        \label{fig:avg_crys_frac_zoomed}
        }
    \subfigure[]
        {
        \includegraphics[width=0.31\hsize]{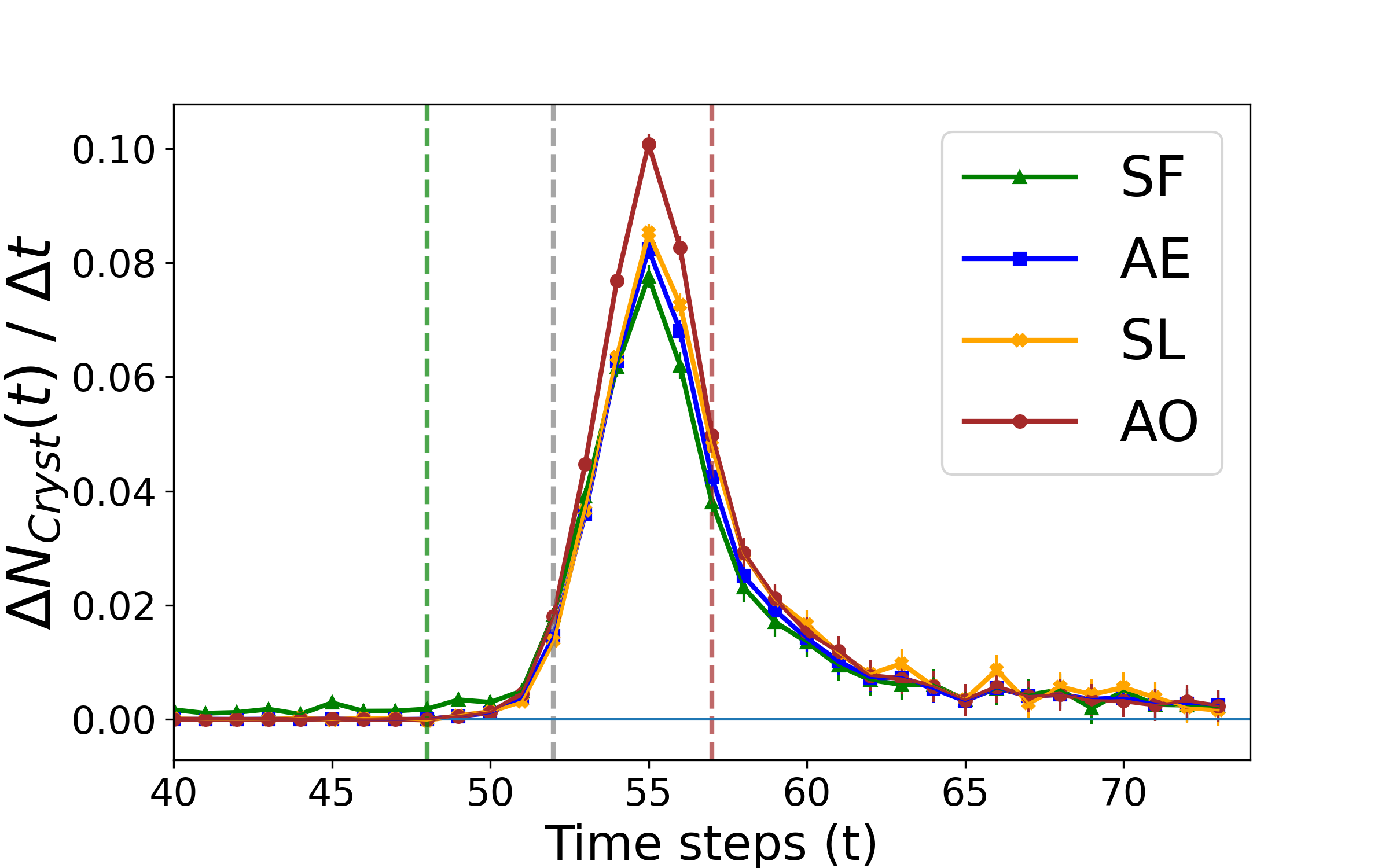}
        \label{fig:dn_dt}
        }
    \subfigure[]
        {
        \includegraphics[width=0.31\hsize]{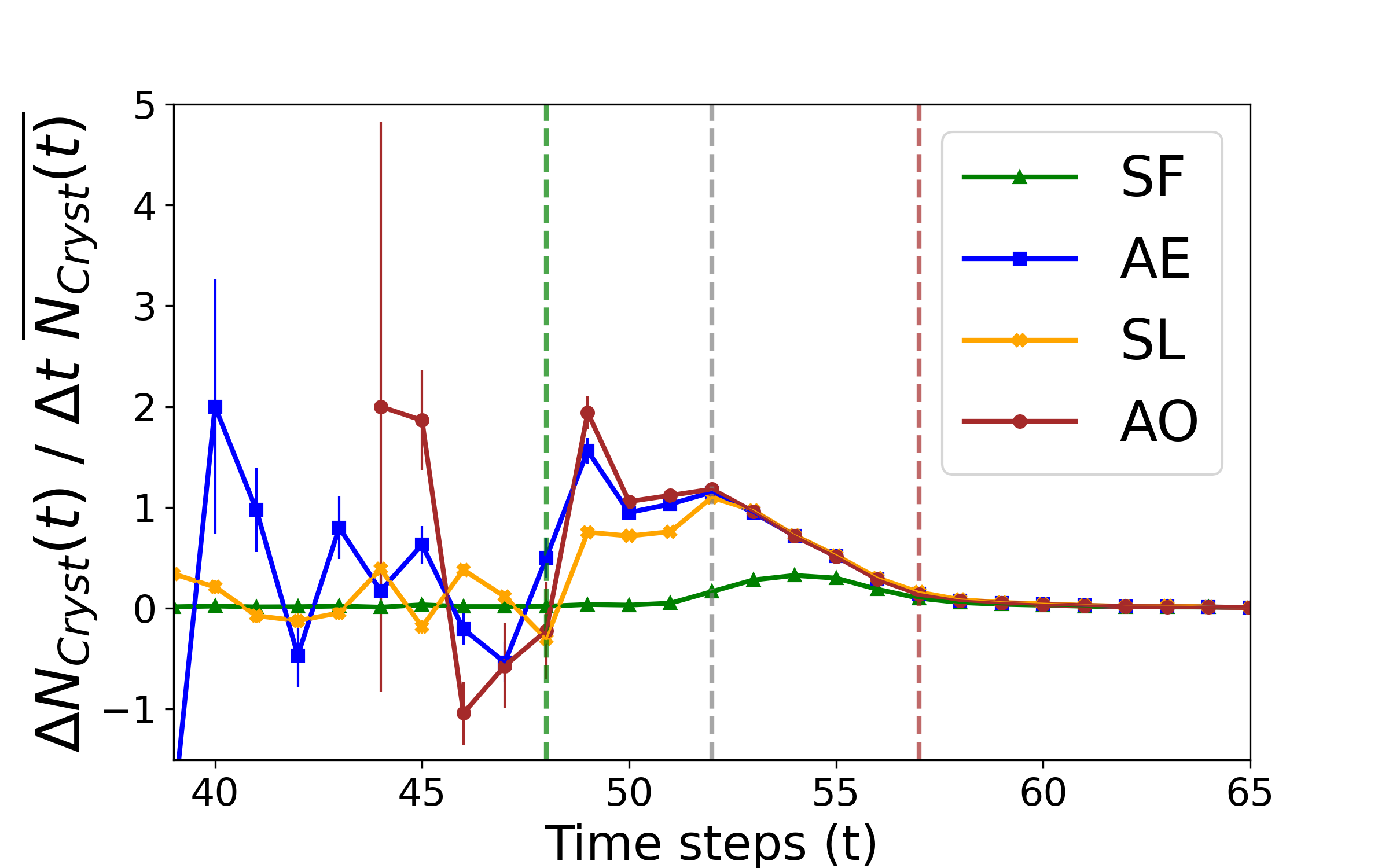}
        \label{fig:dn_dt_by_n}
        }

    \caption{(a) A number of "surviving" monomers, $N_{Surviving}(t,t_{end})$, (see text) as a function of simulation time.
    Vertical dashed lines show characteristic times (table~\ref{tab:times}).
    (b) A fraction of "surviving" monomers (equation~(\ref{eqn:f_sur})),
    (c) a fraction of crystalline labels equivalent to figure~\ref{fig:Timeseries_3} in their logarithmic scale representation and focused on lower temperatures, $N_{Cryst}(t)$,
    (d) the forward difference of the total number of crystalline labels with respect to time, $\Delta N_{Cryst}(t) = N_{Cryst}(t+\Delta t) -N_{Cryst}(t)$,
    (e) a relative derivative of $N_{Cryst}(t)$ between two adjacent time steps.
    $\overline{N_{Cryst}(t)}$ refers to an average of crystalline labels at the two times $t ~\text{and}~ t+1$.
    The characteristic time $t_{pre}$ has been slightly adjusted in (b) to make the error bars visible.
    The error boundaries in (b, c, d, e) were estimated based on the Gaussian error of the number of crystalline labels, represented as $2\sqrt{N_{Cryst}(t)}$.
    Please refer to the text for their respective equations of error bars.
    } \label{fig:surviving_frac}
\end{figure}

In figure~\ref{fig:Nsurv}, we compare the $N_{Surviving} (t, t_{end})$ generated by the different classification methods.
We see a monotonically increasing number of conserved crystalline labels becoming visible beyond the time point $t_{tr}$.
When presented in terms of the surviving fraction in figure~\ref{fig:fsurv}, we observe a significant increase to $f_{Surviving}(t,t_{end})>1$ already at $t_{pre} \leq t \leq t_{tr}$ for the \textit{AE}, \textit{SL}, and \textit{AO labels}, as well as a local decline at later times, $t_{tr} < t \leq t_{min}$.
The results indicate that the crystalline labels generated close to $t_{tr}$ exhibit higher stability with respect to the future crystalline state compared to incoming monomers within the time window $55 \leq t \leq 62$.
Near $t=t_{pre}$, several local maxima and minima are visible, which are presumably characteristic of the series of early nucleation events in the analyzed simulation run.
We anticipate that the shape of the curves to vary between different simulation runs in the region $t<t_{tr}$ and independent simulations are necessary to enhance the resolution.
Nevertheless, the signature observed in the current trajectory is consistent among the \textit{AE, SL} and \textit{AO labels}, showing a local maximum near $t_{pre}$ and a second local maximum near $t_{tr}$.
The error bars for figure~\ref{fig:fsurv} are estimated to facilitate a comparison between classification schemes as $f_{Surviving}(t,t_{end})/\sqrt{N_{Surviving}(t,t_{end})}$.
The error estimate is notably larger for the case of \textit{AO labels} before and near $t_{pre}$ than for \textit{AE labels} due to the smaller crystalline counts for \textit{AO} as illustrated in figure~\ref{fig:avg_crys_frac_zoomed}.

At a later time, $t_{min}\approx 57$ (T = $0.787$), we observe in figure~\ref{fig:fsurv}, a local minimum for all the three labels \textit{AE, SL} and \textit{AO}.
The local minimum in $f_{Surviving}(t, t_{end})$ near $t_{min}$ results from a cross-over towards ripening as $t_{end}$ is approached.
This transition requires $N_{Surviving}(t,t_{end})/N_{Cryst}(t)=1$ at $t=t_{end}$, and the common fraction must decay when moving backward in time due to the mobility of the stems.

We shall examine the features seen in the surviving fraction $f_{Surviving}(t, t_{end})$ within the context of the initial increase in the overall number of crystalline monomers $N_{Cryst}(t)$ near $t_{tr}$, as depicted in figure~\ref{fig:Timeseries_3} and its logarithmic representation in figure~\ref{fig:avg_crys_frac_zoomed}.
For this purpose, we define the forward difference 
\begin{equation}
    \Delta N_{Cryst}(t)=N_{Cryst}(t+\Delta t)-N_{Cryst}(t)~~,
\end{equation}
that is shown in figure~\ref{fig:dn_dt} for $\Delta t=1$.
Here, a maximum is visible between $t_{tr}$ and $t_{min}$ for all the classification schemes that marks the inflection point in crystalline fraction at $t=55$ ((please refer to figure~\ref{fig:Timeseries_3} for comparison).

In order to obtain a relative measure of $\Delta N_{Cryst}(t)$ with respect to the current crystalline number analogous to equation~(\ref{eqn:f_sur}), we define the relative increase of crystalline number as
\begin{equation}
\Delta f_{Cryst}(t)=\frac{N_{Cryst}(t+\Delta t)-N_{Cryst}(t)}{\Delta t \times \frac{1}{2} [N_{Cryst}(t+\Delta t)+N_{Cryst}(t)]}~~.\label{eq:df_cryst}
\end{equation}
In equation (\ref{eq:df_cryst}), the denominator corresponds to an average over two time steps.
The results are shown in figure~\ref{fig:dn_dt_by_n}.
More details emerge in $\Delta f_{Cryst}(t)$ at times $t<t_{tr}$, and the peak of the relative increase of the crystalline fraction is now located at a time before $t_{tr}$.
We estimated the error bars in figure~\ref{fig:dn_dt_by_n} by assuming the statistical independence of crystalline labels, leading to a standard deviation of $\sigma_{N_{Cryst}}=\sqrt{N_{Cryst}(t)}$, and a (Gaussian) combined standard error of $\Delta f_{Cryst}(t)$, which is given by $2\sqrt{N_{Cryst}^3(t+\Delta t)+N_{Cryst}^3(t)}/[N_{Cryst}(t+\Delta t)+N_{Cryst}(t)]^2$.
The assumed independence of the labels may lead to an underestimation of the error, especially when clusters grow after $t_{tr}$.
However, the estimate helps to distinguish fluctuations before transition from later collective changes in crystallinity in figure~\ref{fig:dn_dt_by_n}.

While the curves before $t_{pre}$ are primarily governed by fluctuations, there is a noticeable increase in $\Delta f_{Cryst}(t)$ shortly after $t_{pre}$, followed by a decline towards $\Delta f_{Cryst}(t) \to 0$ after $t_{min}$.
This pattern closely corresponds to the observed region of enhanced surviving fraction $f_{Surviving}(t,t_{end})$ between $t_{pre}$ and $t_{min}$ (see figure~\ref{fig:fsurv}).
For the \textit{AE} and \textit{AO labels}, the shapes of the curves $\Delta f_{Cryst}(t)$ (figure~\ref{fig:dn_dt_by_n}), including local maxima near $t_{pre}$ and $t_{tr}$, closely correspond to the features seen in $f_{Surviving}(t,t_{end})$ (figure~\ref{fig:fsurv}).
The monomers contributing to $\Delta f_{Cryst}(t)$ and $f_{Surviving}(t, t_{end})$ are not necessarily identical.
Thus, determining the causality of the apparent relationship between $\Delta f_{Cryst}(t)$ and $f_{Surviving}(t,t_{end})$ is challenging.
On one hand, the rapid relative increase in the crystalline fraction makes it more likely for monomers to be preserved in the crystalline state near $t_{tr}$.
For instance, the escape from the center of a crystalline cluster can be suppressed by other incoming monomers at the crystalline/amorphous interface.
Conversely, the appearance of conserved monomers can serve as a driving force for the later growth of the nucleus.
To address the question of time order and hence, the causality of the signals in figures~\ref{fig:fsurv} and \ref{fig:dn_dt_by_n}, an enhanced time resolution of stored conformations is required near the transition.

As seen in figure~\ref{fig:avg_crys_frac_zoomed}, both the alignment and conformational order start to develop concurrently, but this occurs only from $t=50$, which is near the local maxima in $f_{Surviving}(t,t_{end})$.
This indicates a transitioning step that is not well resolved in the overall thermodynamic quantities such as specific volume or conformation order parameters.

For the \textit{AO labels}, we observe the largest specificity of crystalline labels before the transition $t_{tr}$ in terms of the steepest slope in figure~\ref{fig:avg_crys_frac_zoomed} at $t_{pre}<t<t_{tr}$ and the lowest fraction of crystalline labels at times $t<t_{pre}$.
On the other hand, as $t$ approaches $t_{end}$, the \textit{AO labels} yield the highest counts, similar to the \textit{SF labels}, demonstrating their high sensitivity to established crystalline regions.
For instance, the crystallinity fraction for the \textit{AO labels} reaches over $50\%$ at $t_{end}$, while it is $43\%$ for the \textit{AE labels}.

The second most specific labels in these terms are provided by the \textit{AE labels}.
They may show higher counts before the transition since they do not necessitate the entire environment under consideration to display orientational order.
Instead, they rely on a combination of conformation and orientation order, as discussed in the previous section.
The higher sensitivity during the initial nucleation times as compared to the \textit{AO} may lead to reduced statistical uncertainty in the count of crystalline labels.
This is illustrated by the error bars shown in figures \ref{fig:avg_crys_frac_zoomed} and \ref{fig:dn_dt}.

When considering conformation order alone, as in the case of \textit{SL labels}, part of the specificity is compromised.
This can be observed in the fact that there are approximately ten times as many \textit{SL} crystalline labels before $t_{pre}$ in figure~\ref{fig:avg_crys_frac_zoomed} as compared to \textit{AE}.

For the \textit{SF labels}, we do not see a local maximum near $t=t_{tr}$ or $t=t_{pre}$ in figure~\ref{fig:fsurv}.
This can be explained by the lower specificity of \textit{SF labels} for the early nuclei, which are outnumbered by false positives.
In other words, there appears to be an approximate $10\%$ crystalline fraction before the transition as evident in figure~\ref{fig:avg_crys_frac_zoomed}.
Hence, the signal is diminished by the larger denominator in equation~(\ref{eqn:f_sur}).

\subsection{Phase change kinetics of individual monomers}

\begin{figure}[!ht]
    \centering
    \includegraphics[width=\hsize]{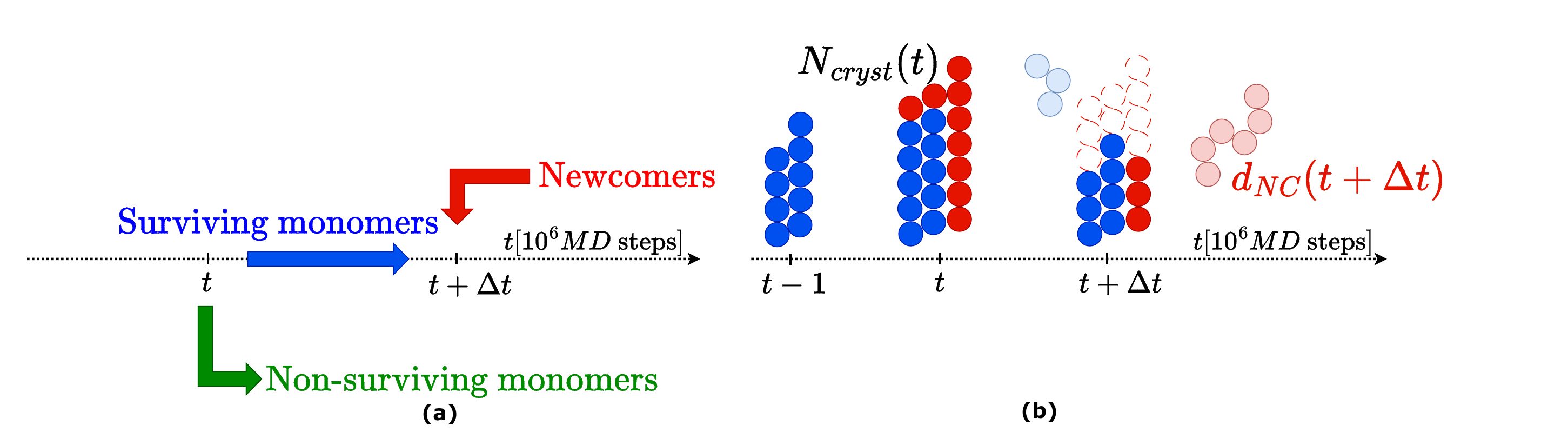}
    \caption{(a) A schematic diagram illustrating the exchange of monomers between the crystalline and amorphous phases at two-time points, \(t\) and \(t + \Delta t\).
    Some crystalline monomer indices are common at both time steps ("surviving" monomers).
    In contrast, some monomers lose their crystallinity and become a part of the amorphous phase at \(t + \Delta t\) ("non-surviving" monomers).
    Since it is a cooling cycle, there are newly added monomers to the crystalline phase at \(t + \Delta t\) (newcomers).
    \label{fig:clabels} 
    (b) Let us consider that at time step $t$, the solid red-colored monomers are newcomers to the crystalline phase compared to the previous time step, $t-1$.
    Now, we are interested in determining how many of these newcomers, at time $t + \Delta t$, return to the amorphous phase and are no longer in a crystalline state.
    The process of returning to the amorphous phase is referred to as the 'decay of newcomers,' denoted as $d_{NC}(t + \Delta t)$.
    \label{fig:new_comers}}
\end{figure}

To investigate the early nucleation behavior further, we studied the exchange between the amorphous and the crystalline monomers by tracking the indices of phase-changing monomers throughout the time series.
For this purpose, we divide the crystalline monomers as shown in figure~\ref{fig:clabels} further into the following subgroups: consider two-time steps separated by \(\Delta t\).
The number of common crystalline monomer indices between \(t\) and \(t + \Delta t\) are shown as "surviving" monomers, $N_{Surviving}(t, t+\Delta t)$ generalizing the definition at equation~(\ref{eqn:f_sur}) above.
The number $N_{Decay}(t,t+\Delta t)$, refers to the crystalline monomers at \(t\) that are not crystalline at \(t + \Delta t\) and join the amorphous phase within a time difference of \(\Delta t\).
At \(t + \Delta t\), the total crystalline monomers $N_{Cryst}(t+\Delta t)$, would be the sum of "surviving" monomers from \(t\) and some newly added crystalline monomers shown as newcomers "NC" according to:
\begin{equation}\label{eqn:clabels}
    \begin{split}
    N_{Cryst}(t) & = N_{Surviving}(t,t+\Delta t) + N_{Decay}(t,t+\Delta t) \\
    N_{Cryst}(t+\Delta t) & = N_{Surviving}(t,t+\Delta t) + N_{NC}(t,t+\Delta t)~~.
    \end{split}
\end{equation}

Now, we are interested in the survival of the newcomers to the crystalline state.
In other words, we investigate when in the future times would these newcomers to the crystalline phase return to the amorphous phase.
We call it the decay of newcomers, $d_{NC} (t-\Delta t_-,t+\Delta t)$.
This refers to the current time step, $t$, with a past time step, $\Delta t_-$.
It is equivalent to the equation~\ref{eqn:clabels} as: $N_{Cryst}(t)=N_{Surviving}(t-\Delta t_-,t) + N_{NC}(t-\Delta t_-,t)$.
This means that we study the decay behavior of newcomers at $t$ after a time interval of $\Delta t$ in the future $d_{NC} (t, t+\Delta t)$, i.e. how many monomers joined to the crystalline phase at $t$ will not stay crystalline at $t + \Delta t$ and will become amorphous.
For this, we simply count the number of common amorphous indices at $t + \Delta t$ and the indices of newcomers at $t$.
In the following, we use $\Delta t_-=1$ ($1\times10^6$MD steps in the past) to define the number of newcomers at $t$.

\begin{figure}[!ht]
    \centering
    \includegraphics[width=0.6\hsize, height = 0.7\hsize]{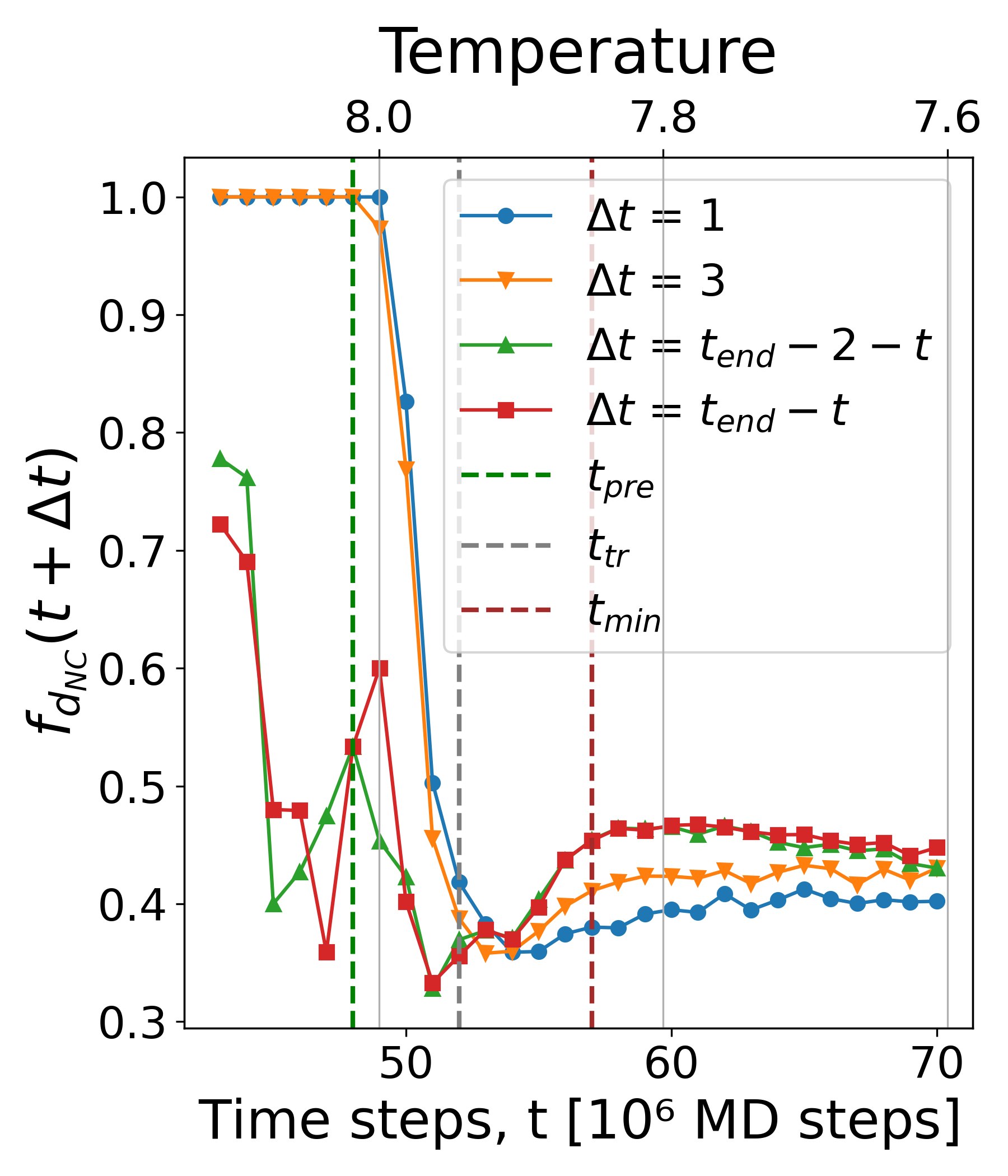}
    \caption{A decay in the newcomers $d_{NC} (t + \Delta t)$, divided by the total newcomers at that time step $N_{NC}(t)$, is denoted as $f_{d_{NC}}(t + \Delta t)$.
    It is analyzed after different time intervals in the future \(\Delta t\), of the \textit{AE labels}.
    The brown dashed line is at the characteristic time step, $t_{min}$ as in figure~\ref{fig:fsurv}, where, the period of high "surviving" fraction of the crystalline labels $f_{Surviving}(t,t_{end})$ after a local maximum settles down to a local minimum.
    $\Delta t = t_{end} - t = 74 - t$ in the legend refers to the comparison with the last time step.
    We can observe that the newcomers added to the crystalline phase around the transition point $t_{tr}$, (grey dashed line) have a low decay fraction.
    This refers to their minimum tendency to join the amorphous phase.
    After a few time steps, i.e. after $t_{min}$), the decay fraction starts to rise again.
    The steep slopes in the $f_{d_{NC}}(t+\Delta t)$ lines right before $t_{tr}$ indicate the presence of some strong alignment patterns near transition.} \label{fig:decay_newcomers}
\end{figure}

In figure~\ref{fig:decay_newcomers}, we present the relative decay $f_{d_{NC}}(t+\Delta t)$ of newcomers for the case of \textit{AE labels} defined as
\begin{equation}\label{eqn:decayNC}
    \begin{split}
    f_{d_{NC}}(t,t+\Delta t) = \frac{d_{NC} (t,t+\Delta t)}{N_{NC}(t)}~~.
    \end{split}
\end{equation}
In equation~(\ref{eqn:decayNC}), we omit the reference to $t-\Delta t_-$ because $\Delta t_-$ is not varied in the analysis.
The value of $f_{d_{NC}}(t, t+\Delta t)$ is studied for various future time points including the immediate future of their arrival to the crystalline phase, \(\Delta t = 1\), \(\Delta t = 2\), and concerning the last time step \(\Delta t = t_{end} - t = 74 - t\).

We see in figure~\ref{fig:decay_newcomers} that for small values of $\Delta t$, there is a rapid decrease of $f_{d_{NC}}(t+\Delta t)$ before the transition point $t_{tr}$.
Hence, we designate this time point as a new characteristic time step, denoted as $t_{pre}$ (see table~\ref{tab:times}), during which a fraction of the newcomers abruptly starts to be conserved.
Before $t_{pre}$, virtually all newcomers decay back to the amorphous phase within $\Delta t$.
For all values of $\Delta t$, a local minimum is observed in $f_{d_{NC}}(t+\Delta t)$, indicating greater stability of newcomers approaching early nuclei compared to those at earlier or later time points.
After $t_{tr}$, and near the local minima of the overall "surviving" fraction $t_{min}$ (from figure~\ref{fig:fsurv}), the decay curve starts to increase again reaching a plateau of $\approx 48\%$ in case of $\Delta t = t_{end} - t$.
At the onset of the transition, more stable crystalline labels appear, and as the system cools further, we observe more exchanges between the crystalline and amorphous monomers.

One possible explanation for the existence of local minima is that early nuclei encounter relatively little resistance when recruiting additional monomers in their immediate environment, as long as the growing crystallites remain smaller than the constraints related to connectivity or entanglements.
Therefore, early monomers are more likely to be buried beneath the subsequently added crystalline layers (also observed in the local maxima near $t_{tr}$) in figure~\ref{fig:decay_newcomers}.
As we pass beyond $t_{min}$, the connectivity constraint, along with the emergence of several competing neighboring domains, comes into play.
This makes the survival of newcoming monomers less likely.

The rapid exchange of monomers between crystalline domains and the amorphous phase facilitates the transfer between domains, a process known as Ostwald ripening, which occurs beyond the nucleation point.
In that context, we anticipate a monotonous increase in the average domain size, with monomers within the larger domain gaining bulk volume.
The monomers in larger domains are supposed to contribute more likely to the surviving fraction as indicated by the steady positive slope of $f_{Surviving}(t,t_{end})$ after the local minimum $t_{min}$ in figure~\ref{fig:fsurv}.
Please note that the exchange between the domains occurs only at longer time scales, typically after the early nucleation phase when inter-domain distances narrow, allowing for monomer exchange through chain relaxation.
At $t_{tr}$, the local maximum in $f_{Surviving}(t,t_{end})$ and accordingly a minimum in $f_{d_{NC}}(t+\Delta t)$, indicate a sudden concentration of (surviving) crystalline order in the early nuclei after they are formed near $t_{tr}$.
The markers $t_{tr}$ and $t_{min}$, therefore, label a distinct stage of rapid growth between nucleation and slowed-down growth or ripening.
The steep slope of $f_{d_{NC}}(t+\Delta t)$ with a sudden onset near $t_{pre}$ is an interesting feature in this context as it indicates the presence of a distinct rearrangement pattern before the transition is visible in the overall crystalline fraction or specific volume at $t_{tr}$ in figure~\ref{fig:Timeseries_3}.

An open question remains regarding the extent to which the overall high mobility of stems, a characteristic known in the CG-PVA model, influences the observed effect.
Specifically, does it primarily govern prefactors such as the exchange time scale and the overall onset of $f_{d_{NC}}$?
To address this problem, it will be worthwhile to investigate the exchange dynamics in other CG polymer models and compare them to the crystallization of atoms or small compounds~\cite{freitas2020uncovering}.

\section{Conclusion}\label{sec:conclusion}

We demonstrate that data-driven methods can detect the crystalline phase in polymers based on the local structural fingerprint in contrast to "classical" methods, where decision boundaries are positioned manually such as by defining a minimal stem length (SL) or a minimal orientational order in the local environment.
The challenge is to reduce the information from the local environment of a monomer into binary data indicating whether it belongs to the amorphous or crystalline phase.
Because of chain connectivity, describing both the local conformation and the local environment within a small number of correlation shells is sufficient to reach a decision.
However, near interfaces or internal defects in crystal domains, this decision is not unique.
This is in contrast to atom, colloidal, or compound-based crystals (even crystallized proteins), where the crystal is defined on the scale of the unit cell.

We define local fingerprints that consider the alignment within neighboring bond vectors and the local chain conformation descriptor based on the scalar products of bond vectors along the chain.
Their concatenated version is referred to as structural fingerprints (SF).
An employed autoencoder (AE) compresses the fingerprint to a two-dimensional latent space.
The distribution of the compressed fingerprints in latent space exhibits structures that can be linked to the emergence of the crystalline phase.
This organization is based on the degrees of local conformation and orientation order.
We use the Gaussian mixture model (GMM) to cluster monomers in the latent space to the amorphous and crystalline phases.
The in-plane hexagonal packing observed between stems in the classified crystal domain, as confirmed by a fast Fourier transform of the projected domain image, validates the attribution of the identified classes to both the amorphous and crystalline portions.
It is worth noting that a simple hierarchical clustering of the SF tends to overestimate the fraction of crystalline monomers in the early stages.
In contrast, the additional information about the relevant features of the SF, as contained in the bottleneck neurons of the AE, allows for a higher resolution in characterizing the early stages.

The obtained crystalline fraction aligns with a previously established classification based on the SL, provided the threshold is raised slightly to a number of $17$ consecutive monomers.
The SL is a collective measure along the chain contour.
This means that to determine the phase of a monomer, it must be assessed in the context of the environment if it is part of the stem.
Whereas, the GMM classification based on AE compressed fingerprints is applied on individual monomer level.

This enabled us to investigate nucleation and growth dynamics with high resolution at the monomer index level.
Based on the data-driven binary labeling of the monomers, we follow the phase-change kinetics of individual monomers.
In particular, we quantify the number of monomers joining the crystalline phase that is conserved (or common) at a future time step such as the next time step or the end of the simulation.
Near the transition point, we observe a local maximum in the fraction of conserved monomers.
Such consistency is not seen at later stages of the cooling cycle, indicating higher crystallization efficiency during those times.
The local maximum of the conserved fraction is consistent with the minimum observed in the fraction of incoming monomers that will leave the crystalline state in the future.
We quantify the "decay of newcomers" to the crystalline state and observe a sudden drop in this value before the transition becomes visible, for instance, as seen in the specific volume.
The differences in the development of alignment and conformation order at the point of sudden stabilization of crystallites indicate a complex nucleation behavior in polymer melts.

\section*{Data Availability}

The processed data used to generate the plots, along with the original simulation snapshots and trained machine learning models, can be found in an online repository on Zenodo at \url{https://doi.org/10.5281/zenodo.8383061}~\cite{bhardwajZenodo2023}.

\section*{Acknowledgements}

J.U.S. acknowledges financial support from the German Research Foundation (Grant No. SO 277/19).
The authors acknowledge the Center for High-Performance Computing (ZIH) Dresden for granting simulation time, Ankush Checkervarty for the technical support in performing the simulations, as well as Huzaifa Shabbir for inspiring discussions.

\printbibliography

\newpage

\section{Appendix}

\subsection{Artificial neural networks}

\begin{table}[h]
    \begin{tabular}{ | Sl | Sl | }
	\hline
        \multicolumn{2}{|Sl|}{\textbf{Hyperparameters used for training an AE}}\\
	\hline
	Length (number of hidden dimensions) & \(14 \) \\
        \hline
	Depth (total number of layers) & \( 14 \) \\
	\hline
        Activation function on every layer & elu  \\
	\hline
	Bottleneck neurons (compressed latent space) & \( 2 \) \\
	\hline
	Number of epochs &  \(600-700\) \\
	\hline
	Batch size &  \(20\) \\
	\hline
        Validation split &  \(0.2\) \\
        \hline
        Learning rate (initial) & \(10e^{-5}\) \\
        \hline
        $L_2$ regularizer & \(10e^{-7}\) \\
        \hline
        Optimizer for compiling the model  & adam \\
        \hline
    \end{tabular}
    \caption{These hyperparameters are chosen to train an autoencoder (AE) on the concatenated list of structural fingerprints (SF) for this work.
    The idea is to compress the large input information to the most crucially required latent data.
    The learning rate was kept variable to avoid overfitting as discussed in section~\ref{sec:ae_description}.
    \label{tab:ae_hyper}}
\end{table}

\begin{table}
    \begin{tabular}{ | Sl | Sl | }
	\hline
        \multicolumn{2}{|Sl|}{\textbf{Hyperparameters used for training an MLP}}\\
	\hline
 	Number of input neurons & \( 17 \) \\
        \hline
	Length & \(15 \) \\
        \hline
	Depth (number of layers) & \( 2 \) \\
	\hline
	Activation function on every layer & tanh  \\
	\hline
        Activation function on the last layer      & sigmoid  \\
	\hline
	Number of epochs &  \(200 - 250\) \\
	\hline
	Batch size &  \(10\) \\
	\hline
        Validation split &  \(0.2\) \\
        \hline
        Learning rate (initial) & \(10e^{-3}\) \\
        \hline
        $L_2$ regularizer & \(10e^{-5}\) \\
        \hline
        Optimizer for compiling the model  & adamax \\
        \hline
        Loss function & binary cross-entropy \\
        \hline
    \end{tabular}
    \caption{These hyperparameters are used to train a multilayer perceptron (MLP) for supervised learning: to train the concatenated list of structural fingerprints (SF) with the labels produced by hierarchical clustering.
    The length of the MLP represents the number of hidden dimensions and the depth represents the number of hidden layers.} \label{tab:mlp_hyper}
\end{table}

\begin{figure}
    \centering
    \includegraphics[width=0.5\hsize]{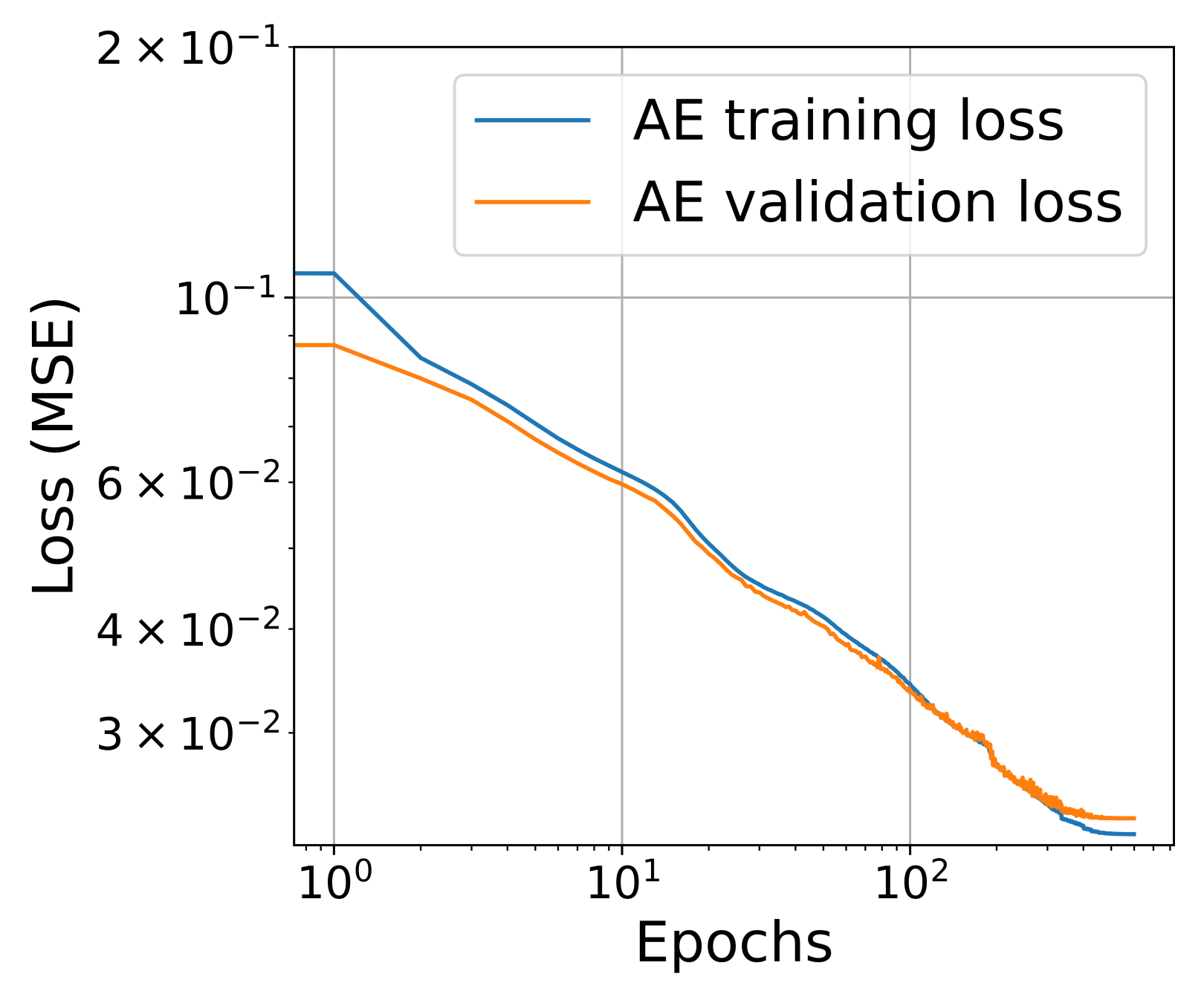}
    \caption{The training curve of an autoencoder (AE) with a variable learning rate (LR) that depends on an increase in the validation loss such that the LR is decreased by a factor of 0.5 after every 25th consecutive increase in the validation loss.
    The losses are calculated as a mean square error (MSE) as given in equation~\ref{eq:loss} and seem to stabilize after around 650 epochs.} \label{fig:AE_training}
\end{figure}

\subsection{Choice of training dataset}\label{sec:choiceoftrainingdataset}

The trained models: an autoencoder (AE), and a Gaussian mixture model (GMM), are slightly dependent on the choice of the dataset.
One can, in principle, choose any cross-section of the simulation and use its data points (or coarse-grained monomers) to train an AE on their structural fingerprints (SF) and compress them to their latent space vectors.
A GMM can be trained on compressed fingerprints to classify every monomer to the crystalline and amorphous phases.
The table below, \ref{tab:dataset}, contains several cross-sections of the simulation box and a couple of cases where data points are chosen at random from the simulation box.
Considering the column-wise selection (shown in bold) of each kind of dataset, we start with the 14th slice and use it to train the models.
Then, we predict all the crystalline labels at the last time step, $t_{end} = 74$, and call them $N_{current}$.
We will get slightly different predictions in crystalline label indices if we use a different cross-section to train the models.
Let’s call them $N_{alternative}$.
One can calculate the deviation found in both sets of crystalline labels $N_{current}$ and $N_{alternative}$ and refer to it as $N_{deviation}$.
This percentage of $N_{deviation}$ is calculated as, 
\begin{equation}
    \frac{N_{deviation}}{N_{current}} \times 100
\end{equation}

\begin{table}
    \begin{tabular}{ | Sc | Sc | Sc | Sc | Sc | Sc | Sc | Sc |}
	\hline
        \textbf{Dataset} & \textbf{Slice 14} & \textbf{Slice 15} & \textbf{Slice 56} & \textbf{Slice 62} & \textbf{Slice 68} & \textbf{10k} & \textbf{20k}\\
	\hline
        Slice 14  & 99.95 & 1.47  & 1.22  & 0.27  & 2.51  & 1.91 & 1.97 \\
        \hline
        Slice 15  & 0.19  & 100.0 & 0.25  & 0.09  & 1.18  & 0.75 & 0.62 \\
	\hline
	Slice 56  & 0.29  & 0.61  & 100.0 & 0.15  & 1.46  & 1.01 & 0.91 \\
	\hline
	Slice 62  & 2.19  & 3.24  & 2.96  & 100.0 & 4.23  & 3.56 & 3.69 \\
	\hline
        Slice 68  & 0.35  & 0.26  & 0.19  & 0.23  & 99.93 & 0.21 & 0.13 \\
	\hline
	10k       & 0.67  & 0.79  & 0.68  & 0.46  & 1.16  & 99.8 & 0.70 \\
	\hline
	20k       & 0.51  & 0.43  & 0.35  & 0.37  & 0.85  & 0.47 & 99.6 \\
	\hline
    \end{tabular}
    \caption{\label{tab:dataset} A slight deviation is seen in the crystalline label indices while choosing a particular dataset to train the ML models over others.
    In the table, the non-diagonal elements are calculated as $N_{deviation}/N_{current}$.
    The bold datasets are chosen as $N_{current}$, one at a time, and compared with all others down in the column, $N_{deviation}$.
    The diagonal entries are replaced by $N_{Overlapping}(t_{end})/N_{Cryst}(t_{end})$ (refer to the text of equation~\ref{eq:overlap}).
    All the entries are shown in percentages.
    The last two rows and columns contain the cases of choosing $10^4~(10\mathrm{k})$ and $2\times10^4~(20\mathrm{k})$ data points at random from the simulation box.}
\end{table}

We repeat the process by choosing all the different datasets mentioned in table~\ref{tab:dataset} to train the models.
The diagonal elements of these would always be zero (no non-matching crystalline labels when compared to itself).
These elements are replaced by another important criterion that plays an important role in choosing a meaningful training dataset.
We look for the percentage overlap of crystalline labels between the application of GMM on the compressed data (referred to as $N_{Cryst}(t_{end})$) and the application of HC on the SF.
Since the HC applied on the SF is independent of the point cloud generated by the AE, we look if the crystalline labels via GMM lie in a similar area as them.
We name the overlapping indices as $N_{overlapping}(t_{end})$.
One can make this comparison through grids or on the index level.
We have used the monomer indices to make this comparison.
This percentage is calculated as 
\begin{equation}
    \frac{N_{Overlapping}(t_{end})}{N_{Cryst}(t_{end})} \times 100
    \label{eq:overlap}
\end{equation}

\subsection{Crystal domains}\label{sec:domains}

The crystalline labels from all strategies \textit{SF, AE, SL} and \textit{AO} can be further sub-classified into their respective crystal domains, to verify their crystalline structure.
We define a function that visits all the crystalline labels and categorizes each domain separately.
There can be several ways to do this step.
For instance, one can do this by identifying the immediate surrounding neighbors of each monomer within a defined cutoff distance.
An appropriate cutoff distance can be decided after several trials.
Then, one can start with the first monomer and save its immediate neighbors into one domain.
Each of those immediate neighbors is visited again to get their neighbors and the process continues until no new neighbors are added to this domain.
Then, we move on to other non-visited monomers (not a part of the first domain) and find their fellow domain members.
This step can be performed until we have visited all crystalline labeled monomers.

\begin{figure}
    \centering
    \subfigure[]{
        \includegraphics[width=.44\hsize]{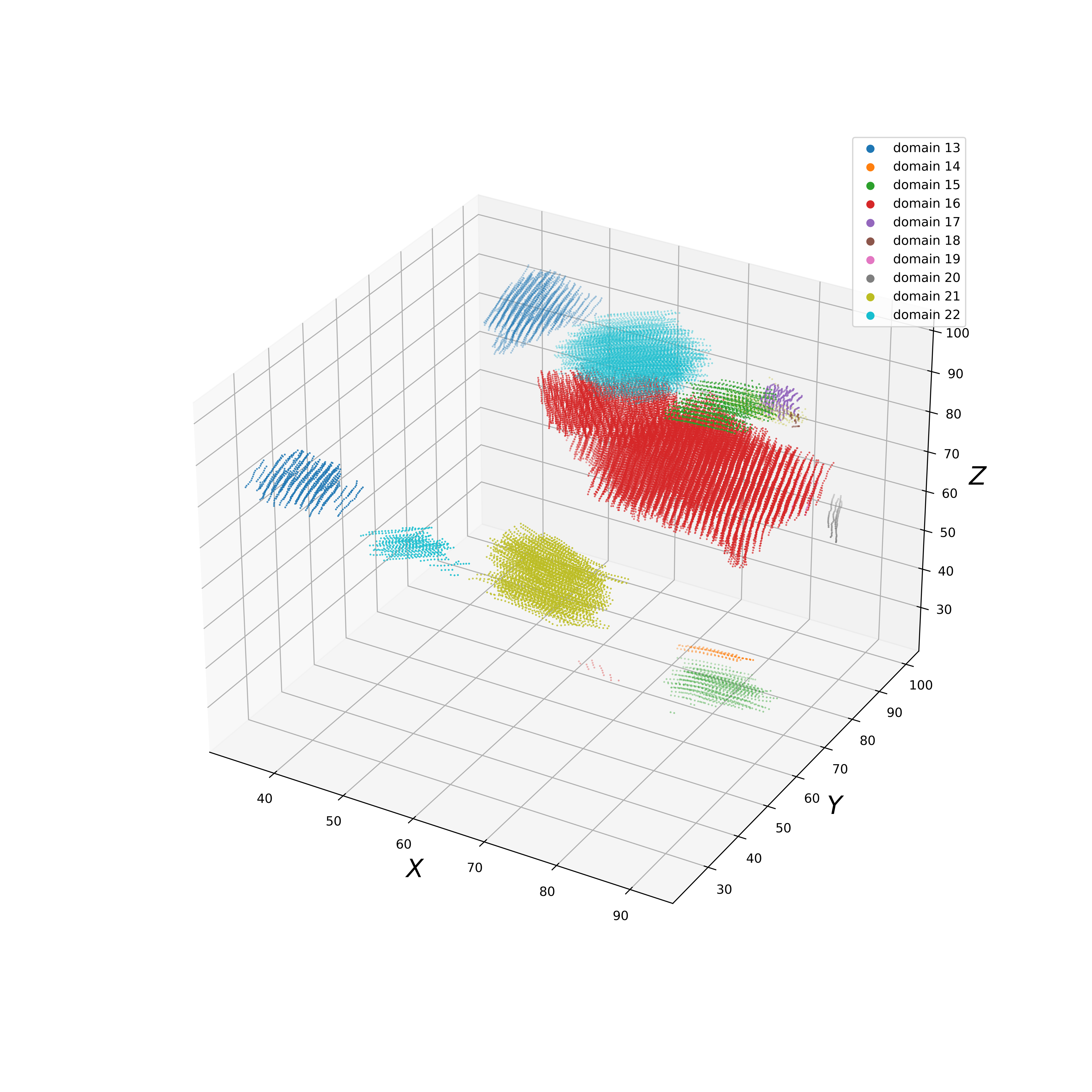}
        \label{fig:Diff_Clusters}
        }
    \subfigure[]
        {
        \includegraphics[width=0.26\hsize]{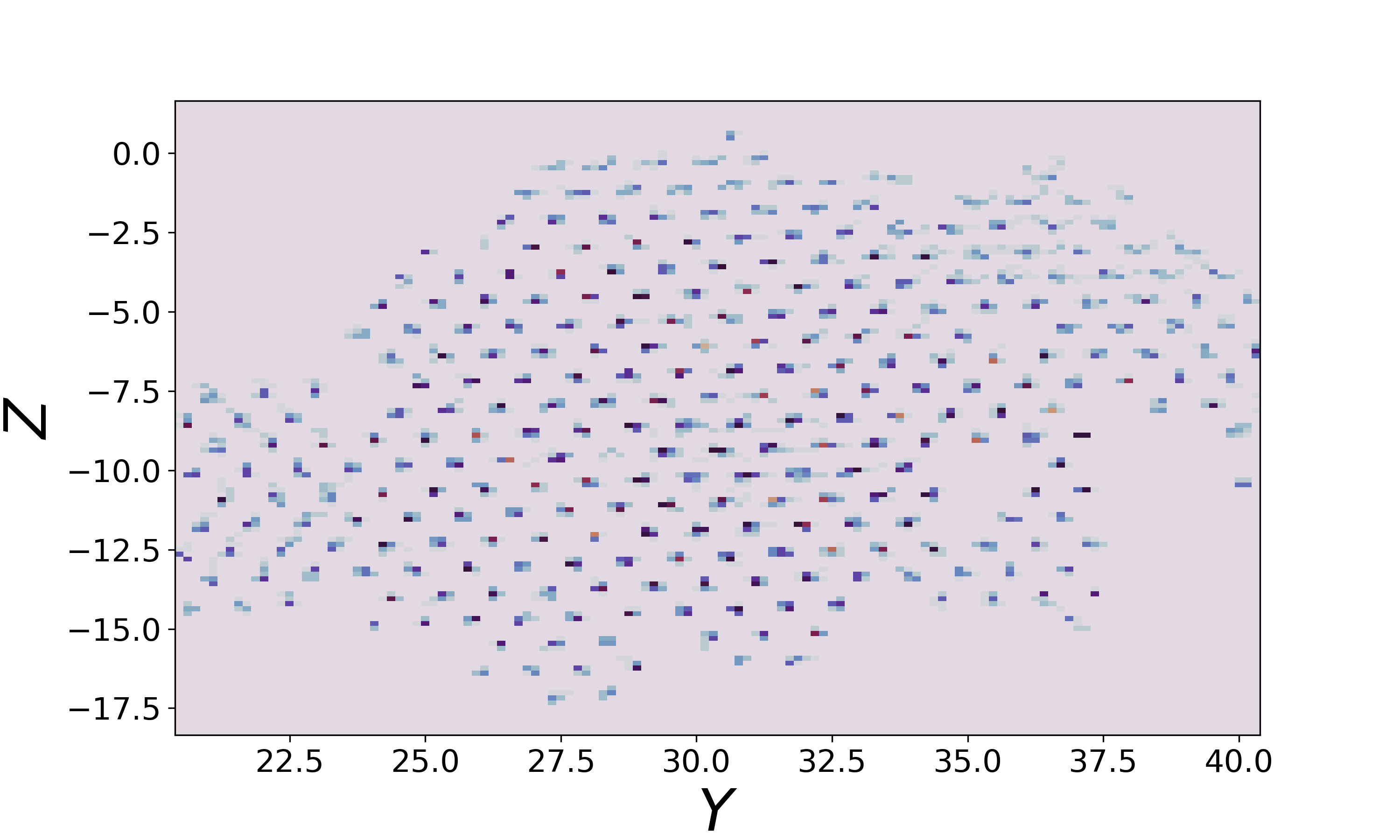}
        \label{fig:2d_hist}
        }
    \subfigure[]
        {
        \includegraphics[width=0.22\hsize]{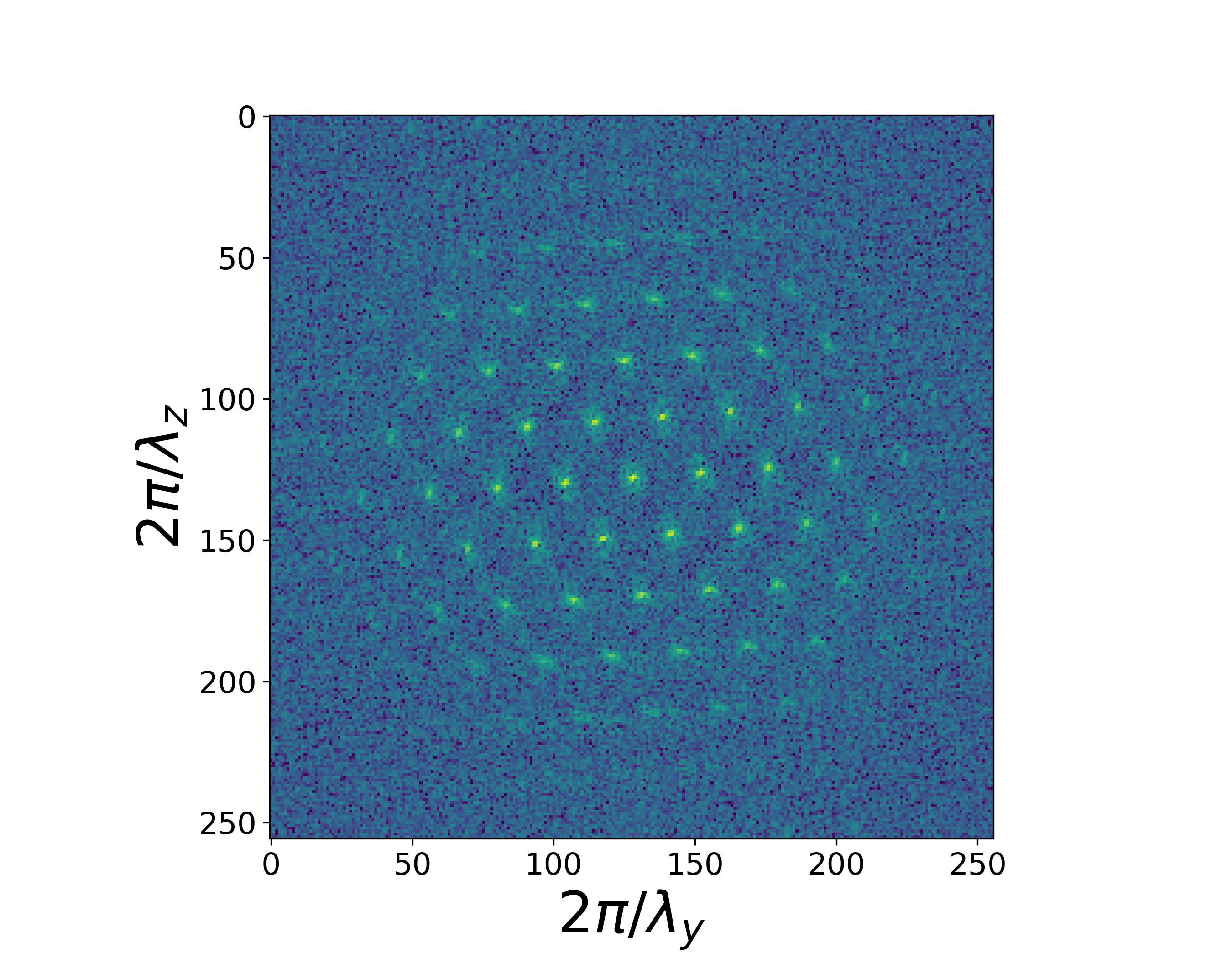}
        \label{fig:Fourier_Tr}
        } 
    \caption{(a) The crystalline labels generated by the application of GMM on AE output data at \(t_{end} = 74\) are divided into their respective crystal domains.
    Out of all the domains obtained \(\sim 1960\), ten are shown here.
    (b) Projection of the members of a crystal domain on a plane perpendicular to the dominant Eigenvector of its \(Q_{\alpha\beta}\) tensor (equation \ref{eqn:qtensor}).
    (c) Fourier transformation of that crystal domain.
    The wave-number, \(\frac{2 \pi}{\lambda} = 0.31\).}
\end{figure} 

\begin{figure}
    \centering
        \includegraphics[width=0.7\hsize]{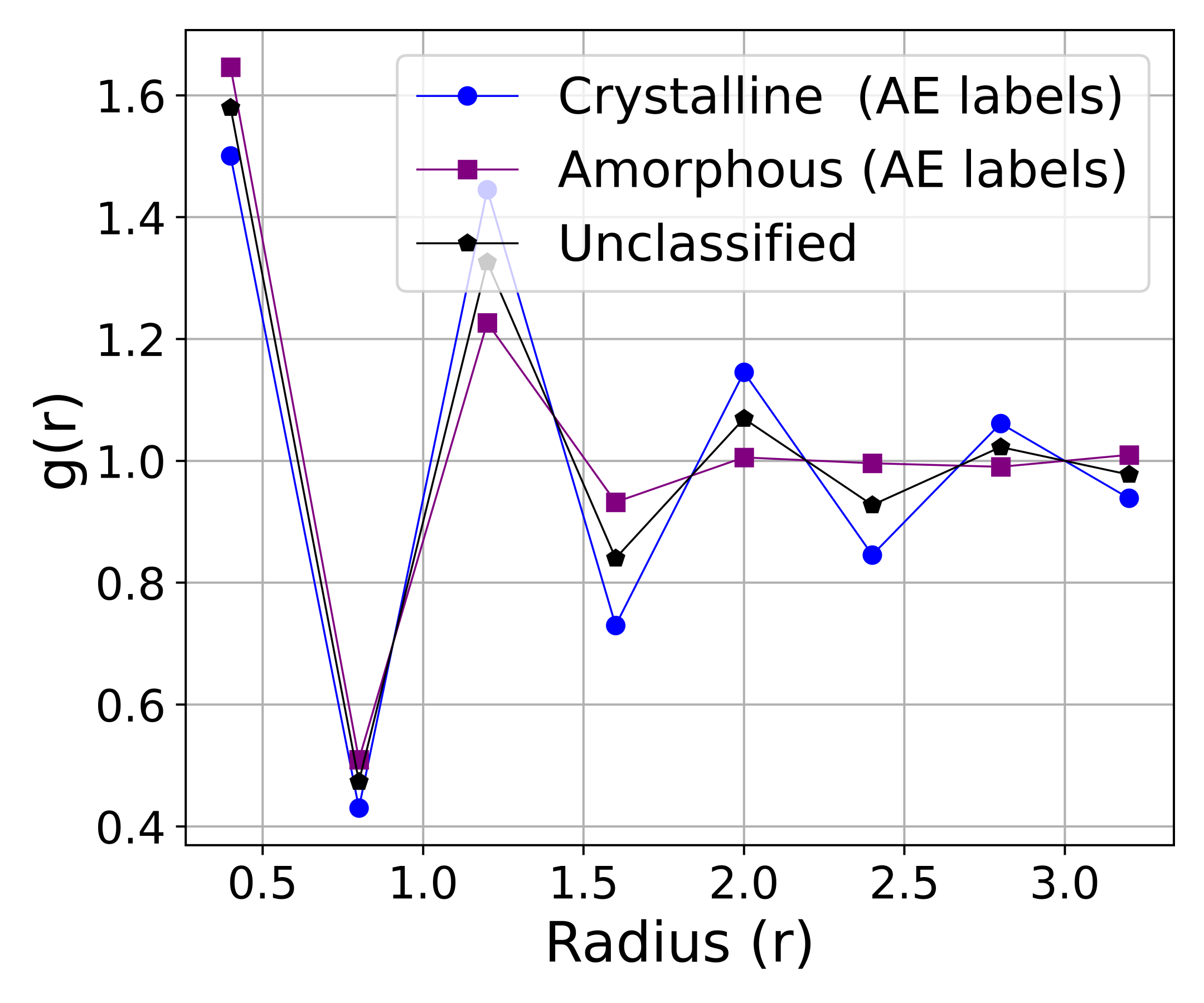}
        \caption{The radial distribution function g(r) shown here contains a bin size (thickness of shells) of \(\Delta r = 0.4\) (\(\widehat{=}~0.26 ~\mathrm{nm}\)).
        The maximum radius taken is \(R_{max} = 3.5\) (\(\widehat{=} ~ 1.82~\mathrm{nm}\)) around each monomer.
        The solid square line in purple represents the g(r) of amorphous labels, which exhibit quickly disappearing or less distinct peaks starting from the third one onwards.
        The solid circle line in blue shows the g(r) of crystalline labels with all the peaks comparatively sharper and the solid pentagon line in black contains a mix of both labels.
        Hence, its g(r) lies somewhere in between the two.}\label{fig:rdf}
\end{figure}

Once the crystal domains have been identified, we evaluate the orientation order of each of them by calculating the tensor order parameter,
\begin{equation} \label{eqn:qtensor}
    Q_{\alpha\beta} = \langle \frac{1}{2N} \sum_{i=1}^{N}(3b_{i\alpha} b_{i\beta} - \delta_{\alpha\beta}) \rangle
\end{equation}
where \(N \) = the number of monomers in a domain, \(b_{i\alpha}\) and \(b_{i\beta}\) = components of bond vectors of each member in a crystal domain.
Each of these tensors will give us a set of Eigenvalues and their respective Eigenvectors.
The Eigenvector corresponding to the most dominant Eigenvalue gives us the orientation direction of the crystal domain.
We project the coordinates of the members of that crystal domain on a plane perpendicular to the dominant Eigenvector.

Figure~\ref{fig:Diff_Clusters} shows a subset of the crystal domains, each with a different color.
The director-defined plane (equation~(\ref{eqn:qtensor})), located in 3 dimensions (x, y, z), is projected on either of the three planes to get a 2-dimensional histogram (figure\ref{fig:2d_hist}) of the target crystal domain.
A shifted fast Fourier transform (FFT) of this graph should give us a hexagonal structure factor in the reciprocal lattice space as expected in the case of polymeric crystalline structures (figure~\ref{fig:Fourier_Tr}).
Any crystal domain can be randomly drawn from the list to expect similar results.

\subsection{$\vec X_{conf}$}\label{sec:pca_conf}

\begin{figure}
    \centering
        \subfigure[]{
        \includegraphics[width=.44\hsize]{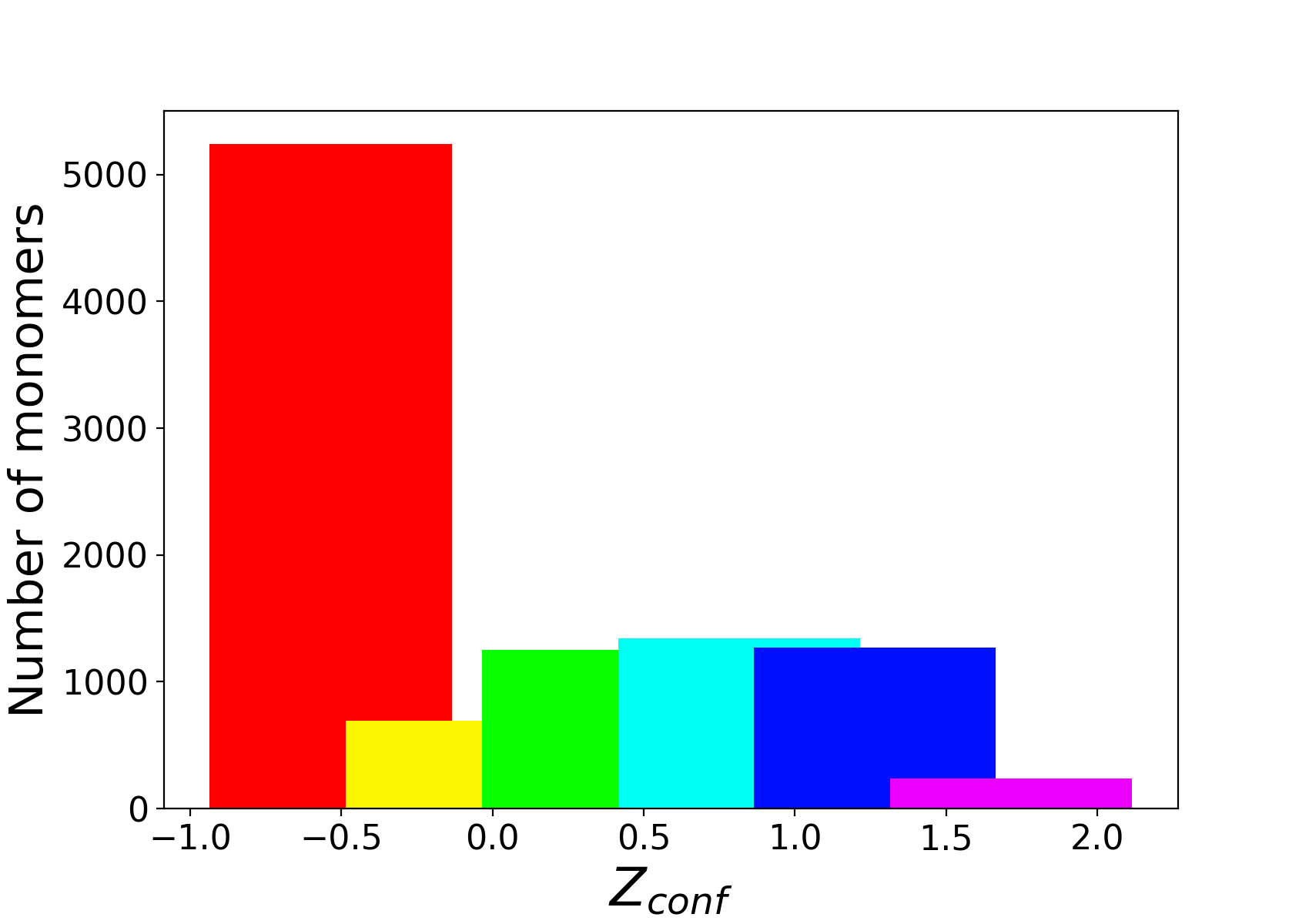}
        \label{fig:conf_hist}
        }
        \subfigure[]
        {
        \includegraphics[width=0.44\hsize]{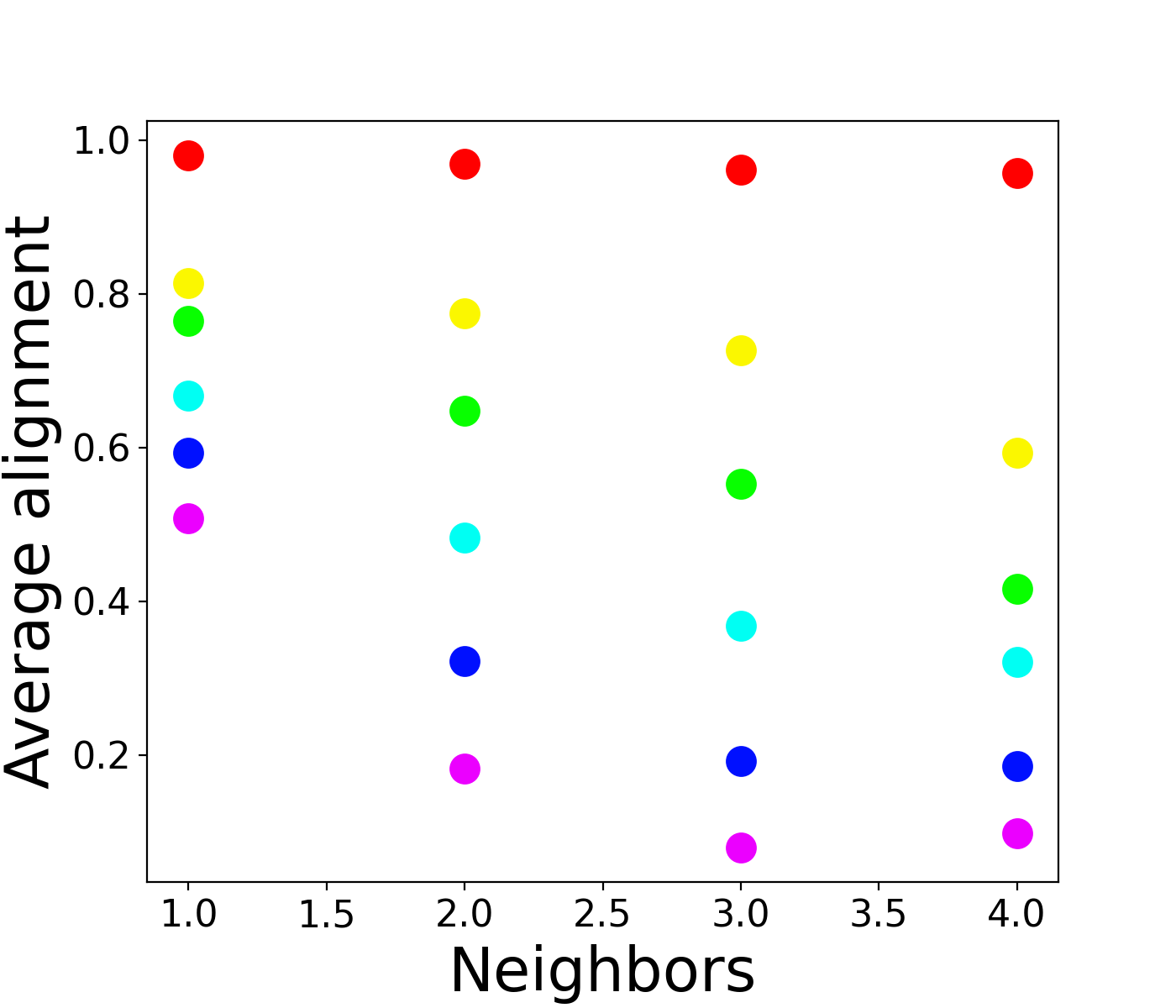}
        \label{fig:conf_neigh}
        }
    \caption{(a) A histogram representation of the six intervals within the range of compressed local conformation fingerprint, $\vec X_{conf}$, via principle component analysis (PCA), $Z_{conf}$.
    A color is assigned to each interval.
    (b) For monomers lying in every such interval, an average of their squared scalar products (equation \ref{eqn:Xconf}) with their first, second, third, and fourth neighbors are shown.
    Hence, the red color interval contains the most aligned monomers.}\label{fig:conf_order}
\end{figure}

The conformational information of the dataset that was chosen to train machine learning models is compressed to 1-dimensional information using principle component analysis (PCA).
The compressed information is referred to as $Z_{conf}$.
It is used to study the latent space generated by the AE as shown in figure~\ref{fig:conf_multi}.
For this, we create six equal partitions within the range of $Z_{conf}$ and associate a unique color to each of the partitions as shown in figure~\ref{fig:conf_hist}.
We assign these colors to the data points on the AE-generated point cloud to see the angular color composition of each of its branches.
For figure~\ref{fig:conf_neigh}, we take an average of the squared scalar products, as shown in equation \ref{eqn:Xconf}, of the coarse-grained monomers that fall in each of the six intervals.
The color code is maintained from figure~\ref{fig:conf_hist}.
The average is taken separately for the first, second, third, and fourth neighbors along the chain conformation, corresponding to contour distance along the chain.
One can see in figure~\label{fig:conf_neigh} that the red color in figure~\ref{fig:conf_hist} represents the most aligned aligned group.

We tested the effect of augmentation of fingerprints at the chain end by creating the same histogram~\ref{fig:conf_hist} while skipping the monomers with augmented fingerprints.
The resulting relative changes of bin heights were $-0.11,-0.31,-0.03, +0.27, +0.27, +0.27$ given in $\%$ in the order of increasing value of $Z_{conf}$.
Thus, when chain ends are disregarded, there appear to be fewer fingerprints associated with ordered conformations (lowest $Z_{conf}$) and more fingerprints associated with disordered conformations (highest $Z_{conf}$).

\begin{figure}[!ht]
    \centering
    \subfigure[]{
        \includegraphics[width=0.31\hsize]{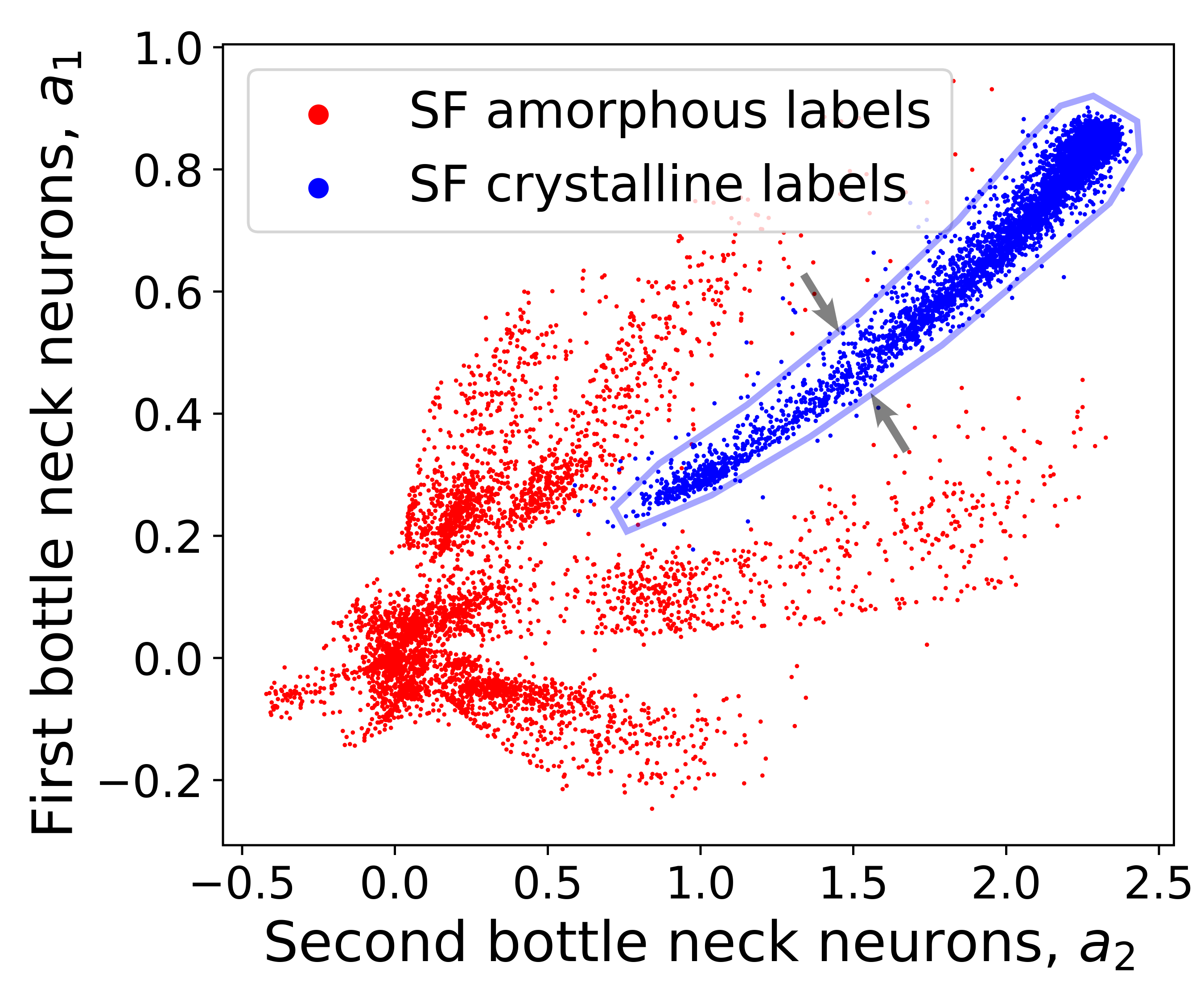}
        \label{fig:SF_PC_74}
        }
    \subfigure[]{
        \includegraphics[width=0.31\hsize]{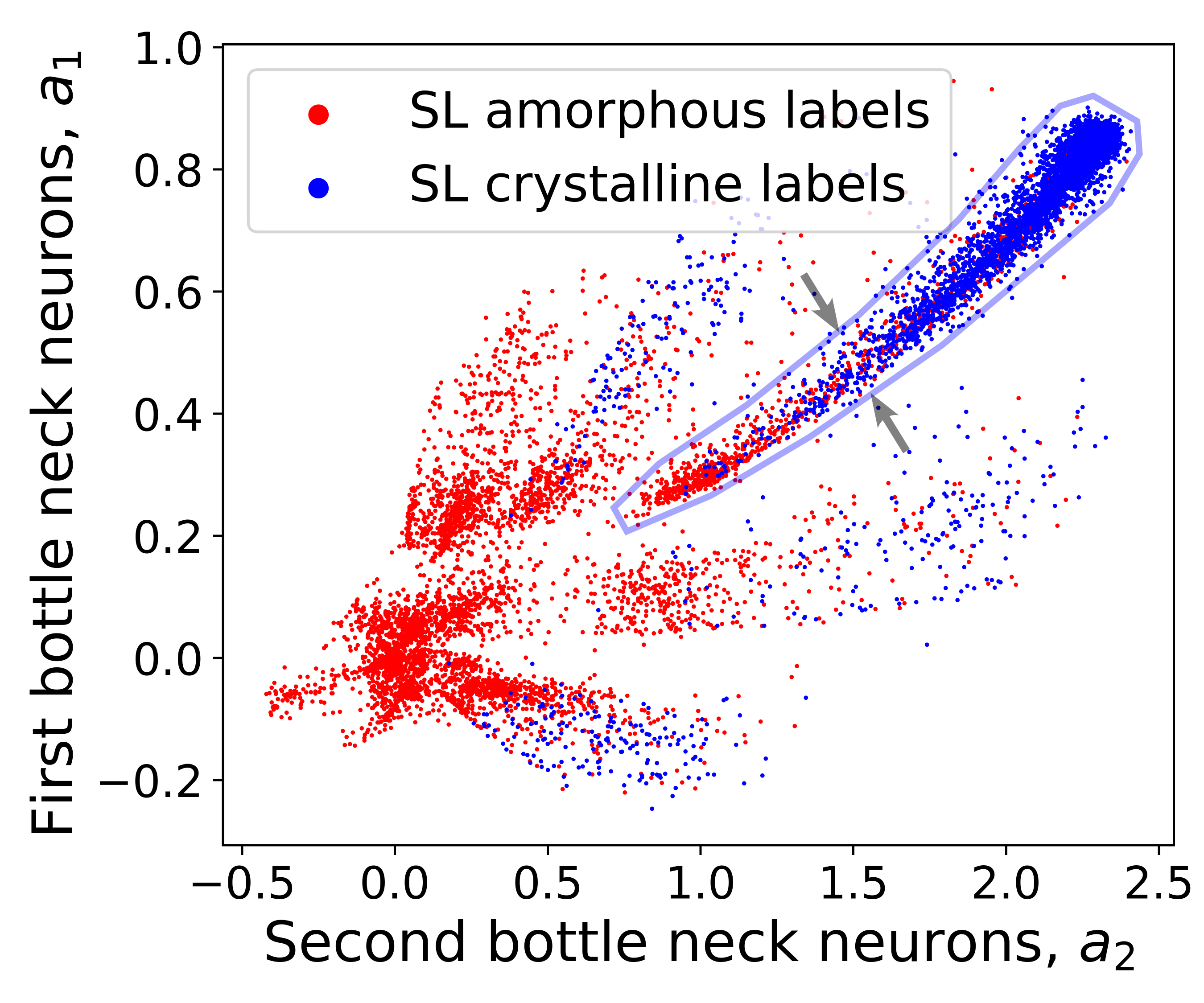}
        \label{fig:SL_PC_74}
        }
    \subfigure[]{
        \includegraphics[width=0.31\hsize]{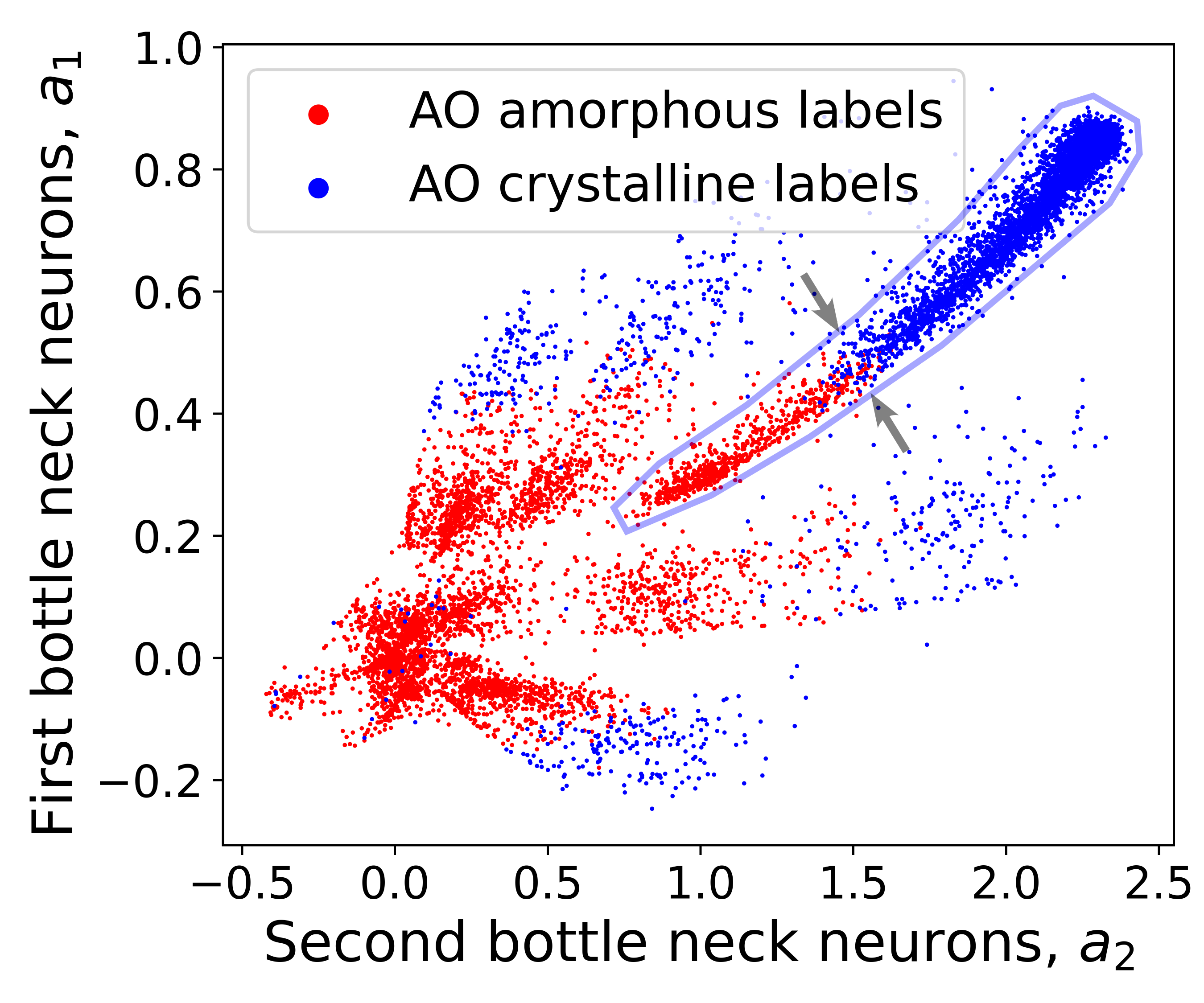}
        \label{fig:AO_PC_74}
        }
    \caption{
    The two phases "crystalline" and "amorphous" according to \textit{SF} (a), \textit{SL} (b), and \textit{AO} (c) \textit{labels} are assigned to the AE-encoded fingerprints $\vec X$ in latent space, $\vec a=\{a_1,a_2\}$ in analogy to figure~\ref{fig:AE_PC_74} for comparison.}\label{fig:PC_74}
\end{figure}

\begin{figure}[!ht]
    \subfigure[]{
        \includegraphics[width=0.31\hsize]{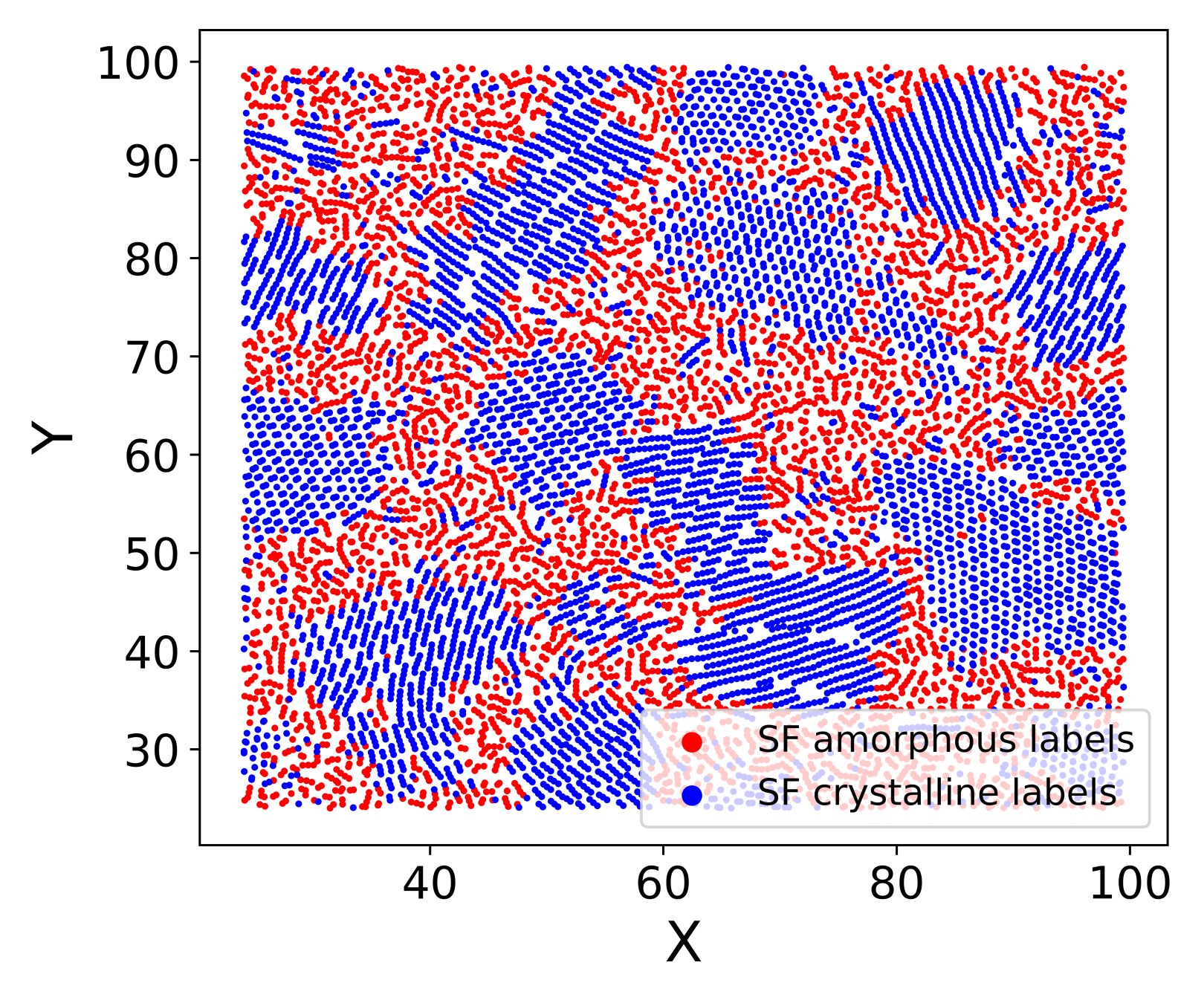}
        \label{fig:SF_Sl_74}
        }
    \subfigure[]{
        \includegraphics[width=0.31\hsize]{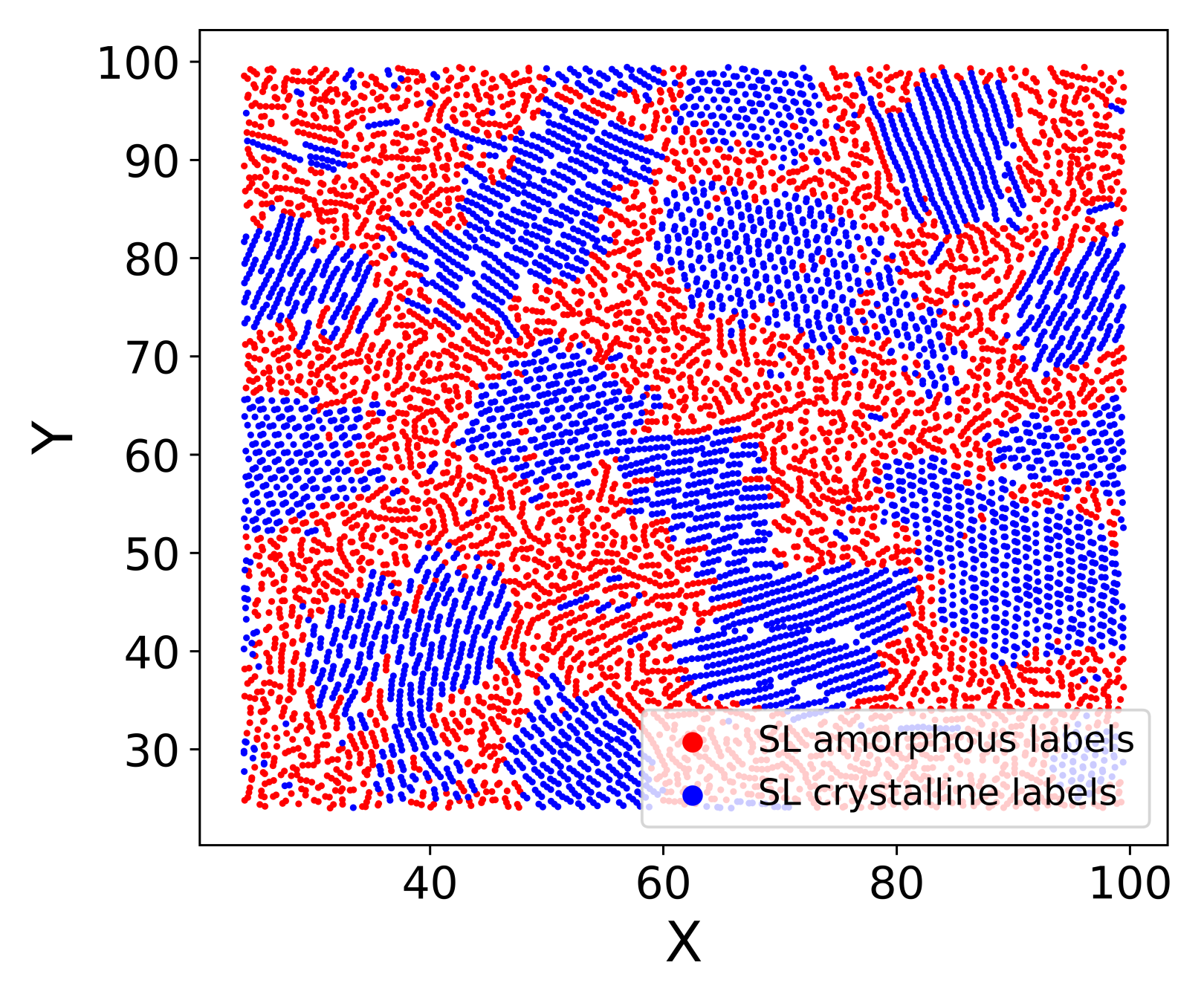}
        \label{fig:SL_Sl_74}
        }
    \subfigure[]{
        \includegraphics[width=0.31\hsize]{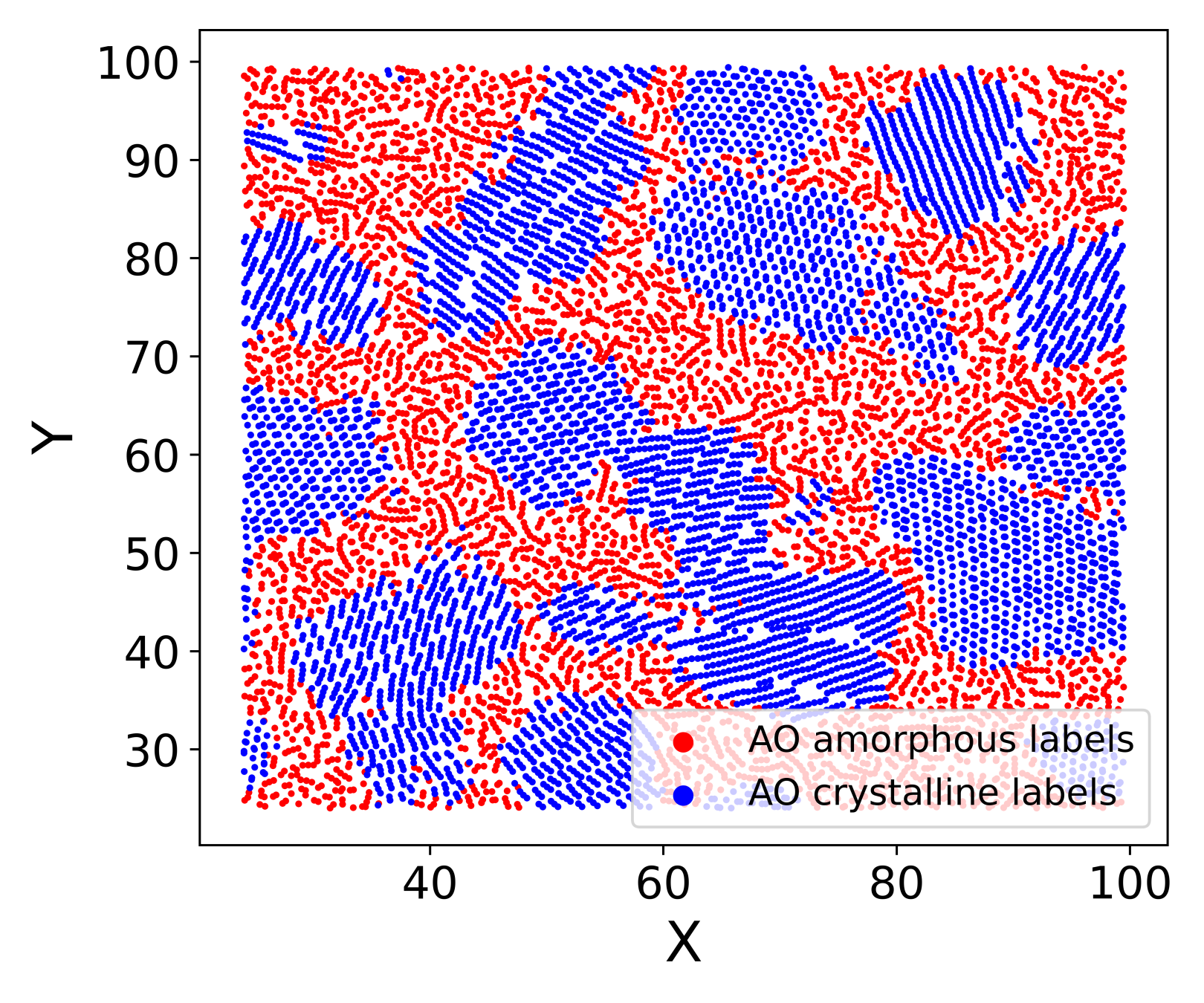}
        \label{fig:AO_Sl_74}
        }
    \caption{The above figures contain 50th cross-section from the simulation box at $t_{end} = 74$.
    (a) The color labels created by applying the Gaussian mixture model (GMM) on the concatenated list of structural fingerprints (SF) are used here.
    (b) The color labels created by applying HC on the latent space (AE) data are used here.
    (c) The color labels created by manually building a boundary in stem lengths (SL) are used here.
    (d) The color labels created by manually building a boundary in the average orientation of monomers with the surroundings (AO) are used here.}
\end{figure}

The three labelling schemes \textit{SF labels} (figure~\ref{fig:SF_PC_74}), \textit{SL labels} (figure~\ref{fig:SL_PC_74}), and \textit{AO labels} (figure~\ref{fig:AO_PC_74}) are presented to compare with \textit{AE labels} shown in figure~\ref{fig:AE_PC_74}.

For \textit{SF labels} (figure~\ref{fig:SF_PC_74}), we see near the center a relatively dense protruding branch-like structure, which we attribute as crystalline (colored blue), the rest of the points are more spread and we attribute this class as amorphous (colored red).
The dense blue \textit{SF} region is relatively narrow as compared to the whole latent space map supporting the association of this region with the ordered phase.
Let us refer to this region as the "\textit{main branch}".
The region is encircled identically in all graphs of figure~\ref{fig:PC_74}).
Since only ~50 of ~5000 \textit{SF} crystalline points are found outside the \textit{main branch}, we use the \textit{SF} crystalline group as a reference to interpret the other labels (\textit{AE, SL, AO}) meaning, according to their respective abundance in the \textit{main branch}.
We see in figures~\ref{fig:AE_PC_74}, ~\ref{fig:SL_PC_74} and ~\ref{fig:AO_PC_74} that the above argument seems less valid for the other labeling schemes when presented in latent space because their "crystalline" labels show fractions of points outside the narrow \textit{main branch} region.
However, in the case of \textit{AE labels}, $\approx 100\%$ of crystalline labels are located within the \textit{main branch} (table~\ref{tab:comparison}), and in the case of \textit{SL} and \textit{AO}, a fraction of $\approx 89\%$ and $\approx 87\%$, respectively.

The results of manually creating classification boundaries as in the case of \textit{SL} and \textit{AO labels} are slightly different from \textit{SF} and \textit{AE labels}.
When we display these labels in the latent space point cloud, it can help us better understand the AE-based latent space.
For example, in figure~\ref{fig:AE_PC_74}, the \textit{main branch} is split conserving the crystallinity only to its far end tip for \textit{AE labels}.
The HC performed on the SF is based on different comparative distances compared to the arrangement of points in the latent space generated by the AE.
We observe that the resulting \textit{AE labels} have almost all of the indices common with \textit{SF labels} which is approximately \(100\%\).
On the other hand, \textit{SF labels} have only \(0.84\) indices in common with \textit{AE labels}.
This proves that the SF-based clustering via HC gives us crystalline labels in addition to the ones given by AE data.
Since the \textit{SF labels} are highly concentrated within the \textit{main branch}, the boundary between 'amorphous' and 'crystalline' within the \textit{main branch} in figure~\ref{fig:SF_PC_74} is critical.
This boundary significantly contributes to the reduced match of $0.84$ in table~\ref{tab:comparison} for the \textit{SF labels} compared to the \textit{AE labels}.
In the case of \textit{AO labels}, a similar boundary exists within the \textit{main branch} between regions that are predominantly amorphous and crystalline.
These regions are positioned near the reference points drawn for \textit{AE labels}, which are consistently placed at identical positions in all the graphs of figure~\ref{fig:PC_74} for comparison.

Vice versa, the \textit{SL labels} show a much more scattered distribution within the \textit{main branch}, and throughout the whole latent space as compared to \textit{AO labels}.
For \textit{SL labels}, the red points seem mixed with the blue ones almost everywhere and one can not see an exclusive blue color-only area.
This observation points to a rather non-local property captured by the SL condition of $15$ consecutive monomers, whereas the fingerprint is resolved on the monomer level.
Likewise, there is a relatively scattered boundary between the crystalline and amorphous \textit{SL labels} within the \textit{main branch} (figure~\ref{fig:SL_PC_74}) as compared to the case of \textit{AO labels} (figure~\ref{fig:AO_PC_74}).
Also, we can see in figure~\ref{fig:SL_PC_74} that the majority of the blue (crystalline) points lie near the tip of the \textit{main branch} area.
We find that \textit{AE labels} have a fraction of \(0.91\) common crystalline monomers with \textit{SL labels} table~\ref{tab:comparison}).
This indicates that the encoded information functions in a similar way to the stem length classification scheme when identifying the crystallinity fraction.
This is the case despite the larger fluctuations in the latent space, which occur due to the less specific association with a particular monomer index (latent space point).

\end{document}